\documentclass[a4paper, 11pt]{article}
\pdfoutput=1

\usepackage{jheppub}
\usepackage{subcaption}
\usepackage{graphicx}
\usepackage{xcolor}
\usepackage{amsmath,amssymb}
\usepackage{mathtools}

\title{First Saturation Correction in High Energy {Proton-Nucleus} Collisions: I.  Time evolution of classical Yang-Mills fields beyond  leading order}
\author{Ming Li}
\author{and Vladimir V. Skokov}
\affiliation{Department of Physics, North Carolina State University, Raleigh, NC 27695, USA}

\emailAdd{mli48@ncsu.edu}
\emailAdd{vskokov@ncsu.edu}

\abstract{
In high energy {proton-nucleus} collisions, the single- and double-inclusive soft gluon productions at the leading order have been calculated and phenomenologically studied in various approaches for many years. These studies do not take into account  the saturation  and multiple rescatterings in the field of the proton. The first saturation correction to these leading order results (the  terms that are enhanced by the combination $\alpha_s^2 \mu^2$, where $\mu^2$ is the proton's color charge squared per unit transverse area)  has not been completely derived despite recent attempts using a diagrammatic approach.  This paper is the first in  a series of papers towards analytically completing the first saturation correction to physical observables in high energy {proton-nucleus} collisions. Our approach is 
to analytically solve the classical Yang-Mills equations in the dilute-dense regime using the Color Glass Condensate effective theory and  compute physical observables constructed from classical gluon fields. In the current paper, the Yang-Mills equations are solved perturbatively in the field of the dilute object  (the proton). Next-to-leading order and next-to-next-to-leading order analytic solutions are explicitly constructed.  A systematic way to obtain all higher order analytic solutions is outlined.
}

\begin{document}
\maketitle
\flushbottom

\section{Introduction}

The two decades worth of experimental measurements at RHIC and, then, the LHC have provided many unexpected
results, including strong evidence for the formation of a strongly coupled
plasma of quarks and gluons in heavy-ion collisions at high energy. This plasma demonstrated
properties of a nearly perfect fluid; this fact facilitated a theoretical description of the collision dynamics in the framework of hydrodynamics starting just about 1~fm/c after the heavy ion impact (see Refs.~\cite{Gale:2013da,Heinz:2013th,Schenke:2019pmk}  and references therein).

The success of the hydrodynamic description, however, cannot be complete without a detailed
understanding of the initial non-equilibrium state. The properties of this state go beyond
the range of applicability of hydrodynamics but are crucial in fitting experimental data; the evolution of this state towards
equilibrated thermal nearly perfect liquid have been one of the open theoretical problems being extensively studied \cite{Baier:2000sb,Kurkela:2015qoa,Kurkela:2018wud, Wu:2017rry, Kovchegov:2017way}. One dominant mechanism
describing the initial phase is based on the saturation framework \cite{Krasnitz:1999wc,Krasnitz:2000gz,  Schenke:2012wb,Gale:2012rq}, also widely known as the
Color Glass Condensate (CGC). According to the framework, the high energy particle production
and scattering processes are dominated by the classical gluon fields providing a background for
systematic weak-coupling computation of quantum correction on top of it.

Under laboratory conditions, collisions of heavy-ions create probably the most optimal environment
for probing quark-gluon plasma near equilibrium, but at the same time they are poorly
suited to study the initial state particle production. This is because most of the observables in
heavy-ion collisions are sensitive not only to initial state, but also to rather strong final sate interactions \cite{Dusling:2015gta,Nagle:2018nvi}. However,
to uniquely map the transport properties of the plasma, it is critical to extract information
about the initial state in collisions where the final state is better understood and the initial state
is expected to play the dominant role. This necessitates probing a nucleus and a nucleon with
the smallest projectiles: proton and ultimately electron. Theoretically, a controlled, first principle
description of such asymmetric collisions (e-A, p-A, heavy-light nuclei) is not as complex as A-A collisions.

In the CGC framework, the key building block of soft gluon production in hadronic collisions is the single inclusive gluon cross section for a fixed configuration of the valence charges. Then 
the multi-gluon productions can be constructed from it iteratively. Analytical calculations of the single inclusive gluon production in asymmetric hadronic  collisions at {\it leading order} in the color charge density of the dilute projectile (e.g. proton) have been done by various groups for more than two decades \cite{Kovchegov:1998bi, Kopeliovich:1998nw, Kovner:2001vi, Dumitru:2001ux,Blaizot:2004wu}. The leading order result takes into account the multiple rescatterings/saturation in the dense nucleus (target) to all orders while treats the collision partner (projectile) as dilute object.    Beyond the leading order result, the first saturation corrections in the projectile to single- and double-inclusive gluons production  were {\it partially}  calculated in Refs.~\cite{Balitsky:2004rr, Chirilli:2015tea, McLerran:2016snu}.  These  incomplete  results were sufficient to convincingly demonstrate that, in the CGC framework,  the first saturation corrections are responsible for the generation of the odd azimuthal anisotropy~\cite{McLerran:2016snu,Kovchegov:2018jun} which was  missing  at the leading order. 
 
 In order to calculate the first saturation corrections to the single gluon production amplitude, the authors of Ref.~\cite{Chirilli:2015tea} used the diagrammatic approach based on  the  light-cone perturbation theory and the eikonal approximation. Reference~\cite{Chirilli:2015tea}  provides a result for  the order-$g^3$ single gluon production amplitude; the order-$g^5$ gluon production amplitude was not evaluated but it is needed to establish the complete first saturation corrections to the single inclusive gluon production in high energy proton-nucleus  collisions.  
 An alternative computational  approach was adopted in Ref.~\cite{McLerran:2016snu}. The authors of  Ref.~\cite{McLerran:2016snu} solve the classical Yang-Mills equations in the dilute-dense regime considering the projectile charge density as parametrically small.  Particle production is then constructed from the classical gluon fields using the Lehmann-Symanzik-Zimmermann (LSZ) reduction formula. Although this approach was proven to be more powerful in helping organize calculations and extract the odd azimuthal anisotropy of double-inclusive gluon production, the authors of Ref.~\cite{McLerran:2016snu} also only computed  order-$g^3$ production amplitude. The goal of this series of papers is to systematically  complete  the effort started in Ref.~\cite{McLerran:2016snu} and more specifically: 1) to solve for the next-to-leading order solutions of the classical Yang-Mills equations; 2) calculate the order-$g^5$ gluon production amplitude; 3) complete the first saturation corrections to the single- and double- inclusive gluon productions; 4) evaluate the early time-dependence of the energy-momentum tensor $\langle T_{\mu\nu}(\tau, x) \rangle $ and its correlation  $\langle T_{\mu\nu}(\tau, x)  T_{\mu'\nu'}(\tau', x') \rangle $.   
 
 Before proceeding with solving the classical Yang-Mills equations,  we want to outline the role of the saturation corrections  in high energy nuclear collisions. 
General perturbative corrections are terms expressed as power series expansions in the strong coupling constant $g$. In high energy nuclear collisions, the colliding hadrons are highly Lorentz contracted and the number of ``valence'' color sources per unit area as a random variable scales as $N_{\perp} \sim \sqrt{A^{1/3}}$ with $A$ the nuclear atomic number for a nucleus. For large $A$,  at each order $g^n$ ($n\geq 1$), there are terms that are enhanced by $N_{\perp}$. The most enhanced term at each perturbative order is the saturation correction we aim to compute.

\begin{figure}[t]
\centering 
\includegraphics[scale = 0.5]{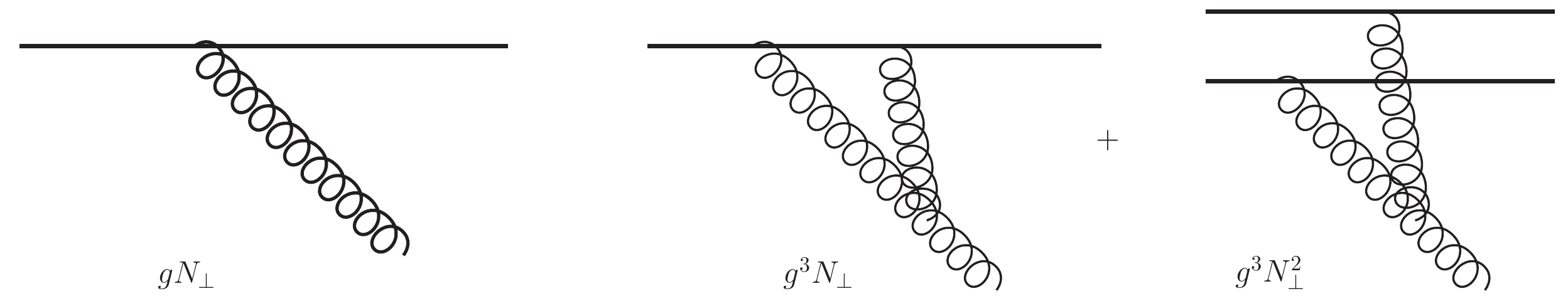}
\caption{Schematic diagrams showing perturbative corrections to small-$x$ gluon distribution at order $g$ and order $g^3$. Saturation correction at order $g^3$ only takes into account the type of diagrams which are parametrically proportional to $g^3N_{\perp}^2$. }
\label{fig:pert_g}
\end{figure}

\begin{figure}
\centering 
\includegraphics[scale = 0.5]{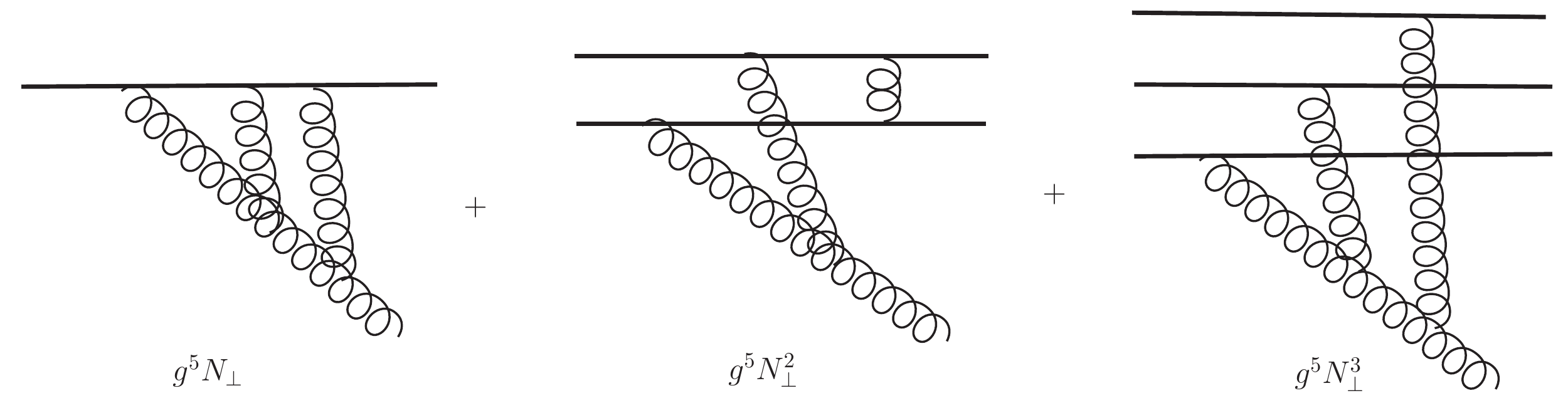}
\caption{Schematic diagrams showing perturbative corrections to small-$x$ gluon distribution at order $g^5$. Saturation correction at order $g^5$ only considers the types of diagrams proportional to $g^5N_{\perp}^3$. }
\label{fig:pert_g3}
\end{figure}

In order to illuminate the meaning of the saturation correction term, we will use small-x gluon distribution of a high energy proton as an example.  The amplitude at order $g$ and $g^3$ are schematically shown in Fig.~\ref{fig:pert_g}. The order-$g^5$ amplitude is illustrated  in Fig.~\ref{fig:pert_g3}.  We start our discussions with the leading order. At order $g$, only one of the color source radiates a gluon, but there are $N_{\perp}$ color sources, thus the amplitude is proportional to $gN_{\perp}$.  The number of produced gluons is proportional to the amplitude squared which is parametrically $ g^2N^2_{\perp} \sim \alpha_s A^{1/3}$.
 At order $g^3$, there are two types of diagrams, see Fig. \ref{fig:pert_g}. For the first kind, the gluon is radiated  from one single color source; in this case, the amplitude is proportional to $g^3 N_{\perp}$. For  the second kind, the gluon originates from interaction of two color sources; in this case,  the amplitude is proportional to $g^3 N_{\perp}^2$.  The saturation correction only takes into account diagrams proportional to $g^3N_{\perp}^2$, as they are most enhanced by the nuclear effects. For this term, the amplitude square is proportional to $g^6N_{\perp}^4 \sim \alpha_s^3 A^{2/3}$. Compared to the leading order contribution, it is higher order in $\alpha_s^2 A^{1/3}$, which is what we have alluded to  as the saturation correction. 
Now, at order $g^5$, there are three types of diagrams,  see Fig.~\ref{fig:pert_g3}. The first type of diagrams only involve one color source to radiate a gluon. The amplitude is proportional to $g^5N_{\perp}$.  The second type comes with two color sources radiating gluons. Its amplitude is proportional to $g^5N_{\perp}^2$.  And, finally, the last type of diagrams with  three color sources emitting gluons leads to the amplitude proportional to $g^5N_{\perp}^3$.  At order $g^5$, the saturation correction only takes into account the diagram that is proportional to $g^5N_{\perp}^3$.  For this term, the amplitude squared is proportional to $g^{10}N_{\perp}^6 \sim \alpha_s^5 A$. Compared to the leading order term, it is parametrically higher order in $(\alpha_s^2A^{1/3})^2$, which is the second order in terms of the  saturation correction. 

The above discussions can only be formally applied to a large nucleus  $A \gg 1$ ensuring that the saturation corrections are leading compared to the other perturbative contributions. Superficially, in case of proton-nucleus collisions, the saturation corrections should not play any special role, since the nuclear atomic number for proton is $A=1$. However, this is not completely right. At high energy, the number of color sources could still be large for at least two  reasons. First, the proton wave-function can be in a rare configuration at the moment of collisions. The configurations like this are believed to be responsible for the high multiplicity events observed in high energy pA collisions in the experiments at the LHC and the RHIC.  Second, the high energy evolution equations (BK and B-JIMWLK, see Refs.~\cite{Balitsky:1995ub,
  Balitsky:1998ya, Balitsky:2001re, Jalilian-Marian:1997jx,
  Jalilian-Marian:1997gr, JalilianMarian:1997dw, Iancu:2001ad,
  Iancu:2000hn, Ferreiro:2001qy, Weigert:2000gi}) predict proliferation of the color charges; this ultimately leads to a universal high energy fixed point at which all hadrons look alike.  To incorporate this general situation, it is more appropriate to reformulate the about counting in terms of the saturation scale $Q_s$ of the projectile instead of the  nuclear atomic number. 
 
In the CGC framework \cite{McLerran:1993ni, McLerran:1993ka}, the color sources responsible for gluon radiations are characterized by the random color charge density $\rho^a(x^-, \mathbf{x})$, which represents ``valence'' partonic degrees of freedom. Specifically, in the   McLerran-Venugopalan model, the color charge density is assumed to follow the Gaussian distribution with width $\mu(x^-, \mathbf{x}) $.
The longitudinally integrated Gaussian width $\mu^2(\mathbf{x}) = \int dx^- \mu^2(x^-, \mathbf{x})$ has the physical meaning of color charge squared per unit transverse area. It represents the Gaussian width of the random variable $\rho^a(\mathbf{x}) =  \int dx^-  \rho^a(x^-, \mathbf{x}) $. 
 As a random variable, the characteristic scale of the color charge density is  $\rho^a(\mathbf{x}) \sim \sqrt{A^{1/3}}$. On the other hand, the saturation scale is shown to be related to $\mu^2$ by $Q_s^2 \sim (g^2 \mu )^2 \sim \alpha_s^2 A^{1/3}$ \cite{Lappi:2007ku}. This highlights the fact that the saturation corrections represent an expansion in terms of the projectile's saturation scale squared. It is not surprising that the power counting is captured exactly by solving the  Yang-Mills equations for the classical gluon fields; this was explicitly  demonstrated by Kovchegov~\cite{Kovchegov:1997pc}. 

To sum up, the single inclusive soft gluon productions in high energy nuclear collisions can be parameterized as a double Taylor series expansion of the saturation scales of the projectile and the target, $Q_{s,P}$ and $Q_{s,T}$, respectively \cite{Chirilli:2015tea, Kovchegov:2018jun, Schlichting:2019bvy}.
Using the notation of Ref.~ \cite{Chirilli:2015tea}, 
\begin{equation}\label{eq:doubleExpansionQs}
\frac{dN}{d^2\mathbf{k} dy}=\frac{1}{\alpha_s}  f\left(\frac{Q_{s,P}^2}{k_{\perp}^2}, \frac{Q_{s,T}^2}{k_{\perp}^2}\right) =\frac{1}{\alpha_s} \sum_{n,m} c_{n,m} \left(\frac{Q_{s,P}^2}{k_{\perp}^2}\right)^n\left(\frac{Q_{s,T}^2}{k_{\perp}^2}\right)^m.
\end{equation}
In the case of a dilute projectile and a dense target, resummation over the target saturation corrections is possible and Eq.~\eqref{eq:doubleExpansionQs} can be written as an expansion in the projectile saturation momentum 
\begin{equation}
\frac{dN}{d^2\mathbf{k} dy}\Big|_{pA}=\frac{1}{\alpha_s} \sum_{n=1}^{\infty}  \left(\frac{Q_{s,P}^2}{k_{\perp}^2}\right)^n f_{n} \left(\frac{Q_{s,T}^2}{k_{\perp}^2}\right).
\end{equation}
The leading order result $f_1(Q_{s,T}^2/k_{\perp}^2)$ is known, see Refs.~\cite{Kovchegov:1998bi,Dumitru:2001ux}.  
Corrections  $f_2$ and $f_{i>2}$ are not known analytically at present.

In case of double-inclusive gluon production, we have, schematically
\begin{equation}
\frac{dN}{d^2\mathbf{k_1} dy_1 d^2\mathbf{k_2} dy_2}\Big|_{pA}=\frac{1}{\alpha_s^2} \sum_{n=1}^{\infty}  \left(\frac{Q_{s,P}^2}{k_{\perp}^2}\right)^{2n} h_{n} \left(\frac{Q_{s,T}^2}{k_{\perp}^2}\right).
\end{equation}
Here for simplicity we consider $|\mathbf{k_1}| = |\mathbf{k_2}|  = k_\perp$. 
The leading order result, $h_1$, was derived in Refs.~\cite{Kovner:2012jm,Kovchegov:2012nd}. The first saturation correction was computed partially -- only the odd component under the transformation  ${\mathbf{k_1}} \to -{\mathbf{k_1}}$  was extracted in Ref.~\cite{McLerran:2016snu,Kovchegov:2018jun}.     

The goal of this series of papers is to compute the complete first saturation corrections, that is $f_2(Q_{s,T}^2/k_{\perp}^2)$ and $h_2(Q_{s,T}^2/k_{\perp}^2)$.  

There are several reasons why saturation corrections are important. On the practical side,
leading order result does not include any final state interactions of the produced gluons. It basically assume the gluons propagate freely once created at proper time $\tau=0$. This might be a reasonable assumption for a dilute system created. The first saturation correction introduces non-trivial  gluon interactions through three-gluon and four-gluon vertices. These interactions might be  responsible for the onset of isotropization, thermalization and hydrodynamization.  Additionally, having an expression for the first saturation corrections provides direct comparison of the relative importance of the initial state vs. final state effects. Furthermore, as alluded before, the most important feature of the final state effects is the generation of odd harmonics in multigluon distributions.  Finally,  the first saturation correction allows one to estimate the role of higher order contributions in the dilute-dense approximation.  In all these cases, the saturation corrections are indispensable for any attempts to compare theory with experimental data. 


On the academic side, rigorously calculating the first saturation correction is a first  step towards including all  saturation corrections. The ultimate goal is to resum all order saturation corrections and thus solve the dense-dense scattering problem analytically \cite{Balitsky:2005we, Hatta:2005rn, Kovner:2020exf}. This is one of the unsolved  problems in high energy QCD.

 It should be mentioned that  the classical Yang-Mills equations can be  and were solved numerically. This approach  was used for calculating the single- and double-inclusive gluon productions to all orders in both projectile and target color charge densities~\cite{Krasnitz:2002mn, Lappi:2003bi, Schenke:2015aqa}. These calculations however rely on truncating the final state interactions at proper finite time.

The paper is organized as follows. After a brief introduction of the CGC framework and the classical Yang-Mills equations in sec. \ref{sec:cgc_review}, we discuss the initial conditions in sec. \ref{sec:expand_initial_conditions}. This includes explicit derivation of high order expansions; we also discuss a few convenient forms of different gauge fixings. The subsequent sections solve the classical Yang-Mills equations at orders $g$, $g^3$ and $g^5$ using the method of variation of parameters and Garf's formula for Bessel functions. For the discussions in sec. \ref{sec:discussion_outlooks}, we review which phsyical quantities can be obtained using our results. 

\section{The Color Glass Condensate Effective Theory}
\label{sec:cgc_review}
The Color Glass Condensate (CGC) effective theory concerns quantum chromodynamics in the hight energy limit \cite{McLerran:2001sr, Iancu:2003xm, Kovchegov:2012mbw}. It is based on a formal separation of large  and small longitudinal momentum modes of partons  inside a hadron. 
The partonic degrees of freedom with large longitudinal momenta (large-x) are effectively described by the color charge density $\rho^a(x)$. The gluons with small longitudinal momentum (small-x) are the dominant degrees of freedom and they are characterized by the classical gluon fields $A^{\mu}(x)$. The color charge density is responsible for the production of the gluon fields through the classical Yang-Mills equations
\begin{equation}
D_{\mu} F^{\mu\nu} = J^{\nu}
\end{equation}
with the covariant derivative $D_{\mu} = \partial_{\mu} - ig A_{\mu}$ and the field strength tensor $F^{\mu\nu}=\partial^{\mu}A^{\nu} -\partial^{\nu}A^{\mu} -ig[A^{\mu}, A^{\nu}]$. For a  right-moving hadron at high energy, the color current $J^{\mu}(x)=\delta^{\mu +}\rho(x^-, \mathbf{x})$ is approximately independent of light-cone time $x^+$ as far as the dynamics of the small-x gluons is concerned. 

\begin{figure}[t]
\centering 
\includegraphics[scale = 0.5]{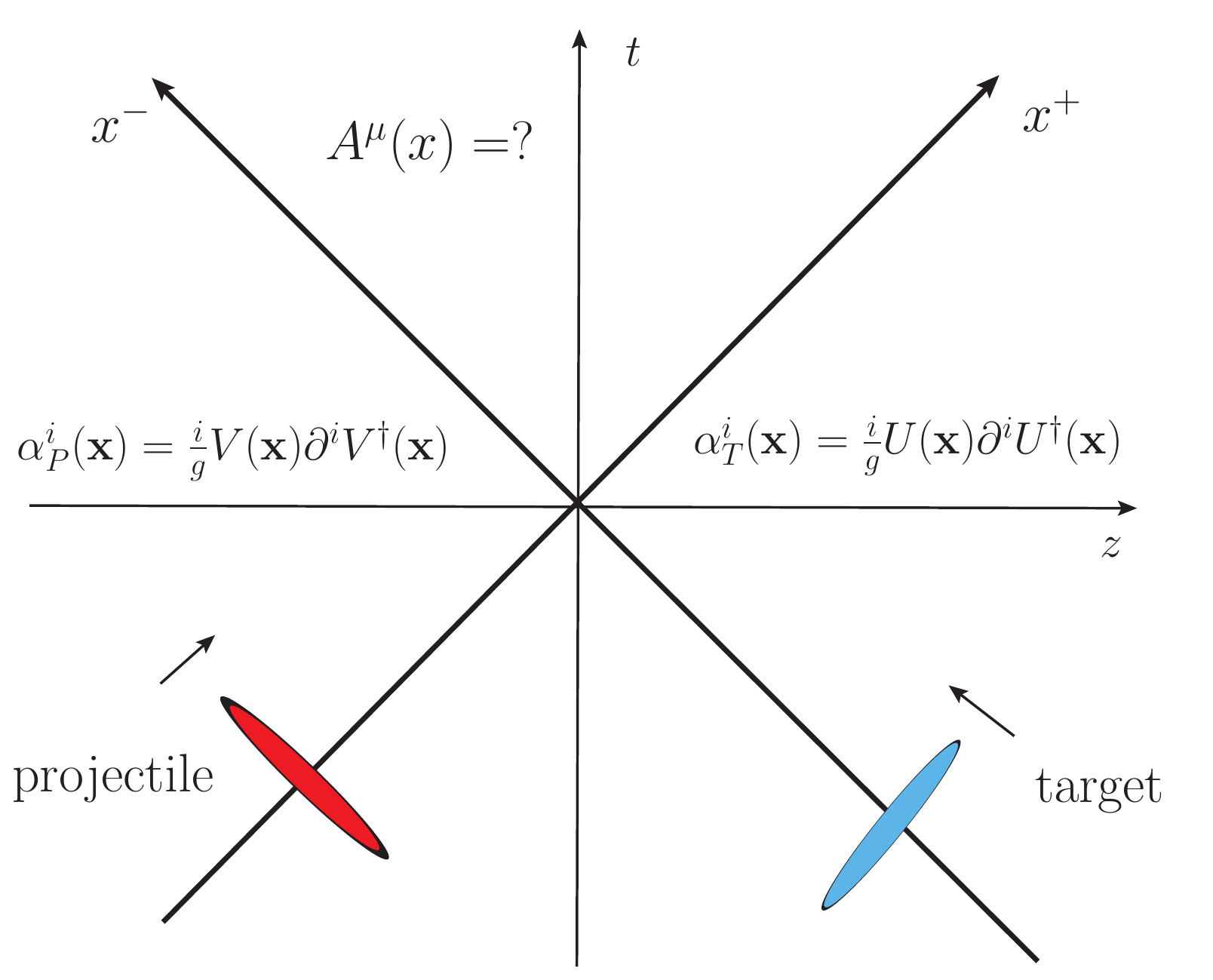}
\caption{Schematic diagram showing high energy nuclear collisions on the spacetime diagram. The Weizsacker-Williams fields of the projectile and targets live in the regions $x^->0, x^+<0$ and $x^- <0, x^+>0$ separately before the collisions. The collisions happen at $x^+=x^-=0$. The goal is to find out the classical gluon fields $A^{\mu}(x)$ produced in the region $x^+>0, x^->0$ after the collisions.  }
\label{fig:spacetime_collisions}
\end{figure}

 In applying CGC to high energy nuclear collisions with a right-moving projectile and a left-moving target, the color current can be approximated as $J^{\mu} = \delta^{\mu +} \rho_P(x^-, \mathbf{x}) + \delta^{\mu -} \rho_T(x^+, \mathbf{x})$. Before the collisions, two sheets of small-x gluons,  generated separately by the projectile and the target, approach each other at the speed of light, see Fig.~\ref{fig:spacetime_collisions}. The collisions happen instantaneously (high-energy approximation). After the collisions, the large-x color charges are still approximately traveling  along the lightcone $x^+=0$ and $x^-=0$ while classical gluon fields are produced in the forward lightcone $x^+>0, x^->0$. The dynamics of the produced gluon fields is governed  by the sourceless Yang-Mills equations. The initial conditions are crucial as they encode the information about the instantaneous collisions. Once this initial value problem for the classical Yang-Mills equations is solved, one can compute physical observables that depend on classical gluon fields. Eventually, through the initial conditions, any  physical observable will be a functional of the color charge densities of the projectile and the target $\mathcal{O}(\rho_P, \rho_T)$.  
 
 The event/initial configuration-averaged results are obtained by evaluating the average over projectile and target color charge densities separately. In the McLerran-Venugopalan model, the color charge densities are assumed to follow Gaussian distributions. Their two-point correlation functions are 
\begin{equation}\label{eq:rho_correlator}
\langle \rho_{P(T)}^a(x^{\mp}, \mathbf{x}) \rho_{P(T)}^b(y^{\mp}, \mathbf{y}) \rangle = \delta^{ab} \delta(x^{\mp}-y^{\mp})\delta^{(2)}(\mathbf{x}-\mathbf{y}) g^2\mu^2(x^{\mp}, \mathbf{x}).
\end{equation}

To solve the sourceless classical Yang-Mills equation in the forward light cone $x^+>0, x^->0$, we follow the literatures \cite{Kovner:1995ja, Dumitru:2001ux} and consider  
the Fock-Schwinger gauge $x^-A^+ + x^+ A^- =0$. In this gauge, the solutions can be parameterized as 
\begin{equation}
\begin{split}
&A^+ = A_- = x^+ \alpha(\tau, \mathbf{x})\,, \\
&A^- = A_+ = -x^- \alpha(\tau, \mathbf{x})\,, \\
&A^i = \alpha^i(\tau, \mathbf{x}).\\
\end{split}
\end{equation}
Note that boost-invariance is assumed so that the classical gluon fields are independent of the rapidity $\eta$.  The coordinate system used is denoted by $(\tau, \eta, \mathbf{x})$ with $\tau= \sqrt{2x^+x^-}$ and $\eta = \frac{1}{2} \ln \frac{x^+}{x^-}$.  In terms of $\alpha(\tau, \mathbf{x}), \alpha^i(\tau, \mathbf{x})$, the classical Yang-Mills equations become
\begin{equation}\label{eq:ym_alpha_alphai}
\begin{split}
& \partial_{\tau}^2 \alpha +\frac{3}{\tau} \partial_{\tau} \alpha  - [D_i, [D_i, \alpha]] =0\,, \\
&-ig[ \alpha, \tau\partial_{\tau} \alpha] + [D_i, \frac{1}{\tau} \partial_{\tau} \alpha_i]=0\,, \\
&\frac{1}{\tau}\partial_{\tau} \alpha_i +\partial^2_{\tau}\alpha_i -ig \tau^2 [ \alpha, [D_i, \alpha]-D_jF_{ji}=0 .\\
\end{split}
\end{equation}
The initial conditions were derived in Refs.~\cite{Kovner:1995ja, Gyulassy:1997vt}:
\begin{equation}\label{eq:initial_conditions}
\begin{split}
&\alpha(\tau=0,\mathbf{x}) = \frac{ig}{2} [\alpha_P^i(\mathbf{x}), \alpha_T^i(\mathbf{x})],\\
&\alpha^i(\tau=0, \mathbf{x}) = \alpha_P^i(\mathbf{x}) + \alpha_T^i(\mathbf{x}).\\
\end{split}
\end{equation}
Here $\alpha_P^i(\mathbf{x})$ and $\alpha_T^i(\mathbf{x})$ are the Weizsacker-Williams gluon fields of the projectile and target, respectively. The fields $\alpha_P^i(\mathbf{x})$ and $\alpha_T^i(\mathbf{x})$ are two dimensional pure gauge fields; they   depend on the transverse coordinate and can be parameterized using Wilson lines
\begin{equation}\label{eq:ww_fields_pt}
\alpha_P^i(\mathbf{x}) = \frac{i}{g} V(\mathbf{x})\partial^i V^{\dagger}(\mathbf{x}), \quad \alpha_T^i(\mathbf{x}) = \frac{i}{g} U(\mathbf{x})\partial^i U^{\dagger}(\mathbf{x})
\end{equation}
with $ V(\mathbf{x}) = \mathrm{exp}\{ ig \Phi_P(\mathbf{x})\}$ and $U(\mathbf{x})=\mathrm{exp}\{ ig \Phi_T(\mathbf{x})\}$.  The relation between $\Phi_{P(T)}$ and $\rho_{P(T)}$ will be explained in more details in the following sections. 


Eqs. \eqref{eq:ym_alpha_alphai}, \eqref{eq:initial_conditions} define an initial value problem for a set of second order partial differential equations. The goal of the paper is to solve the Yang-Mills equations in the dilute-dense regime relevant to high energy proton-nucleus collisions. In the case of proton-nucleus scatterings, the color charge density of the proton is parametrically small $\rho_P \sim g$  while the color charge density of the nucleus is dense $\rho_T \sim 1/g$. We proceed to  solve the classical Yang-Mills equations by expressing the gluon fields as power series expansions in terms of $\rho_P$. The Yang-Mills equations can then be solved order by order perturbatively. The dependence on $\rho_T$ is resummed to all orders through the Wilson line.  In the next section, to be consistent with the expansions of Yang-Mills equations, the initial conditions will also be expressed as power series expansions in $\rho_P$.  It is worth pointing out that the expansion here is not the same as the conventional expansion in the strong coupling constant $g$, as we only keep terms that are enhanced by $\rho_P$ at each order of perturbative expansion in $g$.

\section{Expanding the Initial Conditions}
\label{sec:expand_initial_conditions}
\subsection{Expanding the Weizsacker-Williams field}
From Eq. \eqref{eq:ww_fields_pt}, the field generated by the ``valence'' color charge densities of the proton, also known as the  Weizsacker-Williams  (WW) field, 
can be expanded as 
\begin{equation}\label{eq:alphai_Phi}
\begin{split}
\alpha_P^i(\mathbf{x}) =& \frac{i}{g} V(\mathbf{x})\partial^i V^{\dagger}(\mathbf{x}) \\
=&\partial^i \Phi +\frac{1}{2} ig [\Phi, \partial^i \Phi] -\frac{1}{6} g^2 [\Phi,[\Phi, \partial^i \Phi]]+\mathcal{O}(g^3\Phi^4). 
\end{split}
\end{equation}
To simplify the notation, we dropped the subscript of $\Phi(\mathbf{x})$ for the projectile. 

The structure of the equation suggests that  the WW gluon field  in Eq.~\eqref{eq:alphai_Phi} is a pure gauge field. Additionally, it has  to satisfy the static Yang-Mills equation 
\begin{align}
\partial^i \alpha^i_P(\mathbf{x}) = g \rho_P(\mathbf{x}). 
\label{Eq:SYM}
\end{align}
Note that we have explicitly separated the $g$ dependence in $\rho_P$ and thus parametrically $\rho_P \sim 1$ should be understood in the following discussions.
This equation of motion constrains  $\Phi(\mathbf{x})$,
\begin{equation}\label{eq:Phi_eom}
\partial^2\Phi +\frac{1}{2} ig [\Phi, \partial^2\Phi] -\frac{1}{6} g^2\partial^i [\Phi,[\Phi, \partial^i \Phi]]+\mathcal{O}(g^3\Phi^4) = g\rho_P(\mathbf{x}). 
\end{equation}
We solve for  $\alpha_P^i(\mathbf{x})$ and $\Phi(\mathbf{x})$  perturbatively in $g$.  Only terms with odd powers of coupling constant are nonvanishing 
\begin{equation}
\begin{split}
&\alpha_P^i(\mathbf{x}) = \sum_{m=0}^{\infty}g^{2m+1} \alpha_P^{i, (2m+1)}(\mathbf{x}), \\
&\Phi(\mathbf{x}) = \sum_{m=0}^{\infty}g^{2m+1} \Phi^{(2m+1)}(\mathbf{x}). \\
\end{split}
\end{equation}
 Substituting the expansion of $\Phi(\mathbf{x})$ into Eq.~\eqref{eq:Phi_eom},  we obtain 
\begin{equation}\label{eq:Phi_pert_solutions}
\begin{split}
&\Phi^{(1)} = \phi, \\
&\Phi^{(3)} = -\frac{1}{2} i \frac{1}{\partial^2} [ \Phi^{(1)}, \partial^2 \Phi^{(1)}] = -\frac{1}{2} i \frac{1}{\partial^2} [ \phi, \partial^2 \phi], \\
&\Phi^{(5)} = -\frac{1}{\partial^2} \Big(\frac{1}{2} i [\Phi^{(3)}, \partial^2\Phi^{(1)}] +\frac{1}{2} i [\Phi^{(1)}, \partial^2\Phi^{(3)}]-\frac{1}{6} \partial^i [\Phi^{(1)},[\Phi^{(1)}, \partial^i \Phi^{(1)}]]\Big)\\
&\qquad =- \frac{1}{4}\frac{1}{\partial^2} [\frac{1}{\partial^2}[\phi, \partial^2 \phi], \partial^2 \phi] - \frac{1}{4}\frac{1}{\partial^2}[\phi, [\phi, \partial^2 \phi]] + \frac{1}{6}\frac{ \partial^i}{\partial^2}  [\phi, [\phi, \partial^i \phi]].\\
\end{split}
\end{equation}
To economize notation we introduced $\phi = \frac{1}{\partial^2}\rho_P$. We only need up to order-$g^5$ expansions for the purpose of calculating the first saturation correction to gluon productions in high energy proton-nucleus collisions. 

Now it is straightforward to obtain the perturbative expressions for the WW gluon field order by order. 
Indeed, substituting Eq.~\eqref{eq:Phi_pert_solutions} into Eq.~\eqref{eq:alphai_Phi}, we get  
\begin{equation}
\alpha_P^{(1),i} =\partial^i \Phi^{(1)}= \partial^i \phi
\end{equation}
 at the leading order. 
From this we conclude that the right hand-side of Eq.~\eqref{Eq:SYM}  is saturated automatically at the leading order, i.e. $\partial^i \alpha^{(1),i}_P = \rho_P$.  
Therefore all  higher order contributions $\alpha_P^{(n),i}$ with $(n\geq 3)$ have to have vanishing gradient: 
\begin{equation}
\partial_i  \alpha_P^{(n \geq 3),i} =0. 
\label{Eq:ZeroGrad}
\end{equation}
We use this condition to cross-check the trivial algebra when deriving the higher order terms in the expansion of the WW gluon field.  

The cubic and quintic orders are  
\begin{equation}
\begin{split}
\alpha_P^{(3),i} =\partial^i \Phi^{(3)} + \frac{1}{2} i [\Phi^{(1)}, \partial^i \Phi^{(1)}]
= \frac{1}{2} i \left(\delta^{ij} - \frac{\partial^i \partial^j}{\partial^2}\right) [\phi, \partial^j \phi]
\end{split}
\end{equation}
and 
\begin{equation}
\begin{split}
\alpha_P^{(5),i} 
=&\partial^i \Phi^{(5)} + \frac{1}{2} i[\Phi^{(3)}, \partial^i \Phi^{(1)}] + \frac{1}{2} i[\Phi^{(1)}, \partial^i \Phi^{(3)}] - \frac{1}{6} [\Phi^{(1)}, [\Phi^{(1)}, \partial^i \Phi^{(1)}]]\\
=&\frac{1}{2} \left(\delta^{ij} -\frac{\partial^i\partial^j}{\partial^2}\right)\left [\frac{1}{\partial^2}[\phi, \partial^2 \phi], \partial^j\phi\right] -\frac{1}{6}\left(\delta^{ij}-\frac{ \partial^i\partial^j}{\partial^2} \right)[\phi, [\phi, \partial^j \phi]].
\end{split}
\end{equation}
We factorized out the  projection operator $\delta^{ij}-\partial^i\partial^j/\partial^2$ to make the property~\eqref{Eq:ZeroGrad} manifest. 

We illustrate the expansion using Feynmann diagrams corresponding to terms at each order of the expansion. This is useful to ultimately establish a comparison with results from the diagramatic approach. 
Previously, this has been explicitly done in Ref.~\cite{Kovchegov:1997pc}; however, the discussion was limited to two color sources. Additionally, the expansion in  Ref.~\cite{Kovchegov:1997pc} was conducted in the coupling constant $g$; in the current work, we perform the expansion in terms of the projectile saturation momentum, $\propto \alpha_s \rho_P$.  Therefore our conclusions do not have to agree with  Ref.~\cite{Kovchegov:1997pc}  beyond order $g^3$, as our goal is to account for the saturation correction rather than the perturbative correction!

To proceed with the Feynmann diagrams  it is convenient to explicitly define the inverse Laplacian operator appearing in $ \phi(\mathbf{x}) = \phi^a(\mathbf{x})T^a$,
\begin{equation}
 \phi(\mathbf{x}) = \frac{1}{\partial^2} \rho_P = \frac{1}{2\pi} \int d^2\mathbf{y} \ln{\left(|\mathbf{x}-\mathbf{y}| \Lambda \right)} \rho_P(\mathbf{y})\,.
\end{equation}
Here the scale $\Lambda$ is an IR regularization scale. 
The vector potential is thus 
\begin{equation}
\partial^i  \phi(\mathbf{x}) = \frac{1}{2\pi}\int d^2\mathbf{y} \frac{(\mathbf{x}-\mathbf{y})^i}{|\mathbf{x}-\mathbf{y}|^2} \rho_P(\mathbf{y}). 
\end{equation}
Taking gradient of $\partial^i  \phi(\mathbf{x})$ leads to 
$\partial^2 \phi(\mathbf{x}) = \rho_P(\mathbf{x})$
by construction. 

\begin{figure}[tbp]
\centering 
\includegraphics[scale = 0.5]{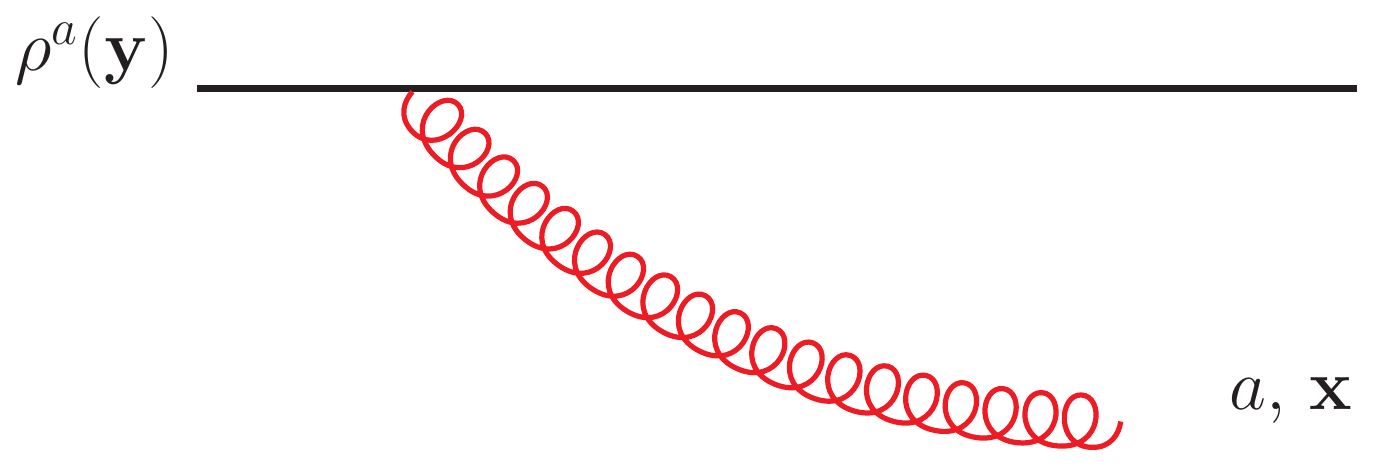}
\caption{The order-$g$ classical gluon field $\alpha_P^{(1),a,i}(\mathbf{x})$ produced at transverse position $\mathbf{x}$ by a color source at transverse position $\mathbf{y}$.}
\label{fig:alphai_g1}
\end{figure}

\begin{figure}[tbp]
\centering 
\includegraphics[scale = 0.45]{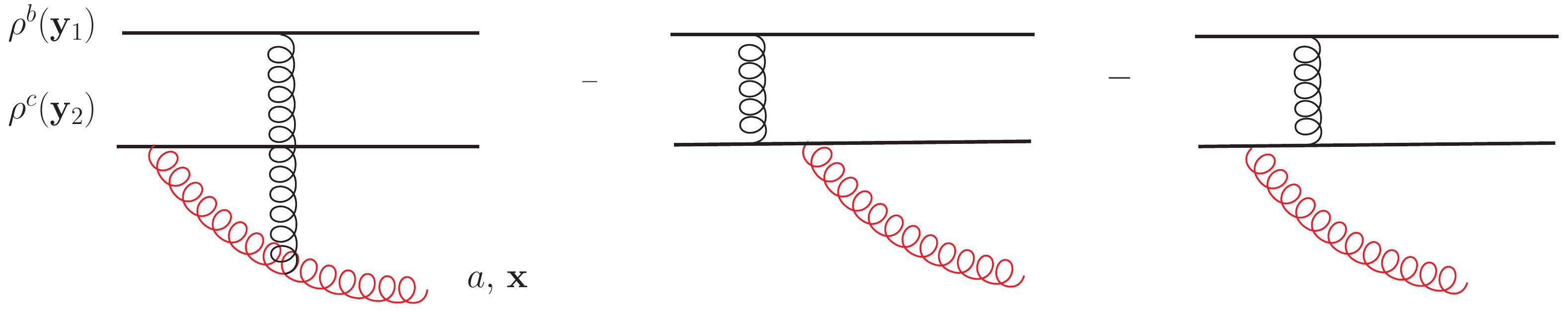}
\caption{The order-$g^3$ WW gluon field $\alpha^{i,a, (3)}_P(\mathbf{x})$ produced at transverse position $\mathbf{x}$ by two color sources at transverse positions $\mathbf{y}_1$ and $\mathbf{y}_2$. }
\label{fig:alphai_g3}
\end{figure}

Now we are ready to proceed with the Feynmann diagrams. 
The leading order  $\alpha^{i, (1)}_P$ is illustrated in  Fig. \ref{fig:alphai_g1}.
The order-$g^3$ WW gluon field  $\alpha^{i, (3)}_P$ involves interactions of two color charges. Written in explicit form, it has two terms   
\begin{equation}
\begin{split}
\alpha^{i, (3)}_P=&\frac{1}{2} i \left(\delta^{ij} - \frac{\partial^i \partial^j}{\partial^2}\right) [\phi, \partial^j \phi] 
\\
=& -\frac{1}{2(2\pi)^2} f^{abc} \int d^2\mathbf{y}_1 \ln{(\Lambda |\mathbf{x}-\mathbf{y}_1|)}\rho_P^b(\mathbf{y}_1) \int d^2\mathbf{y}_2 \frac{(\mathbf{x}-\mathbf{y}_2)_i}{|\mathbf{x}-\mathbf{y}_2|^2} \rho_P^c(\mathbf{y}_2)\\
&+ \frac{1}{2(2\pi)^2} f^{abc} \int d^2\mathbf{y}_2 \frac{(\mathbf{x}-\mathbf{y}_2)_i}{|\mathbf{x}-\mathbf{y}_2|^2} \int d^2\mathbf{y}_1 \ln{(\Lambda |\mathbf{y}_2-\mathbf{y}_1|)} \rho_P^b(\mathbf{y}_1) \rho_P^c(\mathbf{y}_2)\,.
\end{split}
\end{equation}
The Feynmann diagrams corresponding to these two terms are shown in Fig.~\ref{fig:alphai_g3}. In these diagrams, the gluon emission at $\mathbf{x}$ from source at $\mathbf{y}_2$ corresponds to the factor $\frac{(\mathbf{x}-\mathbf{y}_2)_i}{|\mathbf{x}-\mathbf{y}_2|^2} $ while the  factor $\ln{(\lambda |\mathbf{x}-\mathbf{y}_1|)}$ illustrates one-gluon exchange between colored objects  at  $\mathbf{x}$ and $\mathbf{y}_1$.

\begin{figure}[t]
\begin{subfigure}{1.0\textwidth}
\centering 
\includegraphics[scale = 0.4]{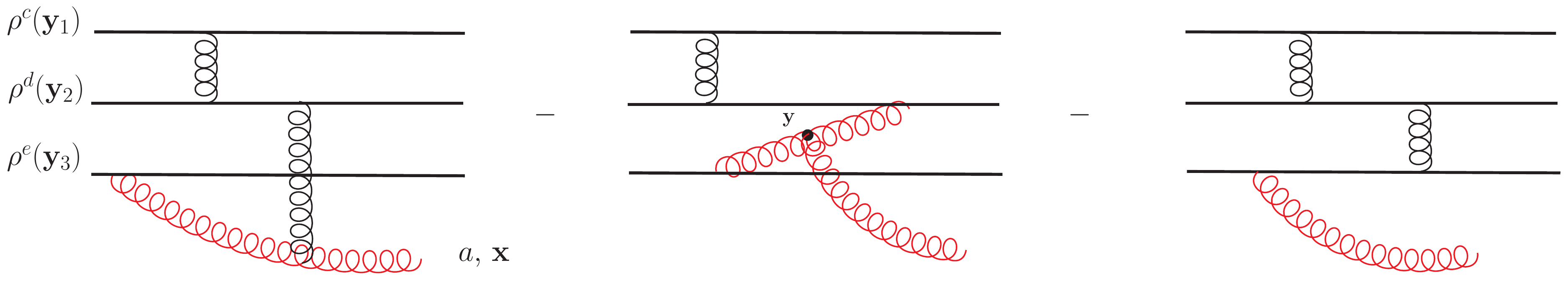}
\caption{}
\label{subfig:a}
\end{subfigure}
\begin{subfigure}{1.0\textwidth}
\centering 
\includegraphics[scale = 0.4]{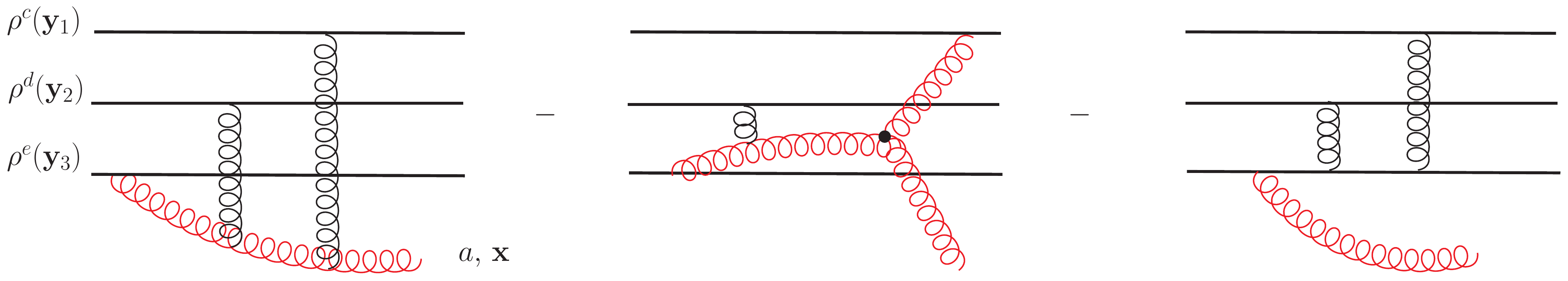}
\caption{}
\label{subfig:b}
\end{subfigure}
\caption{Two sets of diagrams contributing to order-$g^5$ WW field $\alpha_P^{i, a,(5)}(\mathbf{x})$ produced at transverse position $\mathbf{x}$ by three color sources at transverse position $\mathbf{y}_1$, $\mathbf{y}_2$ and $\mathbf{y}_3$. Similar diagrams with different orderings of gluon exchanges are not shown.}
\label{fig:alphai_g5_second3}
\end{figure}

The order-$g^5$ WW gluon field involves interactions of three color charges. In the expression for $\alpha_P^{i,(5)}$, there are six topologically different contributions 
\begin{equation}
\begin{split}
\alpha_P^{i, (5)} 
=&\frac{1}{2} \left(\delta^{ij} -\frac{\partial^i\partial^j}{\partial^2}\right)\left [\frac{1}{\partial^2}[\phi, \partial^2 \phi], \partial^j\phi\right] -\frac{1}{6}\left(\delta^{ij}-\frac{ \partial^i\partial^j}{\partial^2} \right)[\phi, [\phi, \partial^j \phi]]\\
=&\frac{1}{2}\left [\frac{1}{\partial^2}[\phi, \partial^2 \phi], \partial^i\phi\right] -\frac{1}{2}\frac{\partial^i}{\partial^2} \left [\frac{\partial^j}{\partial^2}[\phi, \partial^2 \phi], \partial^j\phi\right]-\frac{1}{2}\frac{\partial^i}{\partial^2} \left [\frac{1}{\partial^2}[\phi, \partial^2 \phi], \partial^2\phi\right]\\
&-\frac{1}{6} [\phi, [\phi, \partial^i \phi]] + \frac{1}{6} \frac{\partial^i}{\partial^2} [\partial^j\phi, [\phi, \partial^j \phi]]+ \frac{1}{6} \frac{\partial^i}{\partial^2} [\phi, [\phi, \partial^2 \phi]].
\end{split}
\end{equation}
We write them out explicitly one by one. The first  term is 
\begin{equation}
\begin{split}
&\frac{1}{2}\left [\frac{1}{\partial^2}[\phi, \partial^2 \phi], \partial^i\phi\right]=-\frac{1}{2} f^{abe}f^{bcd} \frac{1}{\partial^2}(\phi^c \rho_P^d) \partial^i \phi^e\\
=&-\frac{1}{2} f^{abe}f^{bcd} \int d^2\mathbf{y}_2 \ln{(|\mathbf{x}-\mathbf{y}_2|\Lambda)} \left(\int d^2\mathbf{y}_1 \ln{(|\mathbf{y}_2-\mathbf{y}_1|\Lambda)} \rho_P^c(\mathbf{y}_1) \rho_P^d(\mathbf{y}_2)\right) \\
&\qquad \times \int d^2\mathbf{y}_3 \frac{(\mathbf{x}-\mathbf{y}_3)^i}{|\mathbf{x}-\mathbf{y}_3|^2}\rho_P^e(\mathbf{y}_3).
\end{split}
\end{equation}
The subsequent two terms are
\begin{equation}
\begin{split}
&-\frac{1}{2}\frac{\partial^i}{\partial^2} \left [\frac{\partial^j}{\partial^2}[\phi, \partial^2 \phi], \partial^j\phi\right] = \frac{1}{2} f^{abe}f^{bcd} \frac{\partial^i}{\partial^2} \Big(\frac{\partial^j}{\partial^2}(\phi^c \rho_P^d) \partial^j \phi^e\Big)\\
=&\frac{1}{2} f^{abe}f^{bcd} \int d^2 \mathbf{y} \frac{(\mathbf{x}-\mathbf{y})^i}{|\mathbf{x}-\mathbf{y}|^2} \Big(\int d^2\mathbf{y}_2 \frac{(\mathbf{y}-\mathbf{y}_2)^j}{|\mathbf{y}-\mathbf{y}_2|^2} \int d^2\mathbf{y}_1 \ln{(|\mathbf{y}_2-\mathbf{y}_1|\Lambda)} \rho_P^{c}(\mathbf{y}_1 )\rho_P^d(\mathbf{y}_2)\\
&\qquad\qquad \times \int d^2\mathbf{y}_3 \frac{(\mathbf{y}-\mathbf{y}_3)^j}{|\mathbf{y}-\mathbf{y}_3|^2} \rho_P^e(\mathbf{y}_3) \Big)
\end{split}
\end{equation}
and
\begin{equation}
\begin{split}
&-\frac{1}{2}\frac{\partial^i}{\partial^2} \left [\frac{1}{\partial^2}[\phi, \partial^2 \phi], \partial^2\phi\right] = \frac{1}{2} f^{abe}f^{bcd} \frac{\partial^i}{\partial^2}\left(\frac{1}{\partial^2}(\phi^c \rho_P^d) \rho_P^e\right)\\
=&\frac{1}{2} f^{abe}f^{bcd} \int d^2\mathbf{y}_3 \frac{(\mathbf{x}-\mathbf{y}_3)^i}{|\mathbf{x}-\mathbf{y}_3|^2} \int d^2\mathbf{y}_2 \ln{(|\mathbf{y}_3-\mathbf{y}_2|\Lambda)} \\ 
&\qquad\times \int d^2\mathbf{y}_1 \ln{(|\mathbf{y}_2-\mathbf{y}_1|\Lambda)} \rho_P^c(\mathbf{y}_1) \rho_P^d(\mathbf{y}_2) \rho_P^e(\mathbf{y}_3)\,.
\end{split}
\end{equation}
These three terms correspond to the diagrams shown in Fig.~\ref{subfig:a}. Note that in the second diagram, there is an integration over all the possible transverse positions $\mathbf{y}$. 

The remaining  three terms are
\begin{equation}
\begin{split}
&-\frac{1}{6} [\phi, [\phi, \partial^i \phi]] =\frac{1}{6} f^{acb}f^{bde} \phi^c\phi^d \partial^i \phi^e\\
=&\frac{1}{6} f^{acb}f^{bde} \int d^2\mathbf{y}_1 \ln{(|\mathbf{x}-\mathbf{y}_1|\Lambda)} \rho_P^c(\mathbf{y}_1) \int d^2\mathbf{y}_2 \ln{(|\mathbf{x}-\mathbf{y}_2|\Lambda)} \rho_P^d(\mathbf{y}_2) \\ 
 &\qquad \times \int d^2\mathbf{y}_3 \frac{(\mathbf{x}-\mathbf{y}_3)^i}{|\mathbf{x}-\mathbf{y}_3|^2} \rho_P^e(\mathbf{y}_3),\\
 \end{split}
\end{equation}
\begin{equation}
\begin{split}
& \frac{1}{6} \frac{\partial^i}{\partial^2} [\partial^j\phi, [\phi, \partial^j \phi]]=-\frac{1}{6} f^{acb} f^{bde} \frac{\partial^i}{\partial^2} \left(\partial^j \phi^c \phi^d \partial^j \phi^e\right)\\
=&-\frac{1}{6}f^{acb} f^{bde}  \int d^2\mathbf{y} \frac{(\mathbf{x}-\mathbf{y})^i}{|\mathbf{x}-\mathbf{y}|^2} \Big(\int d^2\mathbf{y}_1\frac{(\mathbf{y}-\mathbf{y}_1)^j}{|\mathbf{y}-\mathbf{y}_1|^2} \rho_P^c(\mathbf{y}_1) \int d^2\mathbf{y}_2 \ln{(|\mathbf{y}-\mathbf{y}_2|\lambda)}\rho_P^d(\mathbf{y}_2)\\
&\qquad \times \int d^2\mathbf{y}_3 \frac{(\mathbf{y}-\mathbf{y}_3)^j}{|\mathbf{y}-\mathbf{y}_3|^2} \rho_P^e(\mathbf{y}_3) \Big)
\end{split}
\end{equation}
and 
\begin{equation}
\begin{split}
& \frac{1}{6} \frac{\partial^i}{\partial^2} [\phi, [\phi, \partial^2 \phi]]=-\frac{1}{6} f^{acb}f^{bde} \frac{\partial^i}{\partial^2}  \phi^c \phi^d \rho_P^e\\
=&-\frac{1}{6} f^{acb}f^{bde} \int d^2\mathbf{y}_3 \frac{(\mathbf{x}-\mathbf{y}_3)^i}{|\mathbf{x}-\mathbf{y}_3|^2} \int d^2\mathbf{y}_1 \ln{(|\mathbf{y}_3-\mathbf{y}_1|\Lambda)} \rho_P^c(\mathbf{y}_1) \\
&\qquad \times \int d^2\mathbf{y}_2 \ln{(|\mathbf{y}_3-\mathbf{y}_2|\Lambda)}\rho_P^d(\mathbf{y}_2) \rho_P^d(\mathbf{y}_3).
\end{split}
\end{equation}
The corresponding Feynman diagrams are shown in Fig.~\ref{subfig:b}.


\subsection{Residual gauge fixing and initial conditions}
The classical Yang-Mills equations Eq.~\eqref{eq:ym_alpha_alphai} are written in the Fock-Schwinger (FS) gauge.  
The equations involve  three fields $\alpha^i(\tau, \mathbf{x}), \alpha(\tau, \mathbf{x})$ (here $i=1,2$), but  only two of them are independent degrees of freedom.  In other words,  the FS gauge does not completely fix the gluon fields. There is still  residual freedom to perform gauge transformations which  only depend on the transverse coordinates.  While the form of the classical Yang-Mills equations given in Eqs.~\eqref{eq:ym_alpha_alphai} remain unchanged under residual gauge transformations, the initial conditions in Eq.~\eqref{eq:initial_conditions} depend on sub gauge transformations.  Physical observables, however, are independent of gauge choices. Thus it is beneficial  to select a sub gauge in such a way to simplify the calculations of the physical observables.  
Denoting the sub gauge transformations  by $\Omega(\mathbf{x})$, gluon fields in two sub gauges are related by
\begin{equation}
\begin{split}
&\tilde{A}^i = \Omega^{\dagger} A^i \Omega + \frac{i}{g} \Omega^{\dagger} \partial^i \Omega \,, \\
&\tilde{A}^{+} = \Omega^{\dagger} A^{+}\Omega \, ,\\
&\tilde{A}^{-} = \Omega^{\dagger} A^{-}\Omega \, .\\
\end{split}
\end{equation}
There are a few choices of the sub gauges. We discuss them below.

\subsubsection{Sub gauge transformation by $U(\mathbf{x})$}
\label{sec:subgauge_Ux}

In the literature, when calculating particle production in $pA$ collisions,  $U(\mathbf{x})$ was often  chosen \cite{Dumitru:2001ux, McLerran:2016snu} to define the residual gauge 
fixing. With  $\Omega(\mathbf{x})=U(\mathbf{x})$ the initial condition for transverse fields in Eq. \eqref{eq:initial_conditions} becomes
\begin{equation}
\begin{split}
\zeta^i(\tau=0,\mathbf{x}) =& U^{\dagger} \alpha^i(\tau=0,\mathbf{x}) U + \frac{i}{g} U^{\dagger} \partial^i U =U^{\dagger} \alpha_P^i U\\
 =&\alpha_P^{a,i}(\mathbf{x}) U^{ab}(\mathbf{x}) T^b.\\
\end{split}
\end{equation}
Note that the target field $\alpha_T^i = \frac{i}{g}U\partial^i U^{\dagger}$  is  gauged away. In this form, both the gradient $\partial_i \zeta^i $ and the curl $\epsilon_{ji}\partial_j \zeta_i$ are nonvanishing. The  initial condition for the longitudinal field in Eq. \eqref{eq:initial_conditions} becomes
\begin{equation}\label{eq:betatau0_g1_order}
\begin{split}
\zeta(\tau=0, \mathbf{x}) = &U^{\dagger} \alpha(\tau=0, \mathbf{x}) U =\frac{1}{2} \left(\partial^i (U^{\dagger} \alpha_P^i U) - U^{\dagger} \partial^i \alpha_P^i U\right)\\
=&\frac{1}{2} \alpha_P^{a,i} (\mathbf{x}) \partial^i U^{ab}(\mathbf{x})T^b.\\
\end{split}
\end{equation}

In this sub gauge, the order-$g$, order-$g^3$ and order-$g^5$ initial conditions are 
\begin{equation}
\begin{split}
&\zeta^i_{(1)}(\tau=0, \mathbf{x}) = U^{\dagger} \alpha_{P,(1)}^i U,\\
&\zeta_{(1)}(\tau=0, \mathbf{x}) =\frac{1}{2} \left(\partial^i (U^{\dagger} \alpha_{P,(1)}^i U) - U^{\dagger} \partial^i \alpha_{P,(1)}^i U\right);\\
&\zeta^i_{(3)}(\tau=0, \mathbf{x}) = U^{\dagger} \alpha_{P,(3)}^i U,\\
&\zeta_{(3)}(\tau=0, \mathbf{x}) =\frac{1}{2} \partial^i (U^{\dagger} \alpha_{P,(3)}^i U); \\
&\zeta^i_{(5)}(\tau=0, \mathbf{x}) = U^{\dagger} \alpha_{P,(5)}^i U,\\
&\zeta_{(5)}(\tau=0, \mathbf{x}) =\frac{1}{2} \partial^i (U^{\dagger} \alpha_{P,(5)}^i U) .
\end{split}
\end{equation}

This sub gauge has the advantage that the initial conditions have clear physical meaning. The $\zeta^i$ is the projectile WW gluon field $\alpha_P^i$ eikonally rotated by the target Wilson line $U(\mathbf{x})$.  The $\zeta$ is the difference between the eikonally rotated projectile WW gluon field and the gluon field generated by the eikonally rotated projectile color charge density.  However, in solving the classical Yang-Mills equations for $pA$ collisions beyond the leading order, this sub gauge is not the most convenient one.  In a desirable sub gauge,  either the gradient or the curl of  $\zeta^i(\tau=0, \mathbf{x})$ would vanish. 

\subsubsection{Sub gauge condition $\partial_i\beta_i(\tau=0, \mathbf{x}) =0 $}

 It has been shown in Ref.~\cite{Blaizot:2010kh} that the following sub gauge transformation 
\begin{equation}
\Omega(\mathbf{x})= U(\mathbf{x})\left[ 1 +  ig \frac{\partial^l}{\partial^2}  \left (U^{\dagger} \alpha_{P, (1)}^i U \right) \right]
\end{equation}
guarantees that the lowest order gradient of $\beta_i$ vanishes, i.e. $\partial_i\beta_i^{(1)}(\tau=0,\mathbf{x})=0$. To go beyond the lowest order, we consider the ansatz
\begin{equation}
\Omega(\mathbf{x})= U(\mathbf{x}) \mathcal{W}(\mathbf{x}) 
\end{equation}
with $\mathcal{W}(\mathbf{x}) = e^{ig\Sigma(\mathbf{x})}$. Unitarity condition of  $\Omega$  requires $\Sigma^{\dagger} = \Sigma$.
Under this gauge transformation, the initial condition for the  transverse fields in Eq. \eqref{eq:initial_conditions} becomes
\begin{equation}
\begin{split}
\beta^i(\tau=0, \mathbf{x}) = &\Omega^{\dagger}\alpha_P^i\Omega + \Omega^{\dagger} \alpha_T^i\Omega + \frac{i}{g}\Omega^{\dagger} \partial^i \Omega\\
=&\mathcal{W}^{\dagger}U^{\dagger}\alpha_{P}^iU \mathcal{W} + \frac{i}{g} \mathcal{W}^{\dagger}\partial^i \mathcal{W}\\
=&U^{\dagger}\alpha_{P}^iU - ig[\Sigma, U^{\dagger}\alpha_{P}^iU] + \frac{1}{2} (ig)^2 [\Sigma, [\Sigma, U^{\dagger}\alpha_{P}^iU]]  + \ldots\\
&-\partial^i\Sigma + \frac{1}{2} ig [\Sigma, \partial^i \Sigma] - \frac{1}{6}(ig)^2[\Sigma, [\Sigma, \partial^i \Sigma]]+ \ldots
\end{split}
\end{equation}
In obtaining the last equality, we  used the Baker-Campbell-Hausdorff formula. We express $\Sigma(\mathbf{x})$ as a power series expansion in terms of coupling constant $g$
\begin{equation}
\Sigma (\mathbf{x})= g\Sigma_{(1)} + g^3\Sigma_{(3)} + g^5\Sigma_{(5)} + \ldots
\end{equation}
 and solve $\Sigma(\mathbf{x})$ order by order by imposing the requirement $\partial^i \beta^i (\tau=0, \mathbf{x}) =0$. The results are 
 \begin{equation}\label{eq:sigmas_in_gauge_transformation}
 \begin{split}
&\Sigma_{(1)}=\frac{\partial^l}{\partial^2} (U^{\dagger} \alpha_{P,(1)}^l U),\\
&\Sigma_{(3)} = \frac{\partial^l}{\partial^2} \left(U^{\dagger}\alpha_{P,(3)}^lU -i [\Sigma_{(1)}, U^{\dagger}\alpha_{P,(1)}^l U] + \frac{1}{2} i [\Sigma_{(1)}, \partial^l \Sigma_{(1)}]\right),\\
&\Sigma_{(5)} = \frac{\partial^l}{\partial^2} \Big(U^{\dagger}\alpha_{P,(5)}^lU -i[\Sigma_{(1)}, U^{\dagger}\alpha_{P,(3)}^l U] -i [\Sigma_{(3)}, U^{\dagger}\alpha_{P,(1)}^lU] -\frac{1}{2} [\Sigma_{(1)}, [\Sigma_{(1)}, U^{\dagger} \alpha_{P,(1)}^l U]] \\
& \qquad\qquad\quad+ \frac{1}{2} i[\Sigma_{(1)}, \partial^l \Sigma_{(3)}] + \frac{1}{2} i[\Sigma_{(3)}, \partial^l\Sigma_{(1)}] +\frac{1}{6} [\Sigma_{(1)}, [\Sigma_{(1)}, \partial^l\Sigma_{(1)}]]\Big).
\end{split}
\end{equation}
 The initial conditions for the transverse field at the corresponding orders are
 \begin{equation}\label{eqs:ic_betai_g_g3_g5}
 \begin{split}
 &\beta^i_{(1)}(\tau=0, \mathbf{x})=(\delta^{ij}-\frac{\partial^i \partial^j}{\partial^2}) U^{\dagger} \alpha_{P,(1)}^j U,\\
 &\beta^i_{(3)}(\tau=0,\mathbf{x}) =(\delta^{il} -\frac{\partial^i\partial^l}{\partial^2})\left(U^{\dagger}\alpha_{P,(3)}^lU -i [\Sigma_{(1)}, U^{\dagger}\alpha_{P,(1)}^l U] + \frac{1}{2} i [\Sigma_{(1)}, \partial^l \Sigma_{(1)}]\right),\\
 &\beta^i_{(5)}(\tau=0,\mathbf{x})= (\delta^{il}-\frac{\partial^i\partial^l}{\partial^2}) \Big(U^{\dagger}\alpha_{P,(5)}^lU -i[\Sigma_{(1)}, U^{\dagger}\alpha_{P,(3)}^l U] -i [\Sigma_{(3)}, U^{\dagger}\alpha_{P,(1)}^lU] \\
&\qquad\qquad\qquad\qquad\qquad\qquad\quad-\frac{1}{2} [\Sigma_{(1)}, [\Sigma_{(1)}, U^{\dagger} \alpha_{P,(1)}^l U]] + \frac{1}{2} i[\Sigma_{(1)}, \partial^l \Sigma_{(3)}] \\
&\qquad\qquad\qquad\qquad\qquad\qquad\quad+ \frac{1}{2} i[\Sigma_{(3)}, \partial^l\Sigma_{(1)}] +\frac{1}{6} [\Sigma_{(1)}, [\Sigma_{(1)}, \partial^l\Sigma_{(1)}]]\Big).
 \end{split}
 \end{equation}
We will only need  initial conditions up to order-$g^5$.  However, one can recursively obtain all higher order gauge transformations by imposing the condition $\partial^i\beta^i_{(n)}(\tau=0, \mathbf{x})=0$ for $n\geq 7$. It is not clear to us whether a closed form expression for $\Sigma(\mathbf{x})$ exist or not.

On the other hand, the initial condition for the longitudinal field under the gauge transformation becomes
\begin{equation}
\begin{split}
\beta(\tau=0,\mathbf{x}) = &\frac{ig}{2} \Omega^{\dagger}[\alpha_P^i, \alpha_T^i] \Omega\\
=&\frac{1}{2}\mathcal{W}^{\dagger}\Big(\partial^i (U^{\dagger}\alpha_P^i U)  - U^{\dagger}\partial^i\alpha_P^iU\Big) \mathcal{W}\\
=&\frac{1}{2}(\partial^i (U^{\dagger}\alpha_P^i U)  - U^{\dagger}\partial^i\alpha_P^iU) - ig[\Sigma, \frac{1}{2}(\partial^i (U^{\dagger}\alpha_P^i U)  - U^{\dagger}\partial^i\alpha_P^iU)]\\
&-\frac{1}{2}g^2[\Sigma, [\Sigma, \frac{1}{2}(\partial^i (U^{\dagger}\alpha_P^i U)  - U^{\dagger}\partial^i\alpha_P^iU)]] + \ldots
\end{split}
\end{equation}
From it,  the order-$g$, order-$g^3$, order-$g^5$ initial conditions for the longitudinal field are obtained
\begin{equation}\label{eqs:ic_beta_g_g3_g5}
\begin{split}
&\beta_{(1)}(\tau=0,\mathbf{x})=\frac{1}{2}\Big(\partial^i (U^{\dagger}\alpha_{P,(1)}^i U)  - U^{\dagger}\partial^i\alpha_{P,(1)}^iU\Big),\\
&\beta_{(3)}(\tau=0, \mathbf{x})=\frac{1}{2}\partial^i (U^{\dagger}\alpha_{P,(3)}^i U)  - i[\Sigma_{(1)}, \beta_{(1)}(\tau=0)],\\
&\beta_{(5)}(\tau=0, \mathbf{x})=\frac{1}{2}\partial^i (U^{\dagger}\alpha_{P,(5)}^i U)  - i[\Sigma_{(1)}, \frac{1}{2}\partial^i (U^{\dagger}\alpha_{P,(3)}^i U)  ]-i[\Sigma_{(3)}, \beta_{(1)}(\tau=0)]\\
&\qquad\qquad\qquad\quad-\frac{1}{2}[\Sigma_{(1)}, [\Sigma_{(1)}, \beta_{(1)}(\tau=0)]] .\\
\end{split}
\end{equation}
In obtaining the above results, we have used the fact that $\partial^i \alpha_{P,(n)}^i =0$ for $n\geq 3$. 

The sub gauge condition  $\partial^i \beta^i (\tau=0, \mathbf{x}) =0$ resembles the general Coulomb gauge $\partial^i\beta^i(\tau, \mathbf{x})=0$. 
Previously, in numerically solving the classical Yang-Mills equations, the Coulomb gauge condition was also used. However, instead of imposing it at $\tau=0$, it was imposed at some particularly chosen proper time $\tau_0$, at which physical observables were calculated~\cite{Schenke:2015aqa, Berges:2013fga}.

One may wonder,  whether it is possible to find a sub gauge transformation such  that instead of the gradient of $\beta_i$, the curl of $\beta_i$ is zero $\epsilon_{ij}\partial_i \beta_j (\tau=0, \mathbf{x})=0$. First of all, we want to point out that there is no $\Omega(\mathbf{x})$ that can completely gauge away $\beta^i(\tau=0, \mathbf{x})$.  From 
$\beta^i(\tau=0, \mathbf{x}) = \Omega^{\dagger}\alpha_P^i\Omega + \Omega^{\dagger} \alpha_T^i\Omega + \frac{i}{g}\Omega^{\dagger} \partial^i \Omega  =0$, 
one obtains $\alpha_P^i + \alpha_T^i = \frac{i}{g}\Omega \partial^i \Omega^{\dagger}$.  The right hand side is a pure gauge field. On the other hand, both $\alpha_P^i$ and $\alpha_T^i$ are pure gauge fields, their sum cannot be a pure gauge field.  Not  even at the lowest order.
It is not clear that the following condition on the curl of  $\beta_i$
\begin{equation}\label{eq:gauge_betai=0}
\epsilon_{hi}\partial_h\beta^i(\tau=0, \mathbf{x}) = \epsilon_{hi}\partial_h\Big(\Omega^{\dagger}\alpha_P^i\Omega + \Omega^{\dagger} \alpha_T^i\Omega + \frac{i}{g}\Omega^{\dagger} \partial^i \Omega \Big) =0
\end{equation}
has a perturbative solution for $\Omega(\mathbf{x})$. As it will become clear in the following sections, the gradient of $\beta^i(\tau, \mathbf{x})$ is an auxiliary field while its curl  is a dynamical field. Therefore,  as far as gluon production is concerned, the initial time Coulomb gauge constraining the gradient of $\beta^i(\tau=0, \mathbf{x})$ serves as a convenient gauge choice. 

Our motivation for choosing different sub gauge transformations was purely to simplify computations as  physical observables are gauge invariant. 
However, when discussing the time evolution of the gluon field (a gauge variant object), the concept of initial vs final state effects becomes blurred and, strictly speaking, not well-defined.  A sub-gauge transformation may (and does) shift some final state effects to the realm of initial state effects and vice-versa. As a matter of fact our motivation was exactly to simplify the time evolution  and transfer the computational burden to the initial conditions.  

In the main body of this paper, the classical Yang-Mills equations are solved in the initial time  Coulomb sub gauge $\partial^i\beta^i(\tau=0,\mathbf{x})=0$. In the Appendix \ref{ap:noncoulomb_solutions}, solutions in the sub gauge determined by $U(\mathbf{x})$ are given.  The two sub gauges are related by $\mathcal{W}(\mathbf{x})$. By comparing gluon fields in the two gauges, it will become clear how final state interactions in one gauge become initial state effects in another gauge.

\section{The Dynamical Equations and the Constraint Equation}
As discussed in the previous section, the classical Yang-Mills equations are invariant under the sub gauge transformation. Thus the equations of motion in the initial time Coulomb sub gauge are obtainted by simple replacements of  $\alpha, \alpha^i$ with $\beta, \beta^i$ in Eqs. \eqref{eq:ym_alpha_alphai}. 
\begin{equation}\label{eq:ym_beta_betai_tilde}
\begin{split}
& \tau^2\partial_{\tau}^2 \tilde{\beta} +\tau\partial_{\tau} \tilde{\beta}  -\tilde{\beta}  - \tau^2\partial^2_i \tilde{\beta}+ ig\tau^2\partial_i \left[\beta_i, \tilde{\beta}\right]+ ig\tau^2\left[\beta_i, \partial_i \tilde{\beta}\right] +g^2\tau^2\left[\beta_i,[\beta_i, \tilde{\beta}]\right]  =0\,, \\
& \partial_i\partial_{\tau} \beta_i-ig\left[\beta_i, \partial_{\tau}\beta_i\right]-ig\left[ \tilde{\beta}, \partial_{\tau} \tilde{\beta}\right] =0\,, \\
&\tau^2\partial^2_{\tau}\beta_i+\tau\partial_{\tau} \beta_i - \tau^2(\partial^2 \delta_{ij} - \partial_j \partial_i) \beta_j   -ig \tau^2\left[\tilde{\beta}, \partial_i\tilde{\beta}\right] - g^2 \tau^2 \left[\tilde{\beta}, [\beta_i, \tilde{\beta}]\right] \\
&\qquad + ig\tau^2\partial_j\left[\beta_j, \beta_i\right]  + ig\tau^2\left[\beta_j, \partial_j\beta_i-\partial_i\beta_j\right] + g^2\tau^2\left[\beta_j,  [\beta_j, \beta_i]\right]=0 .\\
\end{split}
\end{equation}
We have written out the detailed expressions for the equations using the covariant derivative $D_i = \partial_i -ig \beta_i$ and the field tensor $F_{ji} = \partial_j \beta_i -\partial_i\beta_j -ig [\beta_j, \beta_i]$.  We also separated the linear  and nonlinear terms in the equations
and introduce $\tilde{\beta} = \tau \beta$ to simplify the notation.

The second equation in Eqs.~\eqref{eq:ym_beta_betai_tilde} is first order in time derivative and thus serves as  a constraint equation. Only two of the three field components $\beta(\tau,\mathbf{x}), \beta_{i=1,2}(\tau, \mathbf{x})$  are independent.  To explicitly demonstrate this, it is convenient to split the transverse field into the gradient part and the curl part
\begin{equation}
\beta_i(\tau, \mathbf{x}) = \epsilon_{il}\partial_l \chi(\tau,\mathbf{x}) + \partial_i \Lambda(\tau, \mathbf{x}).
\end{equation}
Here $\epsilon_{il}$ is the two dimensional Levi-Civita symbol. Using $\epsilon_{ih}\partial_h\beta_i = \partial^2 \chi $ and $\partial_i \beta_i = \partial^2 \Lambda$, one can separate  the third equation in Eqs.~\eqref{eq:ym_beta_betai_tilde} into two equations:
\begin{equation}\label{eq:secondorderLambdaeq}
\begin{split}
\tau^2\partial_{\tau}^2 \partial^2\Lambda + \tau\partial_{\tau}\partial^2\Lambda = &ig\tau^2\left[\tilde{\beta}, \partial^2\tilde{\beta}\right] +ig\tau^2 [\beta_j, (\partial^2\delta_{ji}-\partial_j\partial_i)\beta_i]\\
&+ g^2\tau^2 \partial_i \left[\tilde{\beta},\left [\beta_i, \tilde{\beta}\right]\right] - g^2\tau^2\partial_i\left[\beta_j,[\beta_j,\beta_i]\right].\\
\end{split}
\end{equation}
 \begin{equation}
 \begin{split}
 \tau^2\partial^2_{\tau} \partial^2\chi +\tau\partial_{\tau} \partial^2\chi -\tau^2\partial^2 \partial^2\chi  =& ig\tau^2\epsilon_{ih} \left[\partial_h \tilde{\beta},\partial_i\tilde{\beta}\right]-g^2\tau^2 \epsilon_{ih}\partial_h \left([\tilde{\beta},[\tilde{\beta},\beta_i]] + [\beta_j,[\beta_j,\beta_i]]\right)\\
 &-ig\tau^2 \epsilon_{ih}\partial_h \left([\partial_j\beta_j,\beta_i] +2[\beta_j,\partial_j \beta_i]\right) +ig\tau^2\epsilon_{ih}[\partial_h \beta_j, \partial_i \beta_j].
 \end{split}
 \end{equation}
 The constraint equation only imposes restriction on $\Lambda$
 \begin{equation}\label{eq:Lambda_1storder_eq}
 \partial_{\tau} \partial^2\Lambda = ig [\beta_i, \partial_{\tau}\beta_i]+ig[ \tilde{\beta}, \partial_{\tau} \tilde{\beta}].
 \end{equation}
 The independent degrees of freedom are $\tilde{\beta}(\tau,\mathbf{x})$ and $ \chi(\tau, \mathbf{x})$. The $\Lambda(\tau, \mathbf{x})$ is a non-dynamical field. In the Appendix \ref{sec:nondaynamical_Lambda}, it is  proved perturbatively that   the  second order equation \eqref{eq:secondorderLambdaeq} is just a consequence of the first order equation \eqref{eq:Lambda_1storder_eq}. Thus, although the superficial appearance does not  suggest it,  the second order differential equation for $\Lambda$ is not an independent dynamical equation.
 
In the following sections, we seek solutions of the classical Yang-Mills equations in terms of power series expansion in the coupling constant $g$. 
\begin{equation}
\begin{split}
&\beta(\tau,\mathbf{x}) = \sum_{n=0}^{\infty}g^n \beta^{(n)}(\tau,\mathbf{x}),\\
&\beta_i(\tau, \mathbf{x}) = \sum_{n=0}^{\infty} g^n \beta_i^{(n)} (\tau, \mathbf{x}).\\
\end{split}
\end{equation}
It is obvious that  only odd powers of the expansions $\beta^{(2n+1)}, \beta_i^{(2n+1)}$ are nonvanishing.


\section{Order-$g$ Solutions}
The order-$g$ classical Yang-Mills equations are
\begin{equation}\label{eq:orderg_YM}
\begin{split}
& \tau^2\partial_{\tau}^2 \tilde{\beta}^{(1)} +\tau\partial_{\tau} \tilde{\beta}^{(1)}  -\tilde{\beta}^{(1)}  - \tau^2\partial^2_i \tilde{\beta}^{(1)} =0\,, \\
& \partial_i\partial_{\tau} \beta_i^{(1)} =0\,, \\
&\tau^2\partial^2_{\tau}\beta_i^{(1)}+\tau\partial_{\tau} \beta_i^{(1)} - \tau^2(\partial^2 \delta_{ij} - \partial_j \partial_i) \beta_j^{(1)}   =0 \\
\end{split}
\end{equation}
with the initial conditions 
\begin{equation}
\begin{split}
&\beta_i^{(1)}(\tau=0,\mathbf{x}) = \left(\delta_{ij}-\frac{\partial_i\partial_j}{\partial^2}\right)U^{\dagger} \alpha_{P,(1)}^j U ,\\
&\beta^{(1)}(\tau=0,\mathbf{x})  = \frac{1}{2}\left[\partial^i \left(U^{\dagger} \alpha_{P,(1)}^i U\right) -U^{\dagger}\partial^i \alpha_{P,(1)}^i U\right].
\end{split}
\end{equation}
Performing the decomposition  
$
\beta_i^{(1)} (\tau, \mathbf{x}) = \epsilon_{il}\partial_l \chi^{(1)} (\tau, \mathbf{x}) + \partial_i\Lambda^{(1)}(\tau, \mathbf{x}) 
$, 
 we trivially establish that the constraint equation together with the  initial condition $\partial_i \beta_i^{(1)}(\tau=0, \mathbf{x})=0$  is equivalent  to $\Lambda^{(1)}(\tau, \mathbf{x}) =0$, as expected at this order.

The Yang-Mills equations are easier to solve in transverse momentum space. Using the convention 
\begin{equation}
\begin{split}
&\beta_i(\tau,\mathbf{x}) = \int \frac{d^2\mathbf{k}}{(2\pi)^2} e^{-i\mathbf{k}\cdot\mathbf{x}} \beta_i(\tau, \mathbf{k}) , \qquad \beta(\tau, \mathbf{x}) =  \int \frac{d^2\mathbf{k}}{(2\pi)^2} e^{-i\mathbf{k}\cdot\mathbf{x}} \beta(\tau, \mathbf{k}),
\end{split}
\end{equation}
for the Fourier transformations, we obtain that  
after the projection  $\epsilon_{ih}\partial_h \beta_i^{(1)} = \partial^2\chi^{(1)}$
the two independent equations are 
\begin{equation}
\begin{split}
&s^2 \partial_{s}^2 \tilde{\beta}^{(1)}(\tau, \mathbf{k}) + s\partial_{s} \tilde{\beta}^{(1)} (\tau, \mathbf{k}) + (s^2 -1) \tilde{\beta}^{(1)}(\tau, \mathbf{k}) =0, \\
&s^2\partial_{s}^2 \chi^{(1)}(\tau, \mathbf{k}) + s\partial_{s} \chi^{(1)}(\tau, \mathbf{k}) + s^2 \chi^{(1)}(\tau, \mathbf{k}) =0.
\end{split}
\end{equation}
Here $s \equiv k_{\perp} \tau$ and  $k_{\perp}=|\mathbf{k}|$ is the magnitude of the two dimensional transverse momentum. 
These two equations are easily recognized as standard Bessel equations. We require the solutions to be finite at $\tau=0$, only Bessel functions of first kind satisfy this constraint. 
 We thus have 
\begin{equation}
\begin{split}
&\beta^{(1)}(\tau, \mathbf{k} ) = b_{\eta}(\mathbf{k}) \frac{J_1(k_{\perp}\tau)}{k_{\perp}\tau},\\
&\beta^{(1)}_i (\tau, \mathbf{k}) = \frac{-i\epsilon_{il}\mathbf{k}_l}{k_{\perp}^2} b_{\perp}(\mathbf{k}) J_0(k_{\perp}\tau). \\
\end{split}
\end{equation}
where $b_{\eta, \perp}(\mathbf{k})$ are fixed by the initial conditions
\begin{equation}\label{eq:beta_bperp_coulombgauge}
\begin{split}
&b_{\eta}(\mathbf{k}) = 2\beta^{(1)}(\tau=0, \mathbf{k}),\\
&b_{\perp}(\mathbf{k}) = -i \epsilon_{ij}\mathbf{k}_i \beta_j^{(1)}(\tau=0, \mathbf{k}).\\
\end{split}
\end{equation}
It is easy to check that $\partial_i \beta_i^{(1)}(\tau,\mathbf{x}) =0$ for any $\tau$. 

The order-$g$ solution has already been obtained in Ref.~\cite{Dumitru:2001ux}. The order-$g$ equations in this sub-gauge are free field equations and the solutions are free field solutions. Somewhat nontrivial time dependence characterized by $J_1(k_{\perp}\tau)/k_{\perp}\tau $ and $J_0(k_{\perp}\tau)$ is solely  due to the Milne coordinates $(\tau, \eta, \mathbf{x})$. Now, we turn to higher orders and   for the first time we will obtain order-$g^3$ and order-$g^5$ solutions.


\section{Order-$g^3$ Solutions}

The order-$g^3$ classical Yang-Mills equations are
\begin{equation}\label{eq:ym_beta_betai_g3}
\begin{split}
& \tau^2\partial_{\tau}^2 \tilde{\beta}^{(3)} +\tau\partial_{\tau} \tilde{\beta}^{(3)}  -\tilde{\beta}^{(3)}  - \tau^2\partial^2_i \tilde{\beta}^{(3)}+ i\tau^2\partial_i \left[\beta_i^{(1)}, \tilde{\beta}^{(1)}\right]+ i\tau^2\left[\beta_i^{(1)}, \partial_i \tilde{\beta}^{(1)}\right] =0\,, \\
& \partial_i\partial_{\tau} \beta_i^{(3)}-i\left[\beta_i^{(1)}, \partial_{\tau}\beta_i^{(1)}\right]-i\left[ \tilde{\beta}^{(1)}, \partial_{\tau} \tilde{\beta}^{(1)}\right] =0\,, \\
&\tau^2\partial^2_{\tau}\beta_i^{(3)}+\tau\partial_{\tau} \beta_i^{(3)} - \tau^2(\partial^2 \delta_{ij} - \partial_j \partial_i) \beta_j^{(3)}   -i \tau^2\left[\tilde{\beta}^{(1)}, \partial_i\tilde{\beta}^{(1)}\right] \\
&\qquad + i\tau^2\partial_j\left[\beta_j^{(1)}, \beta_i^{(1)}\right]  + i\tau^2\left[\beta_j^{(1)}, \partial_j\beta_i^{(1)}-\partial_i\beta_j^{(1)}\right] =0 \\
\end{split}
\end{equation}
with the initial conditions
\begin{equation}
\beta_{(3)}(\tau=0, \mathbf{x})
=\frac{1}{2} \partial^i (U^{\dagger}\alpha_{P,(3)}^i U) -i[\Sigma_{(1)}, \beta_{(1)}(\tau=0,\mathbf{x})]
\end{equation}
and 
\begin{equation}
\beta^i_{(3)}(\tau=0,\mathbf{x}) 
=\left(\delta^{ij}-\frac{\partial^i\partial^j}{\partial^2}\right)\Big(U^{\dagger}\alpha_{P,(3)}^jU - i[\Sigma_{(1)}, U^{\dagger} \alpha_{P,(1)}^j U] + \frac{1}{2} i [\Sigma_{(1)}, \partial^j \Sigma_{(1)}]\Big)\,.
\end{equation}

\subsection{Solving for $\beta^{(3)}(\tau, \mathbf{k})$}
The first equation when transformed into momentums space is an inhomogeneous Bessel equation ($s= k_{\perp}\tau$) 
\begin{equation}\label{eq:pde_beta3}
s^2 \partial_s^2 \tilde{\beta}^{(3)}(\tau, \mathbf{k})  + s\partial_s \tilde{\beta}^{(3)} (\tau, \mathbf{k})+ (s^2 -1) \tilde{\beta}^{(3)} (\tau, \mathbf{k}) =S_{\eta}^{(3)}(\tau, \mathbf{k})
\end{equation}
with 
\begin{equation}
\begin{split}
S_{\eta}^{(3)}(\tau, \mathbf{k}) = &-i\tau^3 \int d^2\mathbf{x} e^{i\mathbf{k}\cdot \mathbf{x}} \left([\partial_i\beta_i^{(1)}, \beta^{(1)}]+2[\beta_i^{(1)}, \partial_i \beta^{(1)}] \right)\\
=&i\tau^2  \int \frac{d^2\mathbf{p}}{(2\pi)}  \frac{2\mathbf{k}\times \mathbf{p}}{p_{\perp}^2|\mathbf{k}-\mathbf{p}|}J_0(p_{\perp}\tau)  J_1(|\mathbf{k}-\mathbf{p}|\tau)\Big[b_{\perp}(\mathbf{p}), b_{\eta}(\mathbf{k}-\mathbf{p})\Big]. \\
\end{split}
\end{equation}
In order to solve the inhomogeneous differential equations as Eq.~\eqref{eq:pde_beta3}, one can apply  a well established method which is often referred to as  variation of parameters.  The method is briefly reviewed in the Appendix \ref{ap:method_variation_parameter}.  The two independent solutions for the corresponding homogeneous Bessel equation are $J_1(x)$ and $Y_1(x)$, whose Wronskian is $W(x) = \frac{2}{\pi x}$.  So the general solutions for the inhomogeneous equation can be formally expressed as 
\begin{equation}\label{eq:beta_g3_formal_sol}
\tilde{\beta}^{(3)}(\tau, \mathbf{k}) = C_1 J_1(s) + C_2 Y_1(s) + \frac{\pi}{2}\int_0^s  dz \left[ J_1(z) Y_1(s) - J_1(s) Y_1(z) \right]  \frac{S_{\eta}^{(3)} (z)}{z}\,.
\end{equation}
 The initial condition is finite at $\tau=0$, thus $Y_1(x)$ does not  contribute, i.e. $C_2=0$. 
 The coefficient $C_1$ can be  then determined straightforwardly 
\begin{equation}
\beta^{(3)} (\tau=0, \mathbf{k})= \frac{1}{\tau} \tilde{\beta}^{(3)} (\tau, \mathbf{k})|_{\tau=0} = \frac{k_{\perp}}{2} C_1 \,.
\end{equation} 
Putting everything together we get 
\begin{equation}
C_1 = \frac{2}{k_{\perp}} \beta^{(3)} (\tau=0, \mathbf{k}), \qquad C_2 =0.
\end{equation}

Further evaluations of  the formal solution Eq.~\eqref{eq:beta_g3_formal_sol}  require computing time integrals involving products of three Bessel functions in the integrand
\begin{equation}
\begin{split}
&\int_{0}^s dz z J_1(z) J_0\left(\frac{q_{\perp}}{k_{\perp}} z\right) J_1\left(\frac{|\mathbf{k}-\mathbf{q}|}{k_{\perp}}z \right), \\
&\int_{0}^s dz z Y_1(z) J_0\left(\frac{q_{\perp}}{k_{\perp}} z\right) J_1\left(\frac{|\mathbf{k}-\mathbf{q}|}{k_{\perp}}z\right). \\
\end{split}
\end{equation}
Here we face a difficulty because for integrands involving products of three or more Bessel functions, 
there are no known formula to compute the indefinite integrals. This is in a stark contrast to the case with  
integrands  of only two Bessel functions: 
\begin{equation}
\begin{split}
&\int_0^s dz z J_1(az) J_1(bz) = \frac{1}{a^2-b^2} \Big( bs J_0(bs)J_1(as) - as J_0(as)J_1(bs)\Big),\\
&\int_0^s dz z Y_1(az)J_1(bz) = \frac{2b}{\pi a} \frac{1}{a^2-b^2} + \frac{1}{a^2-b^2}\Big(bsJ_0(bs)Y_1(as) - as Y_0(as)J_1(bs)\Big).\\
\end{split}
\end{equation}
This defines our strategy: we will aim at reducing the number of Bessel functions in the integrand from three (or more) to two in order to use the above equations to evaluate time integrals.  
The key step is to expressing a product of two Bessel functions in terms of an integral of one Bessel function using Graf's formula.  Mathematical details are given in the Appendix \ref{ap:Grafs_formula}.

Substituting the expression of $S_{\eta}^{(3)}$ into the formal solution, one obtains
\begin{equation}
\begin{split}
\tilde{\beta}^{(3)}(\tau,\mathbf{k}) =& C_1 J_1(s) + \frac{\pi}{2} \int_0^s   (Y_1(s) J_1(z) - J_1(s)Y_1(z)) S_{\eta}^{(3)}(z)\frac{dz}{z} \\
=&C_1 J_1(s) +\frac{i\pi}{2} \int \frac{d^2\mathbf{p}}{(2\pi)} \frac{2\mathbf{k}\times \mathbf{p}}{k_{\perp}^2p_{\perp}^2|\mathbf{k}-\mathbf{p}|}\Big[b_{\perp}(\mathbf{p}), b_{\eta}(\mathbf{k}-\mathbf{p})\Big] \\
&\qquad \Bigg(Y_1(s)\int_0^s dz z J_1(z) J_0\left(\frac{p_{\perp}}{k_{\perp}}z\right)  J_1\left(\frac{|\mathbf{k}-\mathbf{p}|}{k_{\perp}}z\right) \\ &\qquad\qquad - J_1(s)\int_0^s dz z Y_1(z) J_0\left(\frac{p_{\perp}}{k_{\perp}}z\right)  J_1\left(\frac{|\mathbf{k}-\mathbf{p}|}{k_{\perp}}z\right)\Bigg)\,. \\
\end{split}
\end{equation}
Introducing the notation  $w_1 =p_{\perp}/k_{\perp}$ and $w_2= |\mathbf{k}-\mathbf{p}|/k_{\perp}$, and
using the formula
\begin{equation}
J_0(w_1z) J_1(w_2z) = \int_{-\pi}^{\pi} \frac{d\phi}{2\pi} e^{i\Psi'}J_1(w z)
\end{equation}
with 
\begin{equation}
w = \sqrt{w_1^2+w_2^2 - 2 w_1 w_2\cos{\phi}},
\end{equation}
and
\begin{equation}
e^{i\Psi'} = \frac{w_2-w_1 \cos{\phi}}{w} + i \frac{w_1\sin{\phi}}{w}  
\end{equation}
we can express the integrals of  products of three Bessel functions as integrals of products of two Bessel functions. We have to pay a price of introducing an  auxiliary angular integral.
Nevertheless, after performing this manipulation,  the structure of the solution simplifies significantly. Finally we will end up with an expression of the following form
\begin{equation}\label{eq:simplification_auxiliary}
\begin{split}
&Y_1(s) \int_0^sdz z J_1(z) J_1(w z) - J_1(s) \int_0^x dz z Y_1(z) J_1(w z)\\
=&Y_1(s) \frac{1}{1-w^2}\left[w s J_0(w s) J_1(s) - sJ_0(s) J_1(w s) \right] \\
&- J_1(s) \frac{1}{1-w^2} \left[w sJ_0(w s)Y_1(s) - sY_0(s)J_1(w s)\right]-J_1(s) \frac{2w}{\pi} \frac{1}{1-w^2}\\
=&\frac{2}{\pi}\frac{1}{1-w^2} \Big(J_1(w s) -w J_1(s)\Big),
\end{split}
\end{equation}
where we performed further simplification by using that fact that  the Wronskian is $J_1(s)Y_0(s)-J_0(s)Y_1(s) = \frac{2}{\pi s}$.  Note that $w=1$ is a removable singularity; indeed, 
\begin{equation}
\frac{2}{\pi}\frac{1}{1-w^2} \Big(J_1(w s) -w J_1(s)\Big) \xrightarrow{w\rightarrow 1}   \frac{1}{\pi} sJ_2(s),
\end{equation}
which is finite and well-defined. 

Collecting everything together we obtain the final solution 
\begin{equation}\label{eq:final_sol_beta_g3}
\begin{split}
\beta^{(3)}(\tau,\mathbf{k}) 
=&2\beta^{(3)}(\tau=0,\mathbf{k}) \frac{J_1(k_{\perp}\tau)}{k_{\perp}\tau}-i \int \frac{d^2\mathbf{p}}{(2\pi)^2} \frac{\mathbf{k}\times \mathbf{p}}{p_{\perp}^2|\mathbf{k}-\mathbf{p}|^2}\Big[ b_{\eta}(\mathbf{p}),b_{\perp}(\mathbf{k}-\mathbf{p})\Big]\\
&\times \int_{-\pi}^{\pi}\frac{d\phi}{2\pi} \left(1+ \frac{2\mathbf{k}\cdot\mathbf{p}}{w_{\perp}^2-k_{\perp}^2}\right) \left(\frac{J_1(w_{\perp}\tau)}{w_{\perp}\tau} - \frac{J_1(k_{\perp}\tau)}{k_{\perp}\tau}\right),\\
\end{split}
\end{equation}
where $w_{\perp} = \sqrt{p_{\perp}^2+|\mathbf{k}-\mathbf{p}|^2 -2p_{\perp}|\mathbf{k}-\mathbf{p}| \cos\phi}$. 

The are a few notable features of the solution. 
\begin{itemize}
\item 
First is that  the time-dependent factors  are completely determined by one type of  Bessel function; in this case, it is  Bessel function of first kind of order one $J_1(\lambda \tau)/\lambda \tau$. However, the Bessel function contributes with different arguments. For the first term, the argument of the Bessel function is completely determined by the external momentum $\mathbf{k}$. The second term is more involved, for given momenta $\mathbf{k}$ and $\mathbf{p}$, the time-dependent factor sums over all possible momentum mode between $w_{\mathrm{max}}=p_{\perp} + |\mathbf{k}-\mathbf{p}|$ and $w_{\mathrm{min}}=|p_{\perp}-|\mathbf{k}-\mathbf{p}||$.  It should be pointed out that the second term has a removable singularity at $w_{\perp}=k_{\perp}$.  It can be checked by performing  Taylor expansions of the difference $J_1(w_{\perp}\tau)/w_{\perp}\tau-J_1(k_{\perp}\tau)/k_{\perp}\tau$; it  starts from $w_{\perp}^2-k_{\perp}^2$. This  combination $J_1(w_{\perp}\tau)/w_{\perp}\tau-J_1(k_{\perp}\tau)/k_{\perp}\tau$ also guarantees that the second term is zero at $\tau=0$.

Naively, it is expected that at asymptotically large $\tau$, the gluon system should  behave like free gas of gluons.
This is not that easy to confirm on the level of the field; the asymptotic behavior of the Bessel function reads 
\begin{equation}
J_1(w_{\perp}\tau) \longrightarrow \frac{1}{\sqrt{2\pi w_{\perp}\tau}} e^{i w_{\perp}\tau }e^{-i \frac{3}{4}\pi}.
\end{equation}
and thus the  summation over all the momentum modes persists even at $\tau\rightarrow \infty$.

\item
The second feature is about the color structure of the solution. It involves interactions of two color charges in the proton. Let us look at each term in the solution in detail. 
First we can recognize from eq. \eqref{eq:sigmas_in_gauge_transformation} that 
\begin{equation}
\Sigma_{(1)}(\mathbf{k}) =\frac{i\mathbf{k}^i}{k_{\perp}^2}\int d^2\mathbf{x} e^{i\mathbf{k}\cdot\mathbf{x}} \alpha^{i,a}_{P,(1)}(\mathbf{x})U^{ad}(\mathbf{x})T^d.\\
\end{equation}
 is the order-$g$ WW gluon field $\alpha^i_{P, (1)}$ eikonally color rotated by the target Wilson line $U^{ad}$ and then projected along the momentum $\mathbf{k}$.
 Next, consider 
 \begin{equation}\label{eq:b_eta_physical_meaning}
 \begin{split}
 &b_{\eta}(\mathbf{k}) = 2\beta_{(1)}(\tau=0, \mathbf{k}) \\
 =& -k_{\perp}^2 \Sigma_{(1)}(\mathbf{k})  - \int d^2\mathbf{x} e^{i\mathbf{k}\cdot\mathbf{x}}\rho_P^a(\mathbf{x}) U^{ad}(\mathbf{x})T^d\\
 =&-i\mathbf{k}^i\left( \int d^2\mathbf{x} e^{i\mathbf{k}\cdot\mathbf{x}} \alpha^{i,a}_{P,(1)}(\mathbf{x})U^{ad}(\mathbf{x})T^d - \frac{i\mathbf{k}^i}{k_{\perp}^2} \int d^2\mathbf{x} e^{i\mathbf{k}\cdot\mathbf{x}}\rho_P^a(\mathbf{x}) U^{ad}(\mathbf{x})T^d\right)
 \end{split}
 \end{equation}
 and 
 \begin{equation}\label{eq:b_perp_physical_meaning}
\begin{split}
b_{\perp}(\mathbf{k})=&-i\epsilon_{ij}\mathbf{k}_i\int d^2\mathbf{x} e^{i\mathbf{k}\cdot\mathbf{x}} \alpha^{j,a}_{P, (1)}(\mathbf{x})U^{ad}(\mathbf{x})T^d\\
=&-i\epsilon_{ij}\mathbf{k}_i\left(\int d^2\mathbf{x} e^{i\mathbf{k}\cdot\mathbf{x}} \alpha^{j,a}_{P, (1)}(\mathbf{x})U^{ad}(\mathbf{x})T^d- \frac{i\mathbf{k}^j}{k_{\perp}^2} \int d^2\mathbf{x} e^{i\mathbf{k}\cdot\mathbf{x}}\rho_P^a(\mathbf{x}) U^{ad}(\mathbf{x})T^d\right).
\end{split}
\end{equation}
In the parenthesis, we have a difference between two terms. One is the order-$g$ projectile WW gluon field eikonally rotated by the target Wilson line $U^{ad}$. The other is the WW field \textit{generated} by the eikonally rotated color density $\rho^a_PU^{ad}$ (it corresponds to the gluon cloud of the receding color charge or to the Fadeev-Kulish state). The net field is projected also along $\mathbf{k}$ for $b_{\eta}(\mathbf{k})$ and projected perpendicular to $\mathbf{k}$ for $b_{\perp}(\mathbf{k})$. In this sense, one can attribute $b_{\eta}(\mathbf{k})$ and $b_{\perp}(\mathbf{k})$ to two polarizations of the order-$g$ gluon field produced in the collisions.

In the solution Eq.~\eqref{eq:final_sol_beta_g3}, the initial field at order-$g^3$ is 
\begin{equation}
\begin{split}
2\beta_{(3)}(\tau=0, \mathbf{k}) 
=&-i\mathbf{k}^i  \int d^2\mathbf{x} e^{i\mathbf{k}\cdot\mathbf{x}} \alpha^{i,a}_{P,(3)}(\mathbf{x})U^{ad}(\mathbf{x})T^d - i\int \frac{d^2\mathbf{p}}{(2\pi)^2}[\Sigma_{(1)}(\mathbf{k}-\mathbf{p}), b_{\eta}(\mathbf{p})]\,.\\
\end{split}
\end{equation}
The first term is the order-$g^3$ WW gluon field $\alpha^i_{P, (3)}$ which was  eikonally color rotated by the target Wilson line $U^{ac}$ and then projected to the momentum $\mathbf{k}$.  The diagrams representing $\alpha^i_{P, (3)}$ have been shown in Fig.~\ref{fig:alphai_g3}. 
\begin{figure}[t]
\centering 
\includegraphics[scale = 0.45]{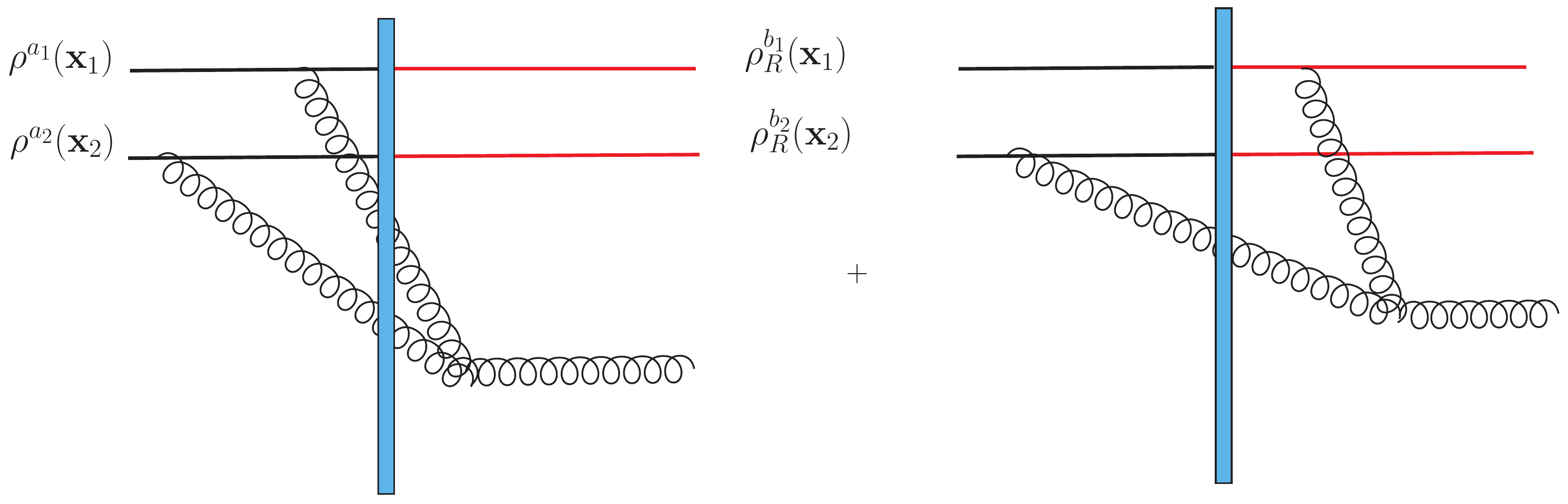}
\caption{Schematic representation of the color structure for the second term in $\beta^{(3)}(\tau, \mathbf{k})$. The shaded bar represents the target nucleus. The eikonally rotated color charge density $\rho_R$ is represented using red lines. }
\label{fig:beta_g3_color_structure}
\end{figure}
The  color structure of the second term in Eq.~\eqref{eq:final_sol_beta_g3} can be schematically illustrated in  Fig.~\ref{fig:beta_g3_color_structure}.

 We note that this discussion was specific for the used sub-gauge, since the color structure depends on sub gauge transformations.

\end{itemize}

\subsection{Solving for $\beta^{(3)}_i(\tau, \mathbf{k})$ } 
As a first step, we want to  demonstrate  explicitly that from the order-$g^3$ constraint equation and the order-$g$ Yang-Mills equations, the second order differential equation for $\partial_i\beta_i^{(3)}$  can be derived. 
This would prove that $\partial_i\beta_i^{(3)}$ is not a dynamical field. 
From the third equation of the set~\eqref{eq:ym_beta_betai_g3},  the second order differential equation for $\partial_i \beta_i^{(3)}$ is
\begin{equation}\label{eq:2ndODE_dibetai}
\tau^2\partial^2_{\tau} \partial_i \beta_i^{(3)} + \tau\partial_{\tau} \partial_i \beta_i^{(3)}  = i\tau^2 [\tilde{\beta}^{(1)}, \partial^2 \tilde{\beta}^{(1)}] - i\tau^2 [\beta_j^{(1)}, (\partial_i\partial_j- \partial^2 \delta_{ij}) \beta_i^{(1)}].
\end{equation}
From the constraint equation of \eqref{eq:ym_beta_betai_g3}, one obtains
\begin{equation}
\begin{split}
&\tau\partial_{\tau}\partial_i \beta_i^{(3)} = i\tau [\beta_i^{(1)}, \partial_{\tau}\beta_i^{(1)}] + i\tau [\tilde{\beta}^{(1)}, \partial_{\tau} \tilde{\beta}^{(1)}],\\
&\tau^2 \partial^2_{\tau} \partial_i \beta_i^{(3)} = i\tau^2 [\beta_i^{(1)}, \partial^2_{\tau} \beta_i^{(1)}] + i\tau^2 [\tilde{\beta}^{(1)}, \partial_{\tau}^2 \tilde{\beta}^{(1)}].\\
\end{split}
\end{equation}
Adding these two equations and substituting the order-$g$ Yang-Mills equations in Eqs.~\eqref{eq:orderg_YM} reproduce Eq.~\eqref{eq:2ndODE_dibetai}.  

In what follows, we will use the decomposition
\begin{equation}
 \beta_i^{(3)}(\tau, \mathbf{x}) = \epsilon_{il}\partial_l\chi^{(3)} (\tau, \mathbf{x})+ \partial_i \Lambda^{(3)}(\tau, \mathbf{x})
 \end{equation}
 to separately solve for $\chi^{(3)}(\tau, \mathbf{x})$ and $\Lambda^{(3)}(\tau, \mathbf{x})$.

\subsubsection{The solution $\Lambda^{(3)}(\tau,\mathbf{k})$}
In order to solve for $\Lambda^{(3)}$, the constraint equation  
 in momentum space is used 
\begin{equation}
-k_{\perp}^2\partial_{\tau} \Lambda^{(3)}(\tau, \mathbf{k}) = S_{\Lambda}^{(3)}(\tau, \mathbf{k}) \,.
\end{equation}
The source term is given by 
\begin{equation}
\begin{split}
&S_{\Lambda}^{(3)}(\tau, \mathbf{k}) \\
=& \int d^2\mathbf{x} e^{i\mathbf{k} \cdot \mathbf{x}} \left(i[\beta_i^{(1)}(\tau, \mathbf{x}), \partial_{\tau}\beta_i^{(1)}(\tau, \mathbf{x})]+i[ \tilde{\beta}^{(1)}(\tau, \mathbf{x}), \partial_{\tau} \tilde{\beta}^{(1)}(\tau, \mathbf{x})] \right)\\
=&i\int\frac{d^2\mathbf{p}}{(2\pi)^2} \bigg( \frac{\mathbf{p}\cdot(\mathbf{k}-\mathbf{p})}{p_{\perp}^2 |\mathbf{k}-\mathbf{p}|} \left[b_{\perp}(\mathbf{p}), b_{\perp}(\mathbf{k}-\mathbf{p})\right]J_0(p_{\perp}\tau) J_1( |\mathbf{k}-\mathbf{p}|\tau)\\
&\qquad\qquad \quad-\frac{1}{p_{\perp}} \left[b_{\eta}(\mathbf{p}), b_{\eta}(\mathbf{k}-\mathbf{p})\right]J_1(p_{\perp}\tau) J_2(|\mathbf{k}-\mathbf{p}|\tau)\bigg)\\
=&i\int\frac{d^2\mathbf{p}}{(2\pi)^2} \frac{\mathbf{p}\cdot(\mathbf{k}-\mathbf{p})}{2p_{\perp}^2|\mathbf{k}-\mathbf{p}|^2} [b_{\perp}(\mathbf{p}), b_{\perp}(\mathbf{k}-\mathbf{p})](|\mathbf{k}-\mathbf{p}|^2-p_{\perp}^2) \int_{-\pi}^{\pi}\frac{d\phi}{2\pi} \frac{1}{w_{\perp}} J_1(w_{\perp}\tau)\\
&+i\int\frac{d^2\mathbf{p}}{(2\pi)^2} \frac{1}{2p_{\perp}|\mathbf{k}-\mathbf{p}|} [b_{\eta}(\mathbf{p}), b_{\eta}(\mathbf{k}-\mathbf{p})](p_{\perp}^2-|\mathbf{k}-\mathbf{p}|^2)\int_{-\pi}^{\pi} \frac{d\phi}{2\pi} \frac{\cos\phi}{w_{\perp}} J_1(w_{\perp}\tau)\\
\end{split}
\end{equation}
 In obtaining the last equality, we changed the integration variable $\mathbf{p}$ to $\mathbf{k}-\mathbf{p}$ and used the fact that $ [b_{\perp}(\mathbf{p}), b_{\perp}(\mathbf{k}-\mathbf{p})]$ and  $ [b_{\eta}(\mathbf{p}), b_{\eta}(\mathbf{k}-\mathbf{p})]$ are antisymmetric under the exchange $\mathbf{p}\leftrightarrow \mathbf{k}-\mathbf{p}$. Additionally, we applied the Bessel function identity $J_2(z) = 2J_1(z)/z- J_0(z)$, and  we  expressed product of two Bessel functions in terms of  angular integral of one Bessel function: 
\begin{equation}
\begin{split}
J_0(p_{\perp}\tau) J_1(|\mathbf{k}-\mathbf{p}|\tau) =&\int_{-\pi}^{\pi} \frac{d\phi}{2\pi} e^{i\Psi'} J_1(w_{\perp}\tau) =\int_{-\pi}^{\pi} \frac{d\phi}{2\pi} \frac{|\mathbf{k}-\mathbf{p}|-p_{\perp}\cos\phi}{w_{\perp}} J_1(w_{\perp}\tau)    \\
\end{split}
\end{equation}
with the definitions $w_{\perp}=\sqrt{p_{\perp}^2+|\mathbf{k}-\mathbf{p}|^2-2p_{\perp}|\mathbf{k}-\mathbf{p}|\cos{\phi}}$ and 
\begin{equation}
e^{i\Psi'} = \frac{|\mathbf{k}-\mathbf{p}|-p_{\perp}\cos{\phi}}{w_{\perp}} + i \frac{p_{\perp} \sin{\phi}}{w}.
\end{equation}
To obtain the solutions, directly integrating $S_{\Lambda}^{(3)}$ involves indefinite integrals of products of two Bessel functions of different orders and with different arguments. We are not aware of if these integrals can be done analytically. That is why we express products of two Bessel functions as an integral of one Bessel function.

The solution for $\Lambda^{(3)}(\tau, \mathbf{k})$ is obtained by direct integration of $S_{\Lambda}^{(3)}$
\begin{equation}\label{eq:Lambda_g3_final_sol}
\begin{split}
&\Lambda^{(3)}(\tau, \mathbf{k}) = \\
&- \frac{i}{k_{\perp}^2} \int\frac{d^2\mathbf{p}}{(2\pi)^2}  \frac{\mathbf{k}\cdot(\mathbf{k}-2\mathbf{p})}{2p_{\perp}^2 |\mathbf{k}-\mathbf{p}|^2}\mathbf{p}\cdot(\mathbf{k}-\mathbf{p}) \left[b_{\perp}(\mathbf{p}), b_{\perp}(\mathbf{k}-\mathbf{p})\right] \int_{-\pi}^{\pi} \frac{d\phi}{2\pi}\frac{1}{w^2_{\perp}}(1- J_0(w_{\perp}\tau))\\
&- \frac{i}{k_{\perp}^2} \int\frac{d^2\mathbf{p}}{(2\pi)^2}  \frac{\mathbf{k}\cdot(\mathbf{k}-2\mathbf{p})}{4p_{\perp}^2 |\mathbf{k}-\mathbf{p}|^2} \left[b_{\eta}(\mathbf{p}), b_{\eta}(\mathbf{k}-\mathbf{p})\right]\int_{-\pi}^{\pi} \frac{d\phi}{2\pi} \left(1-\frac{p_{\perp}^2+|\mathbf{k}-\mathbf{p}|^2}{w_{\perp}^2}\right)  (1- J_0(w_{\perp}\tau))\,.\\
\end{split}
\end{equation}
The time dependent  factors are completely determined by Bessel function of first kind with order zero in the  form $1-J_0(w_{\perp}\tau)$.  Again, for given $\mathbf{k}$ and $\mathbf{p}$, all the momentum modes from $|p_{\perp}-|\mathbf{k}-\mathbf{p}||$ to $p_{\perp} + |\mathbf{k}-\mathbf{p}|$ contribute to the argument of the Bessel function. 
 Interestingly, the two polarization modes $b_{\perp}$ and $b_{\eta}$ do not mix.

\subsubsection{The solution $\beta_{\perp}^{(3)}(\tau, \mathbf{k})$}
The  equation of motion for $\beta^{(3)}_i(\tau, \mathbf{k})$  is
\begin{equation}\label{eq:betai_g3_momentum}
\tau^2\partial^2_{\tau}\beta_i^{(3)}(\tau, \mathbf{k})+\tau\partial_{\tau} \beta_i^{(3)} (\tau, \mathbf{k})+\tau^2(k_{\perp}^2 \delta_{ij} - k_jk_i) \beta_j^{(3)} (\tau,\mathbf{k}) = S_i^{(3)}(\tau, \mathbf{k})
\end{equation}
with the time dependent source term given by 
\begin{equation}
\begin{split}
S_i^{(3)}(\tau, \mathbf{k})=&i\tau^2\int d^2\mathbf{x} e^{i\mathbf{k}\cdot\mathbf{x}}  \left([\tilde{\beta}^{(1)}, \partial_i\tilde{\beta}^{(1)}]  -\partial_j[\beta_j^{(1)}, \beta_i^{(1)}]  - [\beta_j^{(1)}, \partial_j\beta_i^{(1)}-\partial_i\beta_j^{(1)}] \right)\\
=&\tau^2 \int \frac{d^2\mathbf{p}}{(2\pi)^2} (\mathbf{k}-\mathbf{p})_i [\tilde{\beta}^{(1)}(\tau, \mathbf{p}), \tilde{\beta}^{(1)}(\tau, \mathbf{k}-\mathbf{p})] -(2\mathbf{k}-\mathbf{p})_j[\beta_j^{(1)}(\tau, \mathbf{p}), \beta_i^{(1)}(\tau, \mathbf{k}-\mathbf{p})]\\
&+ (\mathbf{k}-\mathbf{p})_i[\beta_j^{(1)}(\tau, \mathbf{p}),  \beta_j^{(1)}(\tau, \mathbf{k}-\mathbf{p})]\,.\\
\end{split}
\end{equation}
Instead of working with $\chi^{(3)}(\tau, \mathbf{k})$, it is convenient to  project out the curl of $\beta_i^{(3)}$ by
\begin{equation}
\beta_{\perp}^{(3)}(\tau, \mathbf{k}) = \frac{i\epsilon_{ih}\mathbf{k}_h}{k_{\perp}} \beta_i^{(3)}(\tau, \mathbf{k})\,.
\end{equation}
The $\beta_{\perp}^{(3)}(\tau, \mathbf{k}) = k_{\perp} \chi^{(3)}(\tau, \mathbf{k})$ by construction has the same dimension as $\beta_i^{(3)}(\tau, \mathbf{k})$.  The source term becomes 
\begin{equation}
\begin{split}
&S_{\perp}^{(3)}(\tau,\mathbf{k}) = \frac{i\epsilon^{ih} \mathbf{k}_h}{k_{\perp}} S_i^{(3)}(\tau, \mathbf{k})\\
=&-i\tau^2 \int \frac{d^2\mathbf{p}}{(2\pi)^2}\Bigg(-\frac{ \mathbf{k}\times \mathbf{p}}{k_{\perp}p_{\perp}|\mathbf{k}-\mathbf{p}|}\Big[b_{\eta}(\mathbf{p}), b_{\eta}( \mathbf{k}-\mathbf{p})\Big] J_1(p_{\perp}\tau) J_1(|\mathbf{k}-\mathbf{p}|\tau)\\
&\qquad+\frac{(\mathbf{k}\times \mathbf{p})(\mathbf{k}\cdot \mathbf{p} -p^2_{\perp}-k_{\perp}^2)}{k_{\perp}p_{\perp}^2|\mathbf{k}-\mathbf{p}|^2} \Big[b_{\perp}(\mathbf{p}), b_{\perp}(\mathbf{k}-\mathbf{p})\Big] J_0(p_{\perp}\tau)J_0(|\mathbf{k}-\mathbf{p}|\tau)\Bigg)\\
\end{split}
\end{equation}
We changed variables  $\mathbf{p} \leftrightarrow \mathbf{k}-\mathbf{p}$, where appropriate, to symmetrize this expression.

After the projection of  Eq.~\eqref{eq:betai_g3_momentum}, the equation of motion for the curl $\beta_{\perp}^{(3)}(\tau, \mathbf{k})$ becomes
\begin{equation}\label{eq:betaperp_g3_eq}
s^2 \partial_{s}^2 \beta_{\perp}^{(3)} + s\partial_{s} \beta_{\perp}^{(3)} + s^2 \beta_{\perp}^{(3)} =  S_{\perp}^{(3)}(\tau, \mathbf{k})
\end{equation}
This is an inhomogeneous differential equation and the corresponding homogeneous part is the the Bessel equation of the first kind. We again use the method of variation of parameters to solve it.  The two independent general solutions for the corresponding homogeneous equation  is $J_0(s)$ and $Y_0(s)$. Their Wronskian is  $W(s) =\frac{2}{\pi s}$. The formal solution of Eq.~\eqref{eq:betaperp_g3_eq} is
\begin{equation}
\beta_{\perp}^{(3)}(s) = D_1 J_0(s) + D_2 Y_0(s) + \frac{\pi}{2} \int_0^s dz \left(J_0(z) Y_0(s) - J_0(s)Y_0(z) \right)\frac{1}{z} S^{(3)}_{\perp}(z) 
\end{equation}
The solution is nonsingular at $\tau=0$, thus $D_2=0$. 
Substituting the explicit expression for $S_{\perp}^{(3)}$ into the formal solution, one obtains
\begin{equation}\label{eq:beta_perp_g3_formal_sol}
\begin{split}
&\beta^{(3)}_{\perp}(\tau, \mathbf{k}) \\
= &D_1J_0(k_{\perp}\tau) -i\frac{\pi}{2}\frac{1}{k_{\perp}^2} \int \frac{d^2\mathbf{p}}{(2\pi)^2}\frac{ \mathbf{p}\times \mathbf{k}}{k_{\perp}p_{\perp}|\mathbf{k}-\mathbf{p}|}
\Big[b_{\eta}(\mathbf{p}), b_{\eta}( \mathbf{k}-\mathbf{p})\Big] \\
&\times\left(Y_0(s)\int_0^s dz z J_0(z)J_1\left(\frac{p_{\perp}}{k_{\perp}}z\right) J_1\left(\frac{|\mathbf{k}-\mathbf{p}|}{k_{\perp}} z\right)-J_0(s)\int_0^x dz z Y_0(z) J_1\left(\frac{p_{\perp}}{k_{\perp}}z\right) J_1\left(\frac{|\mathbf{k}-\mathbf{p}|}{k_{\perp}} z\right)\right)\\
&+\frac{(\mathbf{k}\times \mathbf{p})(\mathbf{p}\cdot \mathbf{k} -p_{\perp}^2-k_{\perp}^2)}{k_{\perp}p_{\perp}^2|\mathbf{k}-\mathbf{p}|^2} \Big[b_{\perp}(\mathbf{p}), b_{\perp}(\mathbf{k}-\mathbf{p})\Big] \\
&\times\left(Y_0(s)\int_0^s dz z J_0(z)J_0\left(\frac{p_{\perp}}{k_{\perp}}z\right)J_0\left(\frac{|\mathbf{k}-\mathbf{p}|}{k_{\perp}}z\right) - J_0(s)\int_0^x dz z Y_0(z)J_0\left(\frac{p_{\perp}}{k_{\perp}}z\right)J_0\left(\frac{|\mathbf{k}-\mathbf{p}|}{k_{\perp}}z\right) \right)\\
\end{split}
\end{equation}
We use the same strategy as described in the previous section for $\beta^{(3)}$, i.e., we proceed by  reducing integrals of three Bessel functions into integrals of two Bessel functions using Graf's formula. In this case, we have 
\begin{equation}
\begin{split}
&J_0(w_1z) J_0(w_2z) = \int_{-\pi}^{\pi} \frac{d\phi}{2\pi} J_0(w z),\\
&J_1(w_1z)J_1(w_2z) = \int_{-\pi}^{\pi} \frac{d\phi}{2\pi} e^{i\phi} J_0(wz).\\
\end{split}
\end{equation}
Here $w_1 = p_{\perp}/k_{\perp}$, $w_2 = |\mathbf{k}-\mathbf{p}|/k_{\perp}$ and  $w^2 = w_1^2+w_2^2-2w_1w_2 \cos\phi$.
After performing this manipulations, the remaining integrals with two Bessel functions can be combined into the form
\begin{equation}
\begin{split}
&Y_0(s)\int_0^s dz z J_0(z) J_0(w z) - J_0(s) \int_{0}^s dz z Y_0(z) J_0(w z)\\
=&Y_0(s)\frac{1}{w^2-1}  \Big(w x J_0(s)J_1(w s) - xJ_0(w s)J_1(s)\Big) \\
&- J_0(s) \frac{1}{w^2-1} \Big(w x J_1(w s)Y_0(s) -x J_0(w s) Y_1(s)\Big)-J_0(s) \frac{2}{\pi} \frac{1}{1-w^2}\\
=&\frac{2}{\pi}\frac{1}{1-w^2} \Big(J_0(w s)  -J_0(s)\Big).
\end{split}
\end{equation}
Using these steps, the parts  involving Bessel functions in Eq.~\eqref{eq:beta_perp_g3_formal_sol} are simplified as  
\begin{equation}
\begin{split}
&Y_0(s) \int_0^s zdz J_0(z) J_1(w_1z) J_1(w_2z)  - J_0\int_0^s zdz Y_0(z) J_1(w_1 z) J_1(w_2z)\\
=&\int_{-\pi}^{\pi} \frac{d\phi}{2\pi} e^{i\phi} \left(Y_0(s) \int_0^s zdz J_0(z) J_0(w z)  - J_0\int_0^s zdz Y_0(z) J_0(wz)\right)\\
=&\int_{-\pi}^{\pi} \frac{d\phi}{2\pi} e^{i\phi}  \frac{2}{\pi} \frac{1}{1-w^2}(J_0(ws) -J_0(s))
\end{split}
\end{equation}
and
\begin{equation}
\begin{split}
&Y_0(x) \int_0^s zdz J_0(z)J_0(w_2z) J_0(w_1z) - J_0(s)\int_0^s zdz Y_0(z) J_0(w_2z) J_0(w_1z)\\
=&\int_{-\pi}^{\pi}\frac{d\phi}{2\pi} \left(Y_0(s) \int_0^s zdz J_0(z)J_0(wz)- J_0(s)\int_0^x zdz Y_0(z) J_0(wz)\right)\\
=&\int_{-\pi}^{\pi}\frac{d\phi}{2\pi} \frac{2}{\pi} \frac{1}{1-w^2} (J_0(ws)-J_0(s)).
\end{split}
\end{equation}

With this, we arrive at the final solution for $\beta_{\perp}^{(3)}(\tau, \mathbf{k})$ 
\begin{equation}\label{eq:beta_perp_g3_final_sol}
\begin{split}
\beta^{(3)}_{\perp}(\tau, \mathbf{k}) 
= &\beta_{\perp}^{(3)}(\tau=0, \mathbf{k})J_0(k_{\perp}\tau) +\frac{i}{k_{\perp}} \int \frac{d^2\mathbf{p}}{(2\pi)^2}\frac{( \mathbf{k}\times \mathbf{p})}{2p^2_{\perp}|\mathbf{k}-\mathbf{p}|^2}\Big[b_{\eta}(\mathbf{p}), b_{\eta}( \mathbf{k}-\mathbf{p})\Big] \\
&\qquad \times\int_{-\pi}^{\pi}\frac{d\phi}{2\pi} \left(1+\frac{2\mathbf{p}\cdot(\mathbf{k}-\mathbf{p})}{w_{\perp}^2-k_{\perp}^2}\right)\left( J_0(w_{\perp}\tau)-J_0(k_{\perp}\tau)\right)  \\
 & +\frac{i}{k_{\perp}} \int \frac{d^2\mathbf{p}}{(2\pi)^2}\frac{( \mathbf{k}\times \mathbf{p})(\mathbf{p}\cdot\mathbf{k}-p_{\perp}^2-k_{\perp}^2)}{p^2_{\perp}|\mathbf{k}-\mathbf{p}|^2} \Big[b_{\perp}(\mathbf{p}), b_{\perp}(\mathbf{k}-\mathbf{p})\Big]\\&\qquad\times\int_{-\pi}^{\pi}\frac{d\phi}{2\pi} \frac{1}{w_{\perp}^2-k_{\perp}^2}\left( J_0(w_{\perp}\tau)-J_0(k_{\perp}\tau)\right)\,.
\end{split}
\end{equation}
The time dependence is completely determined by one type of  Bessel function --  Bessel function of the first kind of zero  order, $J_0(\lambda \tau)$.  The arguments of the Bessel functions are different, ranging from $||\mathbf{k}-\mathbf{p}|-p_{\perp}|$ to $|\mathbf{k}-\mathbf{p}|+p_{\perp}$.

For completeness, we supplement this solution with the  detailed expression of the order-$g^3$ initial field
\begin{equation}
\begin{split}
&\beta_{\perp}^{(3)}(\tau=0, \mathbf{k}) =\frac{i\epsilon_{ih} \mathbf{k}_h}{k_{\perp}}\int d^2\mathbf{x} e^{i\mathbf{k}\cdot\mathbf{x}} \alpha^{i,a}_{P,(3)}(\mathbf{x})U^{ad}(\mathbf{x})T^d  -i\int \frac{d^2\mathbf{p}}{(2\pi)^2} \frac{\mathbf{k}\cdot\mathbf{p}}{k_{\perp} p_{\perp}^2} [\Sigma_{(1)}(\mathbf{k}-\mathbf{p}), b_{\perp}(\mathbf{p})] \\
&\qquad \qquad\qquad-i\int \frac{d^2\mathbf{p}}{(2\pi)^2} \frac{\mathbf{p}\times\mathbf{k}}{2k_{\perp}}[\Sigma_{(1)}(\mathbf{k}-\mathbf{p}), \Sigma_{(1)}(\mathbf{p})].\\
\end{split}
\end{equation}

\section{Order-$g^5$ Solutions}

The solutions presented in previous sections are sufficient to compute the full  first saturation correction to single inclusive gluon production, as will be demonstrated in the second  paper of this series~\cite{Ming:2021b}. However, in order to compute other interesting physical quantites like the energy-momentum tensor of the classical gluon fields,  including the first saturation correction requires going beyond order-$g^3$ and finding order-$g^5$ solutions. Our motivation is to derive the energy-momentum tensor 
in a semi-analytic form to extract information about the energy density, pressure, stresses  and initial flows after the collisions; these quantities can be  as model initial conditions for a subsequent hydrodynamic evolution. 

The order-$g^5$ equations of motion are
\begin{equation}
\begin{split}
&\tau^2\partial^2_{\tau} \tilde{\beta}^{(5)}+\tau\partial_{\tau}\tilde{\beta}^{(5)} -\tilde{\beta}^{(5)}-\tau^2\partial^2\tilde{\beta}^{(5)}\\
=& -ig\tau^2 \left([\partial_i\beta_i^{(3)},\tilde{\beta}^{(1)}] + [\partial_i\beta_i^{(1)}, \tilde{\beta}^{(3)}]\right)   - g^2\tau^2[\beta_i^{(1)}, [\beta_i^{(1)}, \tilde{\beta}^{(1)}]] \\
&-2ig\tau^2\left([\beta_i^{(3)},\partial_i \tilde{\beta}^{(1)}] + [\beta_i^{(1)},\partial_i\tilde{\beta}^{(3)}]\right),
\end{split}
\end{equation}
\begin{equation}\label{eq:constraint_g5}
\begin{split}
&\partial_{\tau}\partial_i\beta_i^{(5)} = ig\left([\beta_i^{(3)},\partial_{\tau}\beta_i^{(1)}]+ [\beta_i^{(1)},\partial_{\tau}\beta_i^{(3)}]\right)+ig\left([\tilde{\beta}^{(1)},\partial_{\tau}\tilde{\beta}^{(3)}]+[\tilde{\beta}^{(3)},\partial_{\tau}\tilde{\beta}^{(1)}]\right),
\end{split}
\end{equation}
and
\begin{equation}\label{eq:betai_g5_eom}
\begin{split}
&\tau^2\partial^2_{\tau}\beta_i^{(5)}+ \tau\partial_{\tau}\beta_i^{(5)}-\tau^2(\partial^2\delta_{ij} -\partial_i\partial_j)\beta_j^{(5)}\\
 =&ig\tau^2\left([\tilde{\beta}^{(1)}, \partial_i \tilde{\beta}^{(3)}]+[\tilde{\beta}^{(3)}, \partial_i \tilde{\beta}^{(1)}]\right) -ig\tau^2\partial_j\left([\beta_j^{(3)},\beta_i^{(1)}]+ [\beta_j^{(1)},\beta_i^{(3)}]\right)\\
 &-ig\tau^2\left([\beta_j^{(1)},\partial_j\beta_i^{(3)}-\partial_i\beta_j^{(3)}]+ [\beta_j^{(3)},\partial_j\beta_i^{(1)}-\partial_i\beta_j^{(1)}]\right)\\
&+g^2\tau^2[\tilde{\beta}^{(1)}, [\tilde{\beta}_i^{(1)}, \tilde{\beta}^{(1)}]] -g^2\tau^2[\beta_j^{(1)}, [\beta_j^{(1)}, \beta_i^{(1)}]].
\end{split}
\end{equation}
with the initial conditions given in Eq.~ \eqref{eqs:ic_betai_g_g3_g5} and Eq.~\eqref{eqs:ic_beta_g_g3_g5}.

Again, as in the previous section for $g^3$-order, one can explicitly show that from the constraint equation for $\partial_i\beta_i^{(5)}$ and all the lower order solutions, the second order differential equation for $\partial_i\beta_i^{(5)}$ can be derived. This leaves only the curl of $\beta_i^{(5)}$ as an independent field, see Appendix \ref{sec:nondaynamical_Lambda}.

We want to write down the equations for the independent fields in momentum space. Performing  the decomposition 
\begin{equation}
\beta^{(5)}_i(\tau, \mathbf{x}) = \epsilon_{il}\partial_l \chi^{(5)} (\tau, \mathbf{x})+\partial_i\Lambda^{(5)}(\tau, \mathbf{x}), 
\end{equation}
we can separate the equations for the curl and the gradient of  $\beta^{(5)}_i$.  In momentum space, we have  
\begin{equation}
\beta_{\perp}^{(5)}(\tau, \mathbf{k}) = k_{\perp} \chi^{(5)}(\tau, \mathbf{k}) = \frac{ i \epsilon_{ih}\mathbf{k}_h}{k_{\perp}}\beta^{(5)}_i(\tau,\mathbf{k}),
\end{equation}
\begin{equation}
\Lambda^{(5)}(\tau, \mathbf{k}) = \frac{i\mathbf{k}_i}{k_{\perp}^2} \beta_i^{(5)}(\tau, \mathbf{k}).
\end{equation}

The  two dynamical equations governing the time evolutions of $\beta_{\perp}^{(5)}(\tau, \mathbf{k})$ and  $\beta^{(5)}(\tau, \mathbf{k})$ are
\begin{equation}
\begin{split}
&\tau^2\partial^2_{\tau} \tilde{\beta}^{(5)}(\tau, \mathbf{k})+\tau\partial_{\tau}\tilde{\beta}^{(5)}(\tau, \mathbf{k})+((k_{\perp}\tau)^2 - 1)\tilde{\beta}^{(5)}(\tau, \mathbf{k}) = S^{(5)}_{\eta}(\tau, \mathbf{k}),  \\
&\tau^2 \partial^2_{\tau}\beta^{(5)}_{\perp}(\tau, \mathbf{k}) + \tau \partial_{\tau} \beta_{\perp}^{(5)}(\tau, \mathbf{k}) + (k_{\perp}\tau)^2 \beta_{\perp}^{(5)}(\tau, \mathbf{k}) = \frac{i\epsilon_{ih}\mathbf{k}_h}{k_{\perp}}S_i^{(5)}(\tau, \mathbf{k}).
\end{split}
\end{equation}
On the other hand,  the constraint equation is sufficient to determine $\Lambda^{(5)}(\tau,\mathbf{k})$
\begin{equation}
\partial_{\tau} \Lambda^{(5)}(\tau, \mathbf{k}) = -\frac{1}{k_{\perp}^2} S^{(5)}_{\Lambda}(\tau, \mathbf{k}).
\end{equation}
The source terms in momentum space are
\begin{equation}\label{eq:S_eta_g5}
\begin{split}
S^{(5)}_{\eta}(\tau,\mathbf{k}) = &-g\tau^3 \int\frac{d^2\mathbf{q}}{(2\pi)^2} (\mathbf{k}+\mathbf{q})_i\left[\beta_i^{(1)}(\tau, \mathbf{k}-\mathbf{q}), \beta^{(3)}(\tau, \mathbf{q})\right]\\
&-g\tau^3 \int\frac{d^2\mathbf{q}}{(2\pi)^2}  (2\mathbf{k}-\mathbf{q})_i\left[\beta_i^{(3)}(\tau, \mathbf{q}), \beta^{(1)}(\tau, \mathbf{k}-\mathbf{q})\right]\\
&-g^2\tau^3\int\frac{d^2\mathbf{q}}{(2\pi)^2} \int\frac{d^2\mathbf{p}}{(2\pi)^2} \left[\beta_i^{(1)}(\tau, \mathbf{k}-\mathbf{q}), \left[\beta_i^{(1)}(\tau, \mathbf{p}), \beta^{(1)}(\tau, \mathbf{q}-\mathbf{p})\right]\right],
\end{split}
\end{equation}
\begin{equation}\label{eq:S_perp_g5}
\begin{split}
S^{(5)}_{\perp}(\tau, \mathbf{k}) =&\frac{i\epsilon^{il}\mathbf{k}_l}{k_{\perp}} \tau^2 \int \frac{d^2\mathbf{q}}{(2\pi)^2}  (\mathbf{k}-2\mathbf{q})_i\left( [\tilde{\beta}^{(3)}(\tau, \mathbf{q}), \tilde{\beta}^{(1)}(\tau, \mathbf{k}-\mathbf{q})]+[\beta_j^{(3)}(\tau, \mathbf{q}), \beta_j^{(1)}(\tau, \mathbf{k}-\mathbf{q})] \right)\\
&\qquad+\left(-(2\mathbf{k}-\mathbf{q})_j [\beta_j^{(3)}(\tau, \mathbf{q}), \beta_i^{(1)}(\tau, \mathbf{k}-\mathbf{q})]+(\mathbf{k}+\mathbf{q})_j[\beta_i^{(3)}(\tau, \mathbf{q}), \beta_j^{(1)}(\tau, \mathbf{k}-\mathbf{q})] \right)\\
&\qquad+ \int \frac{d^2\mathbf{q}}{(2\pi)^2}  \frac{d^2\mathbf{p}}{(2\pi)^2} \Big( [\tilde{\beta}^{(1)}(\tau, \mathbf{k}-\mathbf{q}), [\beta_i^{(1)}(\tau, \mathbf{p}), \tilde{\beta}^{(1)}(\tau, \mathbf{q}-\mathbf{p})]] \\
&\qquad\qquad+ [\beta_j^{(1)}(\tau, \mathbf{k}-\mathbf{q}), [\beta_i^{(1)}(\tau, \mathbf{p}), \beta_j^{(1)}(\tau, \mathbf{q}-\mathbf{p})]]\Big),
\end{split}
\end{equation}
and 
\begin{equation}\label{eq:S_Lambda_g5}
\begin{split}
S_{\Lambda}^{(5)}(\tau, \mathbf{k}) =& ig \int \frac{d^2\mathbf{q}}{(2\pi)^2} \left[\beta_i^{(3)}(\tau, \mathbf{q}), \partial_{\tau} \beta_i^{(1)}(\tau, \mathbf{k}-\mathbf{q})\right]  - \left[\partial_{\tau}\beta_i^{(3)}(\tau, \mathbf{q}), \beta_i^{(1)}(\tau, \mathbf{k}-\mathbf{q})\right]\\
&+\left[\tilde{\beta}^{(3)}(\tau, \mathbf{q}), \partial_{\tau}\tilde{\beta}^{(1)}(\tau, \mathbf{k}-\mathbf{q})\right] - \left[\partial_{\tau}\tilde{\beta}^{(3)}(\tau, \mathbf{q}), \tilde{\beta}^{(1)}(\tau, \mathbf{k}-\mathbf{q})\right]\,.
\end{split}
\end{equation}
We now have everything ready to find  all the components of the gluon fields at order-$g^5$.

\subsection{Solving for $\beta^{(5)}(\tau, \mathbf{k})$}
From eq. \eqref{eq:S_eta_g5}, the explicit expression of the source term $S_{\eta}^{(5)}(\tau, \mathbf{k})$ is
\begin{equation}
\begin{split}
&S^{(5)}_{\eta}(\tau,\mathbf{k}) \\
=&i\tau^2 \int\frac{d^2\mathbf{q}}{(2\pi)^2} \frac{2\mathbf{k}\times \mathbf{q}}{q^2_{\perp}|\mathbf{k}-\mathbf{q}|} \Big([q_{\perp}\beta_{\perp}^{(3)}(\tau=0, \mathbf{q}), b_{\eta}(\mathbf{k}-\mathbf{q})] - \left[ 2\beta^{(3)}(\tau=0,\mathbf{k}-\mathbf{q}) , b_{\perp}(\mathbf{q})\right]\Big)\\
&\qquad\qquad \times J_0(q_{\perp}\tau)J_1(|\mathbf{k}-\mathbf{q}|\tau)\\
&+\tau^2\int\frac{d^2\mathbf{q}}{(2\pi)^2}\frac{d^2\mathbf{p}}{(2\pi)^2}\frac{2\mathbf{q}\times \mathbf{k}}{|\mathbf{k}-\mathbf{q}|^2}\frac{\mathbf{q}\times \mathbf{p}}{p_{\perp}^2|\mathbf{q}-\mathbf{p}|^2}\Big[[b_{\perp}(\mathbf{p}), b_{\eta}(\mathbf{q}-\mathbf{p})], b_{\perp}(\mathbf{k}-\mathbf{q})\Big] \\
&\qquad\times\int_{-\pi}^{\pi}\frac{d\phi}{2\pi}\frac{\mathbf{q}\cdot(\mathbf{q}-2\mathbf{p})+w_{\perp}^2}{q_{\perp}^2-w_{\perp}^2}\left(\frac{J_1(w_{\perp}\tau)}{w_{\perp}}-\frac{J_1(q_{\perp}\tau)}{q_{\perp}}\right)J_0(|\mathbf{k}-\mathbf{q}|\tau) \\
&+\tau^2 \int\frac{d^2\mathbf{q}}{(2\pi)^2}  \frac{d^2\mathbf{p}}{(2\pi)^2}\frac{2\mathbf{k}\times \mathbf{q}}{q^2_{\perp}|\mathbf{k}-\mathbf{q}|}\frac{ \mathbf{q}\times \mathbf{p}}{2p^2_{\perp}|\mathbf{q}-\mathbf{p}|^2}\int_{-\pi}^{\pi}\frac{d\phi}{2\pi}  \frac{1}{q_{\perp}^2-w_{\perp}^2}\Big((-p_{\perp}^2-|\mathbf{q}-\mathbf{p}|^2+w_{\perp}^2)\\
&\qquad\Big[\big[b_{\eta}(\mathbf{p}), b_{\eta}( \mathbf{q}-\mathbf{p})\big],  b_{\eta}(\mathbf{k}-\mathbf{q})\Big]+2(\mathbf{p}\cdot\mathbf{q}-p_{\perp}^2-q_{\perp}^2) \Big[\big[b_{\perp}(\mathbf{p}), b_{\perp}(\mathbf{q}-\mathbf{p})\big],  b_{\eta}(\mathbf{k}-\mathbf{q})\Big] \Big)\\
&\qquad\qquad\times (J_0(w_{\perp}\tau)-J_0(q_{\perp}\tau))J_1(|\mathbf{k}-\mathbf{q}|\tau)\\
&+\tau^2 \int\frac{d^2\mathbf{q}}{(2\pi)^2} \int\frac{d^2\mathbf{p}}{(2\pi)^2}\frac{(2\mathbf{k}-\mathbf{q})\cdot \mathbf{q}}{q_{\perp}^2|\mathbf{k}-\mathbf{q}|}   \frac{\mathbf{q}\cdot(\mathbf{q}-2\mathbf{p})}{4p_{\perp}^2 |\mathbf{q}-\mathbf{p}|^2} \int_{-\pi}^{\pi} \frac{d\phi}{2\pi}\frac{1}{w_{\perp}^2}\Big( (-q_{\perp}^2+w_{\perp}^2+2\mathbf{p}\cdot(\mathbf{q}-\mathbf{p}))\\
&\qquad \Big[\left[b_{\eta}(\mathbf{p}), b_{\eta}(\mathbf{q}-\mathbf{p})\right], b_{\eta}(\mathbf{k}-\mathbf{q})\Big]+2\mathbf{p}\cdot(\mathbf{q}-\mathbf{p}) \Big[\left[b_{\perp}(\mathbf{p}), b_{\perp}(\mathbf{q}-\mathbf{p})\right], b_{\eta}(\mathbf{k}-\mathbf{q})\Big]   \Big)\\
&\qquad\qquad\times (1- J_0(w_{\perp}\tau))J_1(|\mathbf{k}-\mathbf{q}|\tau) \\
&+\tau^2\int\frac{d^2\mathbf{q}}{(2\pi)^2} \int\frac{d^2\mathbf{p}}{(2\pi)^2} \frac{(\mathbf{q}-\mathbf{p})\cdot \mathbf{p}}{p_{\perp}^2|\mathbf{q}-\mathbf{p}|^2|\mathbf{k}-\mathbf{q}|} \Big[b_{\perp}(\mathbf{p}), [b_{\perp}(\mathbf{q}-\mathbf{p}), b_{\eta}(\mathbf{k}-\mathbf{q})]\Big] \\
&\qquad\qquad \times J_0(|\mathbf{q}-\mathbf{p}|\tau)J_0(p_{\perp}\tau) J_1(|\mathbf{k}-\mathbf{q}|\tau).\\
\end{split}
\end{equation}
The last term involves three Bessel functions. The product of the three Bessel functions can be rewritten as 
\begin{equation}
J_0(|\mathbf{q}-\mathbf{p}|\tau)J_0(p_{\perp}\tau) J_1(|\mathbf{k}-\mathbf{q}|\tau) = \int_{-\pi}^{\pi} \frac{d\phi}{2\pi} J_0(w_{\perp}\tau) J_1(|\mathbf{k}-\mathbf{q}|\tau).
\end{equation}

\begin{figure}[t]
\begin{subfigure}{1.0\textwidth}
\centering 
\includegraphics[width=0.8\textwidth]{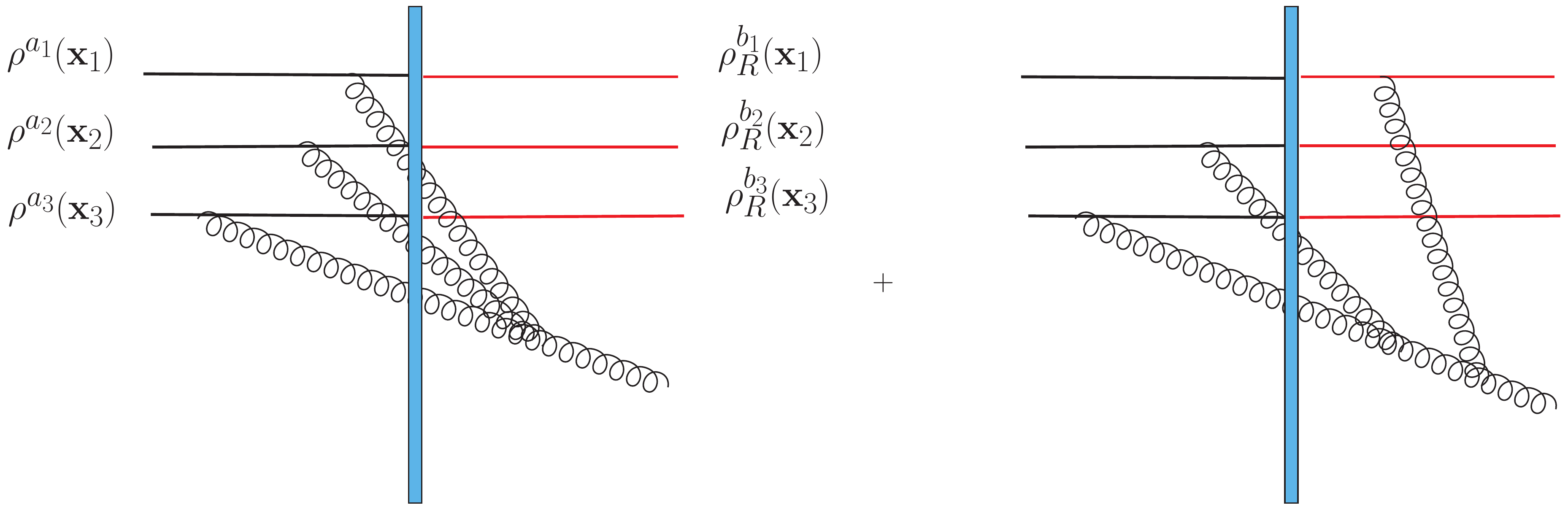}
\caption*{}
\end{subfigure}
\begin{subfigure}{1.0\textwidth}
\centering 
\includegraphics[width=0.8\textwidth]{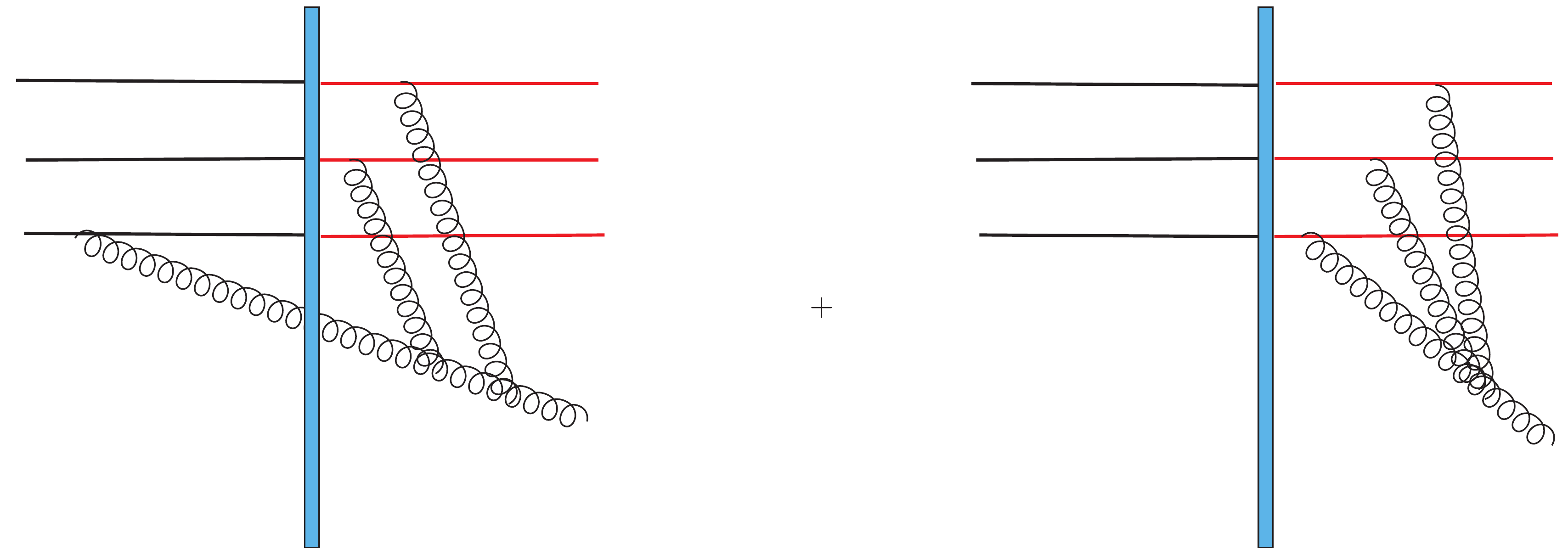}
\caption*{}
\end{subfigure}
\caption{Schematic representation of the color structure for the double commutators in $\beta^{(5)}(\tau, \mathbf{k})$. The shaded bar represents the target nucleus. The eikonally rotated color charge density $\rho_R$ is represented using red lines. }
\label{fig:beta_g5_color_structure_one_two}
\end{figure}

\begin{figure}[t]
\centering 
\includegraphics[width=0.8\textwidth]{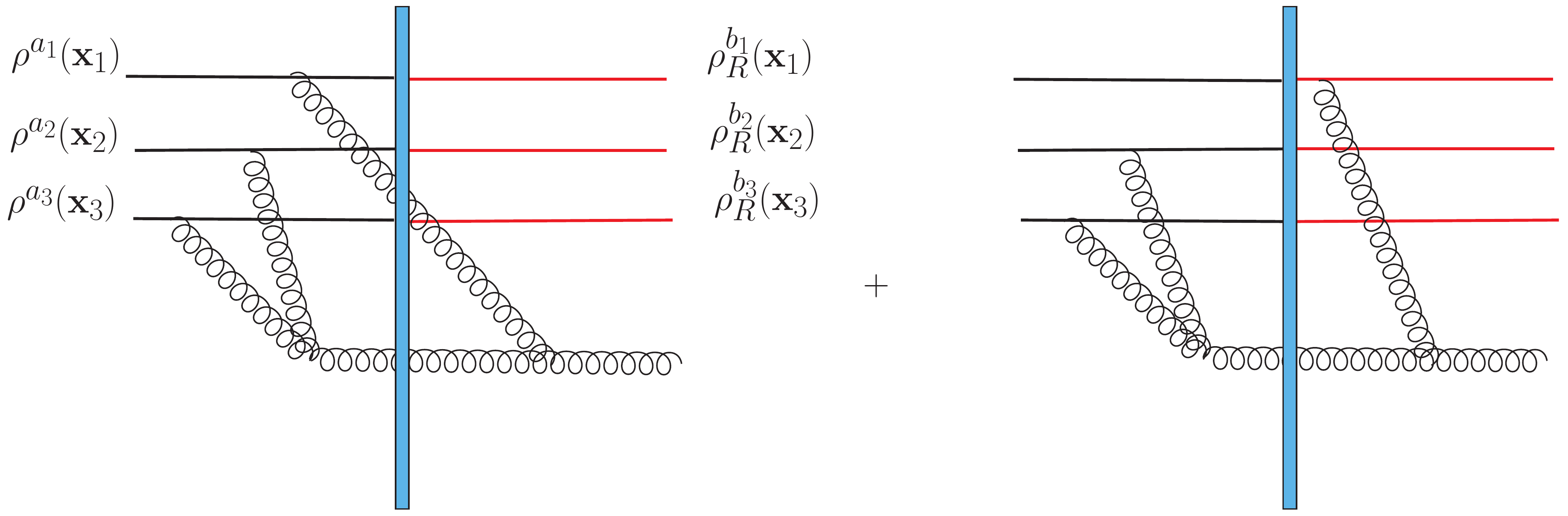}
\caption{Schematic representation of the color structure for the single commutators involving $\beta^{(3)}, \beta_{\perp}^{(3)}$ in $\beta^{(5)}(\tau, \mathbf{k})$. The shaded bar represents the target nucleus. The eikonally rotated color charge density $\rho_R$ is represented using red lines. }
\label{fig:beta_g5_color_structure_three}
\end{figure}

To solve for $\beta^{(5)}(\tau, \mathbf{k})$, we repeat the procedure used when solving for $\beta^{(3)}(\tau, \mathbf{k})$. The formal solution obtained by the method of variation of parameters is
\begin{equation}
\tilde{\beta}^{(5)} (\tau,\mathbf{k}) = C_1^{\prime} J_1(s) + C_2^{\prime} Y_1(s) + \frac{\pi}{2} \int_0^s dz (J_1(z) Y_1(s) - J_1(s)Y_1(z)) \frac{S^{(5)}_{\eta}(z)}{z}
\end{equation}
with the coefficients $C_1^{\prime}$ and $C_2^{\prime}$ fixed by initial conditions at order-$g^5$
\begin{equation}
C_1^{\prime} = \frac{2}{k_{\perp}}\beta^{(5)}(\tau=0, \mathbf{k}), \qquad C_2^{\prime}=0.
\end{equation}
The time-dependent factors in each term of $S_{\eta}^{(5)}(\tau, \mathbf{k})$ can always be reduced to just one single Bessel function, Bessel function of first kind with order one $J_1(c \tau)$, although different  terms might have different values of argument $c$. This reduction is done by using Graf's formula
\begin{equation}
J_0(a\tau) J_1(b\tau) = \int_{-\pi}^{\pi} \frac{d\psi}{2\pi} \frac{b^2-a^2+c^2}{2bc} J_1(c\tau)
\end{equation}
with $c^2 = a^2+b^2-2ab\cos\psi$. 
Next, the formula
\begin{equation}\label{eq:J1ctau_repalce}
\begin{split}
&\frac{\pi}{2} \Big(Y_1(s)\int_0^{s} dz z J_1(z) J_1(\lambda z) - J_1(s) \int_0^s dz z Y_1(z) J_1(\lambda z)\Big)\\
 =& \frac{1}{1-\lambda^2}(J_1(\lambda s)-\lambda J_1(s))\\
 =&\frac{ck_{\perp}^2}{k_{\perp}^2 -c^2}\left(\frac{J_1(c\tau)}{c} - \frac{J_1(k_{\perp}\tau)}{k_{\perp}}\right)\,
\end{split}
\end{equation}
helps to carry out the integration over the source terms in the formal solution. Note that $\lambda = c/k_{\perp}$. 
Schematically, our recipe is to replace each $J_0(a\tau)J_1(b\tau)$ factor in $S_{\eta}^{(5)}(\tau, \mathbf{k})$ by 
\begin{equation}
J_0(a\tau) J_1(b\tau) \longrightarrow \frac{1}{2b} \int_{-\pi}^{\pi} \frac{d\psi}{2\pi} \frac{b^2-a^2+c^2}{k_{\perp}^2-c^2}\left(\frac{J_1(c\tau)}{c} - \frac{J_1(k_{\perp}\tau)}{k_{\perp}}\right).
\end{equation}
To be specific, we need the following replacements
\begin{equation}
\begin{split}
&J_0(q_{\perp}\tau)J_1(|\mathbf{k}-\mathbf{q}|\tau) 
\longrightarrow \frac{1}{2|\mathbf{k}-\mathbf{q}|}\int_{-\pi}^{\pi} \frac{d\phi'}{2\pi} \frac{\mathbf{k}\cdot(\mathbf{k}-2\mathbf{q}) + w_{\perp}^{\prime 2}}{k_{\perp}^2-w_{\perp}^{\prime 2}}\Big(\frac{J_1(w_{\perp}^{\prime}\tau)}{w'_{\perp}}- \frac{J_1(k_{\perp}\tau)}{k_{\perp}}\Big),\\
&J_1(q_{\perp}\tau)J_0(|\mathbf{k}-\mathbf{q}|\tau)
\longrightarrow  \frac{1}{2q_{\perp}}\int_{-\pi}^{\pi} \frac{d\phi'}{2\pi} \frac{-\mathbf{k}\cdot(\mathbf{k}-2\mathbf{q}) + w_{\perp}^{\prime 2}}{k_{\perp}^2-w_{\perp}^{\prime 2}}\Big(\frac{J_1(w_{\perp}^{\prime}\tau)}{w'_{\perp}}- \frac{J_1(k_{\perp}\tau)}{k_{\perp}}\Big),\\
&J_0(w_{\perp})J_1(|\mathbf{k}-\mathbf{q}|\tau)  \longrightarrow 
\frac{1}{2|\mathbf{k}-\mathbf{q}|}\int_{-\pi}^{\pi}\frac{d\theta}{2\pi}\frac{|\mathbf{k}-\mathbf{q}|^2-w_{\perp}^2 + u_{\perp}^2}{k_{\perp}^2-u_{\perp}^2} \Big(\frac{J_1(u_{\perp}\tau)}{u_{\perp}}- \frac{J_1(k_{\perp}\tau)}{k_{\perp}}\Big),\\
&J_1(w_{\perp})J_0(|\mathbf{k}-\mathbf{q}|\tau)  \longrightarrow \frac{1}{2w_{\perp}}\int_{-\pi}^{\pi}\frac{d\theta}{2\pi}\frac{w_{\perp}^2-|\mathbf{k}-\mathbf{q}|^2 + u_{\perp}^2}{ k_{\perp}^2-u_{\perp}^2}\Big(\frac{J_1(u_{\perp}\tau)}{u_{\perp}}- \frac{J_1(k_{\perp}\tau)}{k_{\perp}}\Big)\\
\end{split}
\end{equation}
with 
\begin{equation}\label{eqs:defs_theta_phiprime}
\begin{split}
&w'_{\perp} =\sqrt{q_{\perp}^2+|\mathbf{k}-\mathbf{q}|^2-2q_{\perp}|\mathbf{k}-\mathbf{q}|\cos\phi'},\\
 &u_{\perp}= \sqrt{ w_{\perp}^2+|\mathbf{k}-\mathbf{q}|^2- 2w_{\perp}|\mathbf{k}-\mathbf{q}|\cos\theta}.\\
\end{split}
\end{equation}
Using the above recipe, we obtained the order-$g^5$ solution $\beta^{(5)}(\tau, \mathbf{k})$  
\begin{equation}\label{eq:beta_g5_final_sol}
\begin{split}
&\beta^{(5)}(\tau, \mathbf{k})\\
 = & 2\beta^{(5)}(\tau=0,\mathbf{k})\frac{J_1(k_{\perp}\tau)}{k_{\perp}\tau}+i \int\frac{d^2\mathbf{q}}{(2\pi)^2} \frac{\mathbf{k}\times \mathbf{q}}{q^2_{\perp}|\mathbf{k}-\mathbf{q}|^2} \Big([q_{\perp}\beta_{\perp}^{(3)}(\tau=0, \mathbf{q}), b_{\eta}(\mathbf{k}-\mathbf{q})] \\
&\qquad- \left[ 2\beta^{(3)}(\tau=0,\mathbf{k}-\mathbf{q}) , b_{\perp}(\mathbf{q})\right]\Big) \int_{-\pi}^{\pi} \frac{d\phi'}{2\pi} \frac{\mathbf{k}\cdot(\mathbf{k}-2\mathbf{q}) + w_{\perp}^{\prime 2}}{k_{\perp}^2-w_{\perp}^{\prime 2}}\Big(\frac{J_1(w_{\perp}^{\prime}\tau)}{w'_{\perp}\tau}- \frac{J_1(k_{\perp}\tau)}{k_{\perp}\tau}\Big)\\
&+\int\frac{d^2\mathbf{q}}{(2\pi)^2}\frac{d^2\mathbf{p}}{(2\pi)^2}\frac{2\mathbf{q}\times \mathbf{k}}{|\mathbf{k}-\mathbf{q}|^2}\frac{\mathbf{q}\times \mathbf{p}}{p_{\perp}^2|\mathbf{q}-\mathbf{p}|^2}\Big[[b_{\perp}(\mathbf{p}), b_{\eta}(\mathbf{q}-\mathbf{p})], b_{\perp}(\mathbf{k}-\mathbf{q})\Big] \\
&\qquad \times \int_{-\pi}^{\pi}\frac{d\phi}{2\pi}\frac{\mathbf{q}\cdot(\mathbf{q}-2\mathbf{p})+w_{\perp}^2}{q_{\perp}^2-w_{\perp}^2}\Bigg\{\frac{1}{2w^2_{\perp}}\int_{-\pi}^{\pi}\frac{d\theta}{2\pi}\frac{w_{\perp}^2-|\mathbf{k}-\mathbf{q}|^2 + u_{\perp}^2}{ k_{\perp}^2-u_{\perp}^2}\Big(\frac{J_1(u_{\perp}\tau)}{u_{\perp}\tau}- \frac{J_1(k_{\perp}\tau)}{k_{\perp}\tau}\Big)\\
&\qquad \qquad \qquad - \frac{1}{2q^2_{\perp}}\int_{-\pi}^{\pi} \frac{d\phi'}{2\pi} \frac{-\mathbf{k}\cdot(\mathbf{k}-2\mathbf{q}) + w_{\perp}^{\prime 2}}{k_{\perp}^2-w_{\perp}^{\prime 2}}\Big(\frac{J_1(w_{\perp}^{\prime}\tau)}{w'_{\perp}\tau}- \frac{J_1(k_{\perp}\tau)}{k_{\perp}\tau}\Big)\Bigg\}\\
&+ \int\frac{d^2\mathbf{q}}{(2\pi)^2}  \frac{d^2\mathbf{p}}{(2\pi)^2}\frac{\mathbf{k}\times \mathbf{q}}{q^2_{\perp}|\mathbf{k}-\mathbf{q}|^2}\frac{ \mathbf{q}\times \mathbf{p}}{2p^2_{\perp}|\mathbf{q}-\mathbf{p}|^2}\int_{-\pi}^{\pi}\frac{d\phi}{2\pi}  \frac{1}{q_{\perp}^2-w_{\perp}^2}\Big((-p_{\perp}^2-|\mathbf{q}-\mathbf{p}|^2+w_{\perp}^2)\\
&\qquad\Big[\big[b_{\eta}(\mathbf{p}), b_{\eta}( \mathbf{q}-\mathbf{p})\big],  b_{\eta}(\mathbf{k}-\mathbf{q})\Big]+2(\mathbf{p}\cdot\mathbf{q}-p_{\perp}^2-q_{\perp}^2) \Big[\big[b_{\perp}(\mathbf{p}), b_{\perp}(\mathbf{q}-\mathbf{p})\big],  b_{\eta}(\mathbf{k}-\mathbf{q})\Big] \Big)\\
&\qquad\qquad\times\Bigg\{ \int_{-\pi}^{\pi}\frac{d\theta}{2\pi}\frac{|\mathbf{k}-\mathbf{q}|^2-w_{\perp}^2 + u_{\perp}^2}{k_{\perp}^2-u_{\perp}^2} \Big(\frac{J_1(u_{\perp}\tau)}{u_{\perp}\tau}- \frac{J_1(k_{\perp}\tau)}{k_{\perp}\tau}\Big)\\
&\qquad\qquad \qquad-\int_{-\pi}^{\pi} \frac{d\phi'}{2\pi} \frac{\mathbf{k}\cdot(\mathbf{k}-2\mathbf{q}) + w_{\perp}^{\prime 2}}{k_{\perp}^2-w_{\perp}^{\prime 2}}\Big(\frac{J_1(w_{\perp}^{\prime}\tau)}{w'_{\perp}\tau}- \frac{J_1(k_{\perp}\tau)}{k_{\perp}\tau}\Big)\Bigg\}\\
&+ \int\frac{d^2\mathbf{q}}{(2\pi)^2} \frac{d^2\mathbf{p}}{(2\pi)^2}\frac{(2\mathbf{k}-\mathbf{q})\cdot \mathbf{q}}{q_{\perp}^2|\mathbf{k}-\mathbf{q}|}   \frac{\mathbf{q}\cdot(\mathbf{q}-2\mathbf{p})}{4p_{\perp}^2 |\mathbf{q}-\mathbf{p}|^2} \int_{-\pi}^{\pi} \frac{d\phi}{2\pi}\frac{1}{w_{\perp}^2}\Big( (-q_{\perp}^2+w_{\perp}^2+2\mathbf{p}\cdot(\mathbf{q}-\mathbf{p}))\\
&\qquad \qquad \Big[\left[b_{\eta}(\mathbf{p}), b_{\eta}(\mathbf{q}-\mathbf{p})\right], b_{\eta}(\mathbf{k}-\mathbf{q})\Big]+2\mathbf{p}\cdot(\mathbf{q}-\mathbf{p}) \Big[\left[b_{\perp}(\mathbf{p}), b_{\perp}(\mathbf{q}-\mathbf{p})\right], b_{\eta}(\mathbf{k}-\mathbf{q})\Big]   \Big)\\
&\qquad\qquad \times\Bigg\{\frac{|\mathbf{k}-\mathbf{q}| } {k_{\perp}^2-|\mathbf{k}-\mathbf{q}|^2} \Big(\frac{J_1(|\mathbf{k}-\mathbf{q}|\tau)}{|\mathbf{k}-\mathbf{q}|\tau}- \frac{J_1(k_{\perp}\tau)}{k_{\perp}\tau}\Big)\\
&\qquad\qquad \qquad -\frac{1}{2|\mathbf{k}-\mathbf{q}|}\int_{-\pi}^{\pi}\frac{d\theta}{2\pi}\frac{|\mathbf{k}-\mathbf{q}|^2-w_{\perp}^2 + u_{\perp}^2}{k_{\perp}^2-u_{\perp}^2} \Big(\frac{J_1(u_{\perp}\tau)}{u_{\perp}\tau}- \frac{J_1(k_{\perp}\tau)}{k_{\perp}\tau}\Big)\Bigg\}\\
&+\int\frac{d^2\mathbf{q}}{(2\pi)^2} \frac{d^2\mathbf{p}}{(2\pi)^2} \frac{(\mathbf{q}-\mathbf{p})\cdot \mathbf{p}}{2p_{\perp}^2|\mathbf{q}-\mathbf{p}|^2|\mathbf{k}-\mathbf{q}|^2} \Big[b_{\perp}(\mathbf{p}), [b_{\perp}(\mathbf{q}-\mathbf{p}), b_{\eta}(\mathbf{k}-\mathbf{q})]\Big] \\
&\qquad\qquad \times \int_{-\pi}^{\pi}\frac{d\phi}{2\pi} \int_{-\pi}^{\pi}\frac{d\theta}{2\pi}\frac{|\mathbf{k}-\mathbf{q}|^2-w_{\perp}^2 + u_{\perp}^2}{k_{\perp}^2-u_{\perp}^2} \Big(\frac{J_1(u_{\perp}\tau)}{u_{\perp}\tau}- \frac{J_1(k_{\perp}\tau)}{k_{\perp}\tau}\Big)\,.
\end{split}
\end{equation}
The order-$g^5$ solution $\beta^{(5)}(\tau, \mathbf{k})$ involves interactions of three color charges in the proton as evidenced by the double color commutators. 
From the definitions of $b_{\perp}(\mathbf{k}), b_{\eta}(\mathbf{k})$ in eqs. \eqref{eq:b_perp_physical_meaning} and \eqref{eq:b_eta_physical_meaning}, the color structure of the three double commutators can be diagrammatically illustrated. The $\left[[b_{\perp}(\mathbf{p}), b_{\eta}(\mathbf{q}-\mathbf{p})], b_{\perp}(\mathbf{k}-\mathbf{q})\right] $, $\left[[b_{\eta}(\mathbf{p}), b_{\eta}( \mathbf{q}-\mathbf{p})],  b_{\eta}(\mathbf{k}-\mathbf{q})\right]$ and $\left[b_{\perp}(\mathbf{p}), [b_{\perp}(\mathbf{q}-\mathbf{p}), b_{\eta}(\mathbf{k}-\mathbf{q})]\right]$ are shown in Figs. \ref{fig:beta_g5_color_structure_one_two}. 
As for the single commutators containing $\beta_{\perp}^{(3)}(\tau=0, \mathbf{q})$ and $\beta^{(3)}(\tau=0, \mathbf{q})$, they are represented by many topologically different diagrams. Two of them are shown in Fig. \ref{fig:beta_g5_color_structure_three}. 


The time-dependent factors are expressed solely in terms of one type of Bessel functions $J_1(a\tau)/a\tau$ with the cost of introducing  two auxilliary angular integrals. Typical terms also contain two transverse momentum integrations. These transverse momentum integrations reflect the momentum exchanges between the projectile and the target during the collisions. 

\subsection{Solving for $\beta_{\perp}^{(5)}(\tau, \mathbf{k})$}
From eq. \eqref{eq:S_perp_g5}, the explicit expression of the source term $S_{\perp}^{(5)}(\tau, \mathbf{k})$ is 
\begin{equation}
\begin{split}
&S_{\perp}^{(5)}(\tau,\mathbf{k})\\
=&\frac{i\tau^2}{k_{\perp}}\int \frac{d^2\mathbf{q}}{(2\pi)^2}  \frac{2(\mathbf{k}\times \mathbf{q})}{q_{\perp} |\mathbf{k}-\mathbf{q}|}\Big[2\beta^{(3)}(\tau=0,\mathbf{q}), b_{\eta}(\mathbf{k}-\mathbf{q})\Big]J_1(q_{\perp}\tau) J_1(|\mathbf{k}-\mathbf{q}|\tau)\\
&+\frac{i\tau^2}{k_{\perp}}\int \frac{d^2\mathbf{q}}{(2\pi)^2}\frac{(k_{\perp}^2+q_{\perp}^2-\mathbf{k}\cdot\mathbf{q})(2\mathbf{k}\times \mathbf{q})}{q_{\perp} |\mathbf{k}-\mathbf{q}|^2} \left[\beta_{\perp}^{(3)}(\tau=0, \mathbf{q}),b_{\perp}(\mathbf{k}-\mathbf{q})\right]J_0(q_{\perp}\tau)J_0(|\mathbf{k}-\mathbf{q}|\tau)\\
&-\frac{\tau^2}{k_{\perp}}\int \frac{d^2\mathbf{q}}{(2\pi)^2} \int \frac{d^2\mathbf{p}}{(2\pi)^2} \frac{2(\mathbf{k}\times \mathbf{q})(\mathbf{q}\times \mathbf{p})}{p_{\perp}^2|\mathbf{q}-\mathbf{p}|^2|\mathbf{k}-\mathbf{q}|}\Big[[b_{\perp}(\mathbf{p}), b_{\eta}(\mathbf{q}-\mathbf{p})], b_{\eta}(\mathbf{k}-\mathbf{q})\Big] \\
&\qquad\qquad \times\int_{-\pi}^{\pi}\frac{d\phi}{2\pi}\frac{\mathbf{q}\cdot(\mathbf{q}-2\mathbf{p})+w_{\perp}^2}{q_{\perp}^2-w_{\perp}^2}\left(\frac{J_1(w_{\perp}\tau)}{w_{\perp}}-\frac{J_1(q_{\perp}\tau)}{q_{\perp}}\right)J_1(|\mathbf{k}-\mathbf{q}|\tau)\\
& +\frac{\tau^2}{k_{\perp}}\int \frac{d^2\mathbf{q}}{(2\pi)^2} \frac{d^2\mathbf{p}}{(2\pi)^2}\frac{(k_{\perp}^2+q_{\perp}^2-\mathbf{k}\cdot\mathbf{q})(\mathbf{k}\times \mathbf{q})( \mathbf{q}\times \mathbf{p})}{q^2_{\perp} |\mathbf{k}-\mathbf{q}|^2p^2_{\perp}|\mathbf{q}-\mathbf{p}|^2} \int_{-\pi}^{\pi}\frac{d\phi}{2\pi} \Big((w_{\perp}^2-p_{\perp}^2-|\mathbf{q}-\mathbf{p}|^2)\\
&\qquad\times\Big[[b_{\eta}(\mathbf{p}), b_{\eta}( \mathbf{q}-\mathbf{p})], b_{\perp}(\mathbf{k}-\mathbf{q})\Big]+2(\mathbf{p}\cdot\mathbf{q}-p_{\perp}^2-q_{\perp}^2) \Big[[b_{\perp}(\mathbf{p}), b_{\perp}(\mathbf{q}-\mathbf{p})], b_{\perp}(\mathbf{k}-\mathbf{q})\Big] \Big)\\
&\qquad \qquad\times  \frac{1}{q_{\perp}^2-w_{\perp}^2}(J_0(w_{\perp}\tau)-J_0(q_{\perp}\tau))J_0(|\mathbf{k}-\mathbf{q}|\tau)\\
&+\frac{\tau^2}{k_{\perp}}\int \frac{d^2\mathbf{q}}{(2\pi)^2}\frac{d^2\mathbf{p}}{(2\pi)^2} \frac{(2\mathbf{k}-\mathbf{q})\cdot \mathbf{q}\,\mathbf{k}\cdot(\mathbf{k}-\mathbf{q}) \mathbf{q}\cdot(\mathbf{q}-2\mathbf{p})}{4p_{\perp}^2 |\mathbf{q}-\mathbf{p}|^2|\mathbf{k}-\mathbf{q}|^2q_{\perp}^2} \int_{-\pi}^{\pi} \frac{d\phi}{2\pi}\Big((w_{\perp}^2-q_{\perp}^2+2\mathbf{p}\cdot(\mathbf{q}-\mathbf{p}))\\
&\qquad\times \left[[b_{\eta}(\mathbf{p}), b_{\eta}(\mathbf{q}-\mathbf{p})],b_{\perp}(\mathbf{k}-\mathbf{q})\right] + 2\mathbf{p}\cdot(\mathbf{q}-\mathbf{p})\Big[b_{\perp}(\mathbf{p}), b_{\perp}(\mathbf{q}-\mathbf{p})],b_{\perp}(\mathbf{k}-\mathbf{q})\Big]    \Big)\\
&\qquad\qquad \times\frac{1}{w_{\perp}^2} (1- J_0(w_{\perp}\tau))J_0(|\mathbf{k}-\mathbf{q}|\tau)
\\
&+\frac{\tau^2}{k_{\perp}}  \int \frac{d^2\mathbf{q}}{(2\pi)^2}  \frac{d^2\mathbf{p}}{(2\pi)^2}  \frac{\mathbf{p}\cdot \mathbf{k}}{|\mathbf{q}-\mathbf{p}||\mathbf{k}-\mathbf{q}| p_{\perp}^2} \Big[ b_{\eta}(\mathbf{q}-\mathbf{p}),[b_{\perp}(\mathbf{p}), b_{\eta}(\mathbf{k}-\mathbf{q})]\Big] \\
&\qquad \qquad \times J_1(|\mathbf{q}-\mathbf{p}|\tau)J_0(p_{\perp}\tau)J_1(|\mathbf{k}-\mathbf{q}|\tau) \\
&+ \frac{\tau^2}{k_{\perp}}  \int \frac{d^2\mathbf{q}}{(2\pi)^2}  \frac{d^2\mathbf{p}}{(2\pi)^2} \frac{\mathbf{q}\cdot(\mathbf{q}-\mathbf{p}) (\mathbf{k}-\mathbf{q})\cdot \mathbf{k}}{|\mathbf{q}-\mathbf{p}|^2p_{\perp}^2 |\mathbf{k}-\mathbf{q}|^2} \Big[b_{\perp}(\mathbf{q}-\mathbf{p}), [b_{\perp}(\mathbf{p}), b_{\perp}(\mathbf{k}-\mathbf{q})]\Big] \\
&\qquad\qquad \times J_0(|\mathbf{q}-\mathbf{p}|\tau)J_0(p_{\perp}\tau)J_0(|\mathbf{k}-\mathbf{q}|\tau)\,.\\
\end{split}
\end{equation}
The product of three Bessel functions in the last two terms can be further reduced to a product of two Bessel functions using Graf's formula: 
\begin{equation}
\begin{split}
&J_1(|\mathbf{q}-\mathbf{p}|\tau)J_0(p_{\perp}\tau)J_1(|\mathbf{k}-\mathbf{q}|\tau) = \int_{-\pi}^{\pi} \frac{d\phi}{2\pi} \frac{|\mathbf{q}-\mathbf{p}|^2 -p_{\perp}^2 + w_{\perp}^2}{2|\mathbf{q}-\mathbf{p}| w_{\perp}} J_1(w_{\perp}\tau)J_1(|\mathbf{k}-\mathbf{q}|\tau),\\
&J_0(|\mathbf{q}-\mathbf{p}|\tau)J_0(p_{\perp}\tau)J_0(|\mathbf{k}-\mathbf{q}|\tau) = \int_{-\pi}^{\pi} \frac{d\phi}{2\pi} J_0(w_{\perp}\tau)J_0(|\mathbf{k}-\mathbf{q}|\tau).\\
\end{split}
\end{equation}

To solve for $\beta_{\perp}^{(5)}(\tau,\mathbf{k})$, we again follow the same procedure used when solving for $\beta_{\perp}^{(3)}(\tau,\mathbf{k})$. The formal solution is obtained by using the method of variation of parameters
\begin{equation}
\beta_{\perp}^{(5)}(\tau, \mathbf{k}) = D_1^{\prime} J_0(s) + D_2^{\prime} Y_0(s) + \frac{\pi}{2}\int_0^s dz (J_0(z) Y_0(s) -J_0(s)Y_0(z)) \frac{S_{\perp}^{(5)}(z)}{z}
\end{equation}
with the coefficients determined by the initial condtions at order-$g^5$
\begin{equation}
D_1^{\prime} = \beta_{\perp}^{(5)}(\tau=0, \mathbf{k}), \qquad D_2^{\prime} =0.
\end{equation}
The time-dependent factors in $S_{\perp}^{(5)}(\tau, \mathbf{k})$ can always be reduced to one single type of Bessel function, Bessel function of the first kind with order zero $J_0(c\tau)$, although different terms might have different values of the argument $c$. To be more precise, using Graf's formula, only two possibilities are involved
\begin{equation}
\begin{split}
&J_0(a\tau) J_0(b\tau) =  \int_{-\pi}^{\pi} \frac{d\psi}{2\pi} J_0(c\tau),\\
&J_1(a\tau)J_1(b\tau) = \int_{-\pi}^{\pi} \frac{d\psi}{2\pi} \cos\psi J_0(c\tau)\\
\end{split}
\end{equation}
with $c^2=a^2+b^2-2ab\cos\psi$. The next step is to use the formula ($\lambda = c/k_{\perp}$)
\begin{equation}\label{eq:J0(ctau)_replace}
\begin{split}
&\frac{\pi}{2}\int_0^s zdz (J_0(z) Y_0(s) -J_0(s)Y_0(z)) J_0(\lambda z)\\
 =& \frac{1}{1-\lambda^2}(J_0(\lambda s) -J_0(s))\\
 =&\frac{k_{\perp}^2}{k_{\perp}^2-c^2}(J_0(c \tau) -J_0(k_{\perp}\tau))\\
\end{split}
\end{equation}
It is clearly by now that the recipe is to replace time-dependent factors $J_0(a\tau)J_0(b\tau)$ and $J_1(a\tau)J_1(b\tau)$ in the source term $S_{\perp}^{(5)}(\tau, \mathbf{k})$ by 
\begin{equation}
\begin{split}
&J_0(a\tau)J_0(b\tau) \longrightarrow  \int_{-\pi}^{\pi} \frac{d\psi}{2\pi}\frac{1}{k_{\perp}^2-c^2}(J_0(c \tau) -J_0(k_{\perp}\tau))\\
&J_1(a\tau)J_1(b\tau) \longrightarrow  \int_{-\pi}^{\pi} \frac{d\psi}{2\pi}\frac{\cos\psi}{k_{\perp}^2-c^2}(J_0(c \tau) -J_0(k_{\perp}\tau))\\
\end{split}
\end{equation}
To be specific, we only need the following replacements in the source term 
\begin{equation}
\begin{split}
&J_1(q_{\perp}\tau) J_1(|\mathbf{k}-\mathbf{q}|\tau)  \longrightarrow  \int_{-\pi}^{\pi} \frac{d\phi'}{2\pi} \frac{\cos\phi'}{k_{\perp}^2 - w_{\perp}^{\prime 2}} (J_0(w'_{\perp}\tau) - J_0(k_{\perp}\tau)),\\
&J_0(q_{\perp}\tau) J_0(|\mathbf{k}-\mathbf{q}|\tau)\longrightarrow  \int_{-\pi}^{\pi} \frac{d\phi'}{2\pi} \frac{1}{k_{\perp}^2 - w_{\perp}^{\prime 2}} (J_0(w'_{\perp}\tau) - J_0(k_{\perp}\tau)),\\
&J_1(w_{\perp}\tau) J_1(|\mathbf{k}-\mathbf{q}|\tau)  \longrightarrow \int_{-\pi}^{\pi} \frac{d\theta}{2\pi} \frac{\cos\theta}{k_{\perp}^2 - u_{\perp}^{2}} (J_0(u_{\perp}\tau) - J_0(k_{\perp}\tau)),\\
&J_0(w_{\perp}\tau) J_0(|\mathbf{k}-\mathbf{q}|\tau)\longrightarrow \int_{-\pi}^{\pi} \frac{d\theta}{2\pi} \frac{1}{k_{\perp}^2 - u_{\perp}^{2}} (J_0(u_{\perp}\tau) - J_0(k_{\perp}\tau)),\\
&J_0( |\mathbf{k}-\mathbf{p}|\tau)\longrightarrow \frac{1}{k_{\perp}^2 - |\mathbf{k}-\mathbf{p}|^{2}} (J_0( |\mathbf{k}-\mathbf{p}|\tau) - J_0(k_{\perp}\tau)).\\
\end{split}
\end{equation}
The definitions of $w'_{\perp}, u_{\perp}, \phi^{\prime}, \theta$ are given in Eqs. \eqref{eqs:defs_theta_phiprime}.

The final expression for the order-$g^5$ solution $\beta_{\perp}^{(5)}(\tau, \mathbf{k})$ is 
\begin{equation}\label{eq:betaperp_g5}
\begin{split}
&\beta_{\perp}^{(5)}(\tau, \mathbf{k}) \\
&= \beta_{\perp}^{(5)}(\tau=0, \mathbf{k})J_0(k_{\perp}\tau) + i\frac{1}{k_{\perp}}\int \frac{d^2\mathbf{q}}{(2\pi)^2}  \frac{2(\mathbf{k}\times \mathbf{q})}{q_{\perp} |\mathbf{k}-\mathbf{q}|}\Big[2\beta^{(3)}(\tau=0,\mathbf{q}), b_{\eta}(\mathbf{k}-\mathbf{q})\Big]\\
&\qquad \times \int_{-\pi}^{\pi} \frac{d\phi'}{2\pi} \frac{\cos\phi'}{k_{\perp}^2 - w_{\perp}^{\prime 2}} (J_0(w'_{\perp}\tau) - J_0(k_{\perp}\tau))\\
&+i\frac{1}{k_{\perp}}\int \frac{d^2\mathbf{q}}{(2\pi)^2}\frac{(k_{\perp}^2+q_{\perp}^2-\mathbf{k}\cdot\mathbf{q})(2\mathbf{k}\times \mathbf{q})}{q_{\perp} |\mathbf{k}-\mathbf{q}|^2} [\beta_{\perp}^{(3)}(\tau=0, \mathbf{q}),b_{\perp}(\mathbf{k}-\mathbf{q})] \int_{-\pi}^{\pi} \frac{d\phi'}{2\pi} \frac{J_0(w'_{\perp}\tau) - J_0(k_{\perp}\tau)}{k_{\perp}^2 - w_{\perp}^{\prime 2}} \\
&-\frac{1}{k_{\perp}}\int\frac{d^2\mathbf{q}}{(2\pi)^2}\frac{d^2\mathbf{p}}{(2\pi)^2} \frac{2(\mathbf{k}\times \mathbf{q})(\mathbf{q}\times \mathbf{p})}{p_{\perp}^2|\mathbf{q}-\mathbf{p}|^2|\mathbf{k}-\mathbf{q}|}\Big[[b_{\perp}(\mathbf{p}), b_{\eta}(\mathbf{q}-\mathbf{p})], b_{\eta}(\mathbf{k}-\mathbf{q})\Big] \int_{-\pi}^{\pi}\frac{d\phi}{2\pi}\frac{\mathbf{q}\cdot(\mathbf{q}-2\mathbf{p})+w_{\perp}^2}{q_{\perp}^2-w_{\perp}^2}\\
&\quad\times \left\{\frac{1}{w_{\perp}}\int_{-\pi}^{\pi} \frac{d\theta}{2\pi} \frac{\cos\theta(J_0(u_{\perp}\tau) - J_0(k_{\perp}\tau))}{k_{\perp}^2 - u_{\perp}^{2}}  - \frac{1}{q_{\perp}}  \int_{-\pi}^{\pi} \frac{d\phi'}{2\pi} \frac{\cos\phi'(J_0(w'_{\perp}\tau) - J_0(k_{\perp}\tau))}{k_{\perp}^2 - w_{\perp}^{\prime 2}} \right\}\\
& +\frac{1}{k_{\perp}}\int \frac{d^2\mathbf{q}}{(2\pi)^2}\frac{d^2\mathbf{p}}{(2\pi)^2}\frac{(k_{\perp}^2+q_{\perp}^2-\mathbf{k}\cdot\mathbf{q})(\mathbf{k}\times \mathbf{q})( \mathbf{q}\times \mathbf{p})}{q^2_{\perp} |\mathbf{k}-\mathbf{q}|^2p^2_{\perp}|\mathbf{q}-\mathbf{p}|^2} \int_{-\pi}^{\pi}\frac{d\phi}{2\pi}  \frac{1}{q_{\perp}^2-w_{\perp}^2}\Big((w_{\perp}^2-p_{\perp}^2-|\mathbf{q}-\mathbf{p}|^2)\\
&\qquad\times\Big[[b_{\eta}(\mathbf{p}), b_{\eta}( \mathbf{q}-\mathbf{p})], b_{\perp}(\mathbf{k}-\mathbf{q})\Big]+2(\mathbf{p}\cdot\mathbf{q}-p_{\perp}^2-q_{\perp}^2) \Big[[b_{\perp}(\mathbf{p}), b_{\perp}(\mathbf{q}-\mathbf{p})], b_{\perp}(\mathbf{k}-\mathbf{q})\Big] \Big)\\
&\qquad \qquad\times\left\{\int_{-\pi}^{\pi} \frac{d\theta}{2\pi} \frac{J_0(u_{\perp}\tau) - J_0(k_{\perp}\tau)}{k_{\perp}^2 - u_{\perp}^{2}}  -  \int_{-\pi}^{\pi} \frac{d\phi'}{2\pi} \frac{J_0(w'_{\perp}\tau) - J_0(k_{\perp}\tau)}{k_{\perp}^2 - w_{\perp}^{\prime 2}}  \right\}\\
&+\frac{1}{k_{\perp}}\int \frac{d^2\mathbf{q}}{(2\pi)^2}\frac{d^2\mathbf{p}}{(2\pi)^2} \frac{(2\mathbf{k}-\mathbf{q})\cdot \mathbf{q}\,\mathbf{k}\cdot(\mathbf{k}-\mathbf{q}) \mathbf{q}\cdot(\mathbf{q}-2\mathbf{p})}{4p_{\perp}^2 |\mathbf{q}-\mathbf{p}|^2|\mathbf{k}-\mathbf{q}|^2q_{\perp}^2} \int_{-\pi}^{\pi} \frac{d\phi}{2\pi}\frac{1}{w_{\perp}^2}\Big((w_{\perp}^2-q_{\perp}^2+2\mathbf{p}\cdot(\mathbf{q}-\mathbf{p}))\\
& \qquad\times \Big[[b_{\eta}(\mathbf{p}), b_{\eta}(\mathbf{q}-\mathbf{p})],b_{\perp}(\mathbf{k}-\mathbf{q})\Big] + 2\mathbf{p}\cdot(\mathbf{q}-\mathbf{p})\Big[b_{\perp}(\mathbf{p}), b_{\perp}(\mathbf{q}-\mathbf{p})],b_{\perp}(\mathbf{k}-\mathbf{q})\Big]    \Big)\\
&\qquad \qquad\times\left\{\frac{J_0( |\mathbf{k}-\mathbf{p}|\tau) - J_0(k_{\perp}\tau)}{k_{\perp}^2 - |\mathbf{k}-\mathbf{p}|^{2}}  - \int_{-\pi}^{\pi} \frac{d\theta}{2\pi} \frac{J_0(u_{\perp}\tau) - J_0(k_{\perp}\tau)}{k_{\perp}^2 - u_{\perp}^{2}} \right\}
\\
&+\frac{1}{k_{\perp}}  \int \frac{d^2\mathbf{q}}{(2\pi)^2}  \frac{d^2\mathbf{p}}{(2\pi)^2}  \frac{\mathbf{p}\cdot \mathbf{k}}{|\mathbf{q}-\mathbf{p}||\mathbf{k}-\mathbf{q}| p_{\perp}^2} \Big[ b_{\eta}(\mathbf{q}-\mathbf{p}),[b_{\perp}(\mathbf{p}), b_{\eta}(\mathbf{k}-\mathbf{q})]\Big] \\
&\qquad\qquad  \times \int_{-\pi}^{\pi} \frac{d\phi}{2\pi} \frac{|\mathbf{q}-\mathbf{p}|^2 -p_{\perp}^2 + w_{\perp}^2}{2|\mathbf{q}-\mathbf{p}| w_{\perp}} \int_{-\pi}^{\pi} \frac{d\theta}{2\pi} \frac{\cos\theta}{k_{\perp}^2 - u_{\perp}^{2}} (J_0(u_{\perp}\tau) - J_0(k_{\perp}\tau))\\
& +\frac{1}{k_{\perp}}  \int \frac{d^2\mathbf{q}}{(2\pi)^2}  \frac{d^2\mathbf{p}}{(2\pi)^2}  \frac{\mathbf{q}\cdot(\mathbf{q}-\mathbf{p}) (\mathbf{k}-\mathbf{q})\cdot \mathbf{k}}{|\mathbf{q}-\mathbf{p}|^2p_{\perp}^2 |\mathbf{k}-\mathbf{q}|^2} \Big[b_{\perp}(\mathbf{q}-\mathbf{p}), [b_{\perp}(\mathbf{p}), b_{\perp}(\mathbf{k}-\mathbf{q})]\Big]\\
&\qquad\qquad  \times \int_{-\pi}^{\pi} \frac{d\phi}{2\pi} \int_{-\pi}^{\pi} \frac{d\theta}{2\pi} \frac{1}{k_{\perp}^2 - u_{\perp}^{2}} (J_0(u_{\perp}\tau) - J_0(k_{\perp}\tau))\,.\\
\end{split}
\end{equation}
It is apparent that the transverse gluon field at order-$g^5$ solely depends on one type of Bessel function $J_0(a\tau)$ although different terms  have different arguments. Color structure of the solution can be similarly analyzed as having done for $\beta^{(5)}(\tau, \mathbf{k})$.

\subsection{Solving for $\Lambda^{(5)}(\tau, \mathbf{k})$}
Unlike the solutions $\beta^{(5)}(\tau, \mathbf{k})$ and $\beta^{(5)}_{\perp}(\tau, \mathbf{k})$ , which are determined through the method of variation of parameters, the non-dynamical field  $\Lambda^{(5)}(\tau, \mathbf{k})$ is obtained by direct integration over the  source term $S_{\Lambda}^{(5)}(\tau, \mathbf{k})$. In the following, we reorganize the time-dependent factors in  $S_{\Lambda}^{(5)}(\tau, \mathbf{k})$ so that they only involve one type of Bessel function, Bessel function of the first kind with order one $J_1(c\tau)$.   The expression of $S_{\Lambda}^{(5)}(\tau, \mathbf{k})$ in Eq.~\eqref{eq:S_Lambda_g5} can be further expressed as
\begin{equation}
\begin{split}
S_{\Lambda}^{(5)}(\tau, \mathbf{k})
=&i\int\frac{d^2\mathbf{q}}{(2\pi)^2} \frac{\mathbf{q}\cdot (\mathbf{k}-\mathbf{q})}{q_{\perp}|\mathbf{k}-\mathbf{q}|} \left[\beta_{\perp}^{(3)}(\tau,\mathbf{q}), b_{\perp}(\mathbf{k}-\mathbf{q})\right] J_1(|\mathbf{k}-\mathbf{q}|\tau) \\
&+i\int\frac{d^2\mathbf{q}}{(2\pi)^2} \frac{\mathbf{q}\times \mathbf{k}}{|\mathbf{k}-\mathbf{q}|} \left[\Lambda^{(3)}(\tau,\mathbf{q}), b_{\perp}(\mathbf{k}-\mathbf{q})\right] J_1(|\mathbf{k}-\mathbf{q}|\tau)\\
& +i\int\frac{d^2\mathbf{q}}{(2\pi)^2}\frac{\mathbf{q}\cdot(\mathbf{k}-\mathbf{q})}{q_{\perp} |\mathbf{k}-\mathbf{q}|^2} [\partial_{\tau}\beta_{\perp}^{(3)}(\tau,\mathbf{q}), b_{\perp}(\mathbf{k}-\mathbf{q})] J_0(|\mathbf{k}-\mathbf{q}|\tau) \\
&+i\int\frac{d^2\mathbf{q}}{(2\pi)^2}\frac{\mathbf{q}\times \mathbf{k}}{|\mathbf{k}-\mathbf{q}|^2} [\partial_{\tau} \Lambda^{(3)}(\tau, \mathbf{q}), b_{\perp}(\mathbf{k}-\mathbf{q})] J_0(|\mathbf{k}-\mathbf{q}|\tau)\\
&-i\int\frac{d^2\mathbf{q}}{(2\pi)^2}[\tilde{\beta}^{(3)}(\tau, \mathbf{q}), b_{\eta}(\mathbf{k}-\mathbf{q})] J_2(|\mathbf{k}-\mathbf{q}|\tau)\\
&-i\int\frac{d^2\mathbf{q}}{(2\pi)^2}\frac{1}{|\mathbf{k}-\mathbf{q}|}[\tau \partial_{\tau}\beta^{(3)}(\tau,\mathbf{q}), b_{\eta}(\mathbf{k}-\mathbf{q})] J_1(|\mathbf{k}-\mathbf{q}|\tau)\,.
\end{split}
\end{equation}
We compute each term separately. The first and the third terms can be combined together 
\begin{equation}
\begin{split}
&i\int\frac{d^2\mathbf{q}}{(2\pi)^2} \frac{\mathbf{q}\cdot (\mathbf{k}-\mathbf{q})}{q_{\perp}|\mathbf{k}-\mathbf{q}|} \Big(\left[\beta_{\perp}^{(3)}(\tau,\mathbf{q}), b_{\perp}(\mathbf{k}-\mathbf{q})\right] J_1(|\mathbf{k}-\mathbf{q}|\tau) \\
&\qquad\qquad\qquad\qquad\qquad + \frac{1}{ |\mathbf{k}-\mathbf{q}|} [\partial_{\tau}\beta_{\perp}^{(3)}(\tau,\mathbf{q}), b_{\perp}(\mathbf{k}-\mathbf{q})] J_0(|\mathbf{k}-\mathbf{q}|\tau)\Big) \\
=&i\int\frac{d^2\mathbf{q}}{(2\pi)^2} \frac{\mathbf{q}\cdot (\mathbf{k}-\mathbf{q})}{q_{\perp}|\mathbf{k}-\mathbf{q}|^2} [\beta_{\perp}^{(3)}(\tau=0, \mathbf{q}), b_{\perp}(\mathbf{k}-\mathbf{q})] (|\mathbf{k}-\mathbf{q}|^2-q_{\perp}^2) \int_{-\pi}^{\pi} \frac{d\phi'}{2\pi} \frac{J_1(w_{\perp}^{\prime}\tau)}{w_{\perp}^{\prime}}\\
&+\int\frac{d^2\mathbf{q}}{(2\pi)^2} \frac{\mathbf{q}\cdot (\mathbf{k}-\mathbf{q})}{q^2_{\perp}|\mathbf{k}-\mathbf{q}|^2}\int \frac{d^2\mathbf{p}}{(2\pi)^2}\frac{ \mathbf{q}\times \mathbf{p}}{2p^2_{\perp}|\mathbf{q}-\mathbf{p}|^2}\int_{-\pi}^{\pi}\frac{d\phi}{2\pi}  \frac{1}{q_{\perp}^2-w_{\perp}^2}\Big(-(p_{\perp}^2+|\mathbf{q}-\mathbf{p}|^2-w_{\perp}^2)\\
&\qquad\times \Big[[b_{\eta}(\mathbf{p}), b_{\eta}( \mathbf{q}-\mathbf{p})], b_{\perp}(\mathbf{k}-\mathbf{q})\Big]+2(\mathbf{p}\cdot\mathbf{q}-p_{\perp}^2-q_{\perp}^2) \Big[[b_{\perp}(\mathbf{p}), b_{\perp}(\mathbf{q}-\mathbf{p})], b_{\perp}(\mathbf{k}-\mathbf{q})\Big] \Big)\\
&\qquad\qquad\times \left((|\mathbf{k}-\mathbf{q}|^2-w_{\perp}^2) \int_{-\pi}^{\pi}\frac{d\theta}{2\pi} \frac{J_1(u_{\perp}\tau)}{u_{\perp}}-(|\mathbf{k}-\mathbf{q}|^2-q_{\perp}^2) \int_{-\pi}^{\pi}\frac{d\phi'}{2\pi} \frac{J_1(w'_{\perp}\tau)}{w'_{\perp}}\right) \,.\\
\end{split}
\end{equation}
Here 
\begin{equation}
\begin{split}
&w'_{\perp} =\sqrt{q_{\perp}^2+|\mathbf{k}-\mathbf{q}|^2-2q_{\perp}|\mathbf{k}-\mathbf{q}|\cos\phi'},\\
 &u_{\perp}= \sqrt{ w_{\perp}^2+|\mathbf{k}-\mathbf{q}|^2- 2w_{\perp}|\mathbf{k}-\mathbf{q}|\cos\theta}.\\
 \end{split}
 \end{equation}
We have also used the identity  
 \begin{equation}
\int_{-\pi}^{\pi} \frac{d\phi}{2\pi} \frac{J_1(wz)}{w} = \frac{1}{a^2-b^2} \Big(aJ_0(bz)J_1(az) - bJ_0(az)J_1(bz)\Big)
\end{equation}
with $w=\sqrt{a^2+b^2-2ab\cos\phi}$.

The second and the fourth terms can be combined together
\begin{equation}
\begin{split}
&i\int\frac{d^2\mathbf{q}}{(2\pi)^2} \frac{\mathbf{q}\times \mathbf{k}}{|\mathbf{k}-\mathbf{q}|^2} \Big(\left[\Lambda^{(3)}(\tau,\mathbf{q}), b_{\perp}(\mathbf{k}-\mathbf{q})\right] J_1(|\mathbf{k}-\mathbf{q}|\tau)|\mathbf{k}-\mathbf{q}|\\
&\qquad\qquad\qquad\qquad+ [\partial_{\tau} \Lambda^{(3)}(\tau, \mathbf{q}), b_{\perp}(\mathbf{k}-\mathbf{q})] J_0(|\mathbf{k}-\mathbf{q}|\tau)\Big)\\
=&\int\frac{d^2\mathbf{q}}{(2\pi)^2} \frac{d^2\mathbf{p}}{(2\pi)^2}  \frac{\mathbf{q}\times \mathbf{k}}{|\mathbf{k}-\mathbf{q}|^2q_{\perp}^2} \frac{\mathbf{q}\cdot(\mathbf{q}-2\mathbf{p})}{4p_{\perp}^2 |\mathbf{q}-\mathbf{p}|^2}\int_{-\pi}^{\pi} \frac{d\phi}{2\pi}\frac{1}{w_{\perp}^2}\Big(2\mathbf{p}\cdot(\mathbf{q}-\mathbf{p}) \left[[b_{\perp}(\mathbf{p}), b_{\perp}(\mathbf{q}-\mathbf{p})],b_{\perp}(\mathbf{k}-\mathbf{q})\right]\\
&\qquad\qquad - (q_{\perp}^2-w_{\perp}^2-2\mathbf{p}\cdot(\mathbf{q}-\mathbf{p})) \left[[b_{\eta}(\mathbf{p}), b_{\eta}(\mathbf{q}-\mathbf{p})], b_{\perp}(\mathbf{k}-\mathbf{q})\right]  \Big)\\
&\qquad \times \left\{(w_{\perp}^2-|\mathbf{k}-\mathbf{q}|^2)\int_{-\pi}^{\pi}\frac{d\theta}{2\pi} \frac{J_1(u_{\perp}\tau)}{u_{\perp}}   + J_1(|\mathbf{k}-\mathbf{q}|\tau) |\mathbf{k}-\mathbf{q}|\right\}\,.
\end{split}
\end{equation}
Finally the fifth and the sixth terms  are combined
\begin{equation}
\begin{split}
&-i\int\frac{d^2\mathbf{q}}{(2\pi)^2}\Big([\tilde{\beta}^{(3)}(\tau, \mathbf{q}), b_{\eta}(\mathbf{k}-\mathbf{q})] J_2(|\mathbf{k}-\mathbf{q}|\tau)+\frac{1}{|\mathbf{k}-\mathbf{q}|}[\tau \partial_{\tau}\beta^{(3)}(\tau,\mathbf{q}), b_{\eta}(\mathbf{k}-\mathbf{q})] J_1(|\mathbf{k}-\mathbf{q}|\tau)\Big)\\
=&-i\int\frac{d^2\mathbf{q}}{(2\pi)^2}\frac{q_{\perp}^2-|\mathbf{k}-\mathbf{q}|^2}{q_{\perp}|\mathbf{k}-\mathbf{q}|} [2\beta^{(3)}(\tau=0,\mathbf{q}), b_{\eta}(\mathbf{k}-\mathbf{q})] \int _{-\pi}^{\pi}\frac{d\phi'}{2\pi} \frac{\cos\phi'}{w_{\perp}^{\prime}}J_1(w'_{\perp}\tau)\\
&+\int\frac{d^2\mathbf{q}}{(2\pi)^2}\int \frac{d^2\mathbf{p}}{(2\pi)}  \frac{\mathbf{q}\times \mathbf{p}}{p_{\perp}^2|\mathbf{q}-\mathbf{p}|^2}\Big[[b_{\perp}(\mathbf{p}), b_{\eta}(\mathbf{q}-\mathbf{p})], b_{\eta}(\mathbf{k}-\mathbf{q})\Big] \int_{-\pi}^{\pi}\frac{d\phi}{2\pi}\frac{\mathbf{q}\cdot(\mathbf{q}-2\mathbf{p})+w_{\perp}^2}{q_{\perp}^2-w_{\perp}^2}\\
&\qquad\ \times \left\{\frac{w_{\perp}^2-|\mathbf{k}-\mathbf{q}|^2}{w_{\perp}|\mathbf{k}-\mathbf{q}|} \int _{-\pi}^{\pi}\frac{d\theta}{2\pi} \frac{\cos\theta}{u_{\perp}}J_1(u_{\perp}\tau)-\frac{q_{\perp}^2-|\mathbf{k}-\mathbf{q}|^2}{q_{\perp}|\mathbf{k}-\mathbf{q}|} \int _{-\pi}^{\pi}\frac{d\phi'}{2\pi} \frac{\cos\phi'}{w_{\perp}^{\prime}}J_1(w'_{\perp}\tau)\right\}\,.
\end{split}
\end{equation}
We  used relation $J_2(z) = \frac{2}{z}J_1(z) -J_0(z)$ to obtain
\begin{equation}
\begin{split}
& \frac{1}{q_{\perp}}J_1(q_{\perp}\tau)J_2(|\mathbf{k}-\mathbf{q}|\tau)-\frac{1}{|\mathbf{k}-\mathbf{q}|} J_2(q_{\perp}\tau)J_1(|\mathbf{k}-\mathbf{q}|\tau)\\
=&\frac{1}{q_{\perp}}J_1(q_{\perp}\tau)J_0(|\mathbf{k}-\mathbf{q}|\tau) - \frac{1}{|\mathbf{k}-\mathbf{q}|} J_0(q_{\perp}\tau)J_1(|\mathbf{k}-\mathbf{q}|\tau)\\
=&\frac{1}{q_{\perp}}\int _{-\pi}^{\pi}\frac{d\phi'}{2\pi} \frac{ q_{\perp} -|\mathbf{k}-\mathbf{q}|\cos\phi'}{w_{\perp}^{\prime}} J_1(w'_{\perp}\tau)  - \frac{1}{|\mathbf{k}-\mathbf{q}|} \int _{-\pi}^{\pi}\frac{d\phi'}{2\pi} \frac{|\mathbf{k}-\mathbf{q}|- q_{\perp} \cos\phi'}{w_{\perp}^{\prime}} J_1(w'_{\perp}\tau)\\
=&\frac{q_{\perp}^2-|\mathbf{k}-\mathbf{q}|^2}{q_{\perp}|\mathbf{k}-\mathbf{q}|} \int _{-\pi}^{\pi}\frac{d\phi'}{2\pi} \frac{\cos\phi'}{w_{\perp}^{\prime}}J_1(w'_{\perp}\tau)\,.\\
\end{split}
\end{equation}
Using the explicit expression of $S^{(5)}_{\Lambda}(\tau,\mathbf{k})$, the solution $\Lambda^{(5)}(\tau, \mathbf{k})$ is 
\begin{equation}\label{eq:Lambda_g5_final_sol}
\begin{split}
&\Lambda^{(5)}(\tau, \mathbf{k})= -\frac{1}{k_{\perp}^2}\int_0^{\tau} d\tau' S^{(5)}_{\Lambda}(\tau', \mathbf{k})\\
=&i\int\frac{d^2\mathbf{q}}{(2\pi)^2} \frac{\mathbf{q}\cdot (\mathbf{k}-\mathbf{q})(|\mathbf{k}-\mathbf{q}|^2-q_{\perp}^2)}{q_{\perp}|\mathbf{k}-\mathbf{q}|^2}\left [\beta_{\perp}^{(3)}(\tau=0, \mathbf{q}), b_{\perp}(\mathbf{k}-\mathbf{q})\right]  \int_{-\pi}^{\pi} \frac{d\phi'}{2\pi} \frac{1}{w_{\perp}^{\prime 2}}(1-J_0(w_{\perp}^{\prime}\tau))\\
&-i\int\frac{d^2\mathbf{q}}{(2\pi)^2}\frac{q_{\perp}^2-|\mathbf{k}-\mathbf{q}|^2}{q_{\perp}|\mathbf{k}-\mathbf{q}|} \left[2\beta^{(3)}(\tau=0,\mathbf{q}), b_{\eta}(\mathbf{k}-\mathbf{q})\right] \int _{-\pi}^{\pi}\frac{d\phi'}{2\pi} \frac{\cos\phi'}{w_{\perp}^{\prime 2}}(1-J_0(w'_{\perp}\tau))\\
&+\int\frac{d^2\mathbf{q}}{(2\pi)^2}\frac{d^2\mathbf{p}}{(2\pi)^2} \frac{\mathbf{q}\cdot (\mathbf{k}-\mathbf{q})}{q^2_{\perp}|\mathbf{k}-\mathbf{q}|^2}\frac{ \mathbf{q}\times \mathbf{p}}{2p^2_{\perp}|\mathbf{q}-\mathbf{p}|^2}\int_{-\pi}^{\pi}\frac{d\phi}{2\pi}  \frac{1}{q_{\perp}^2-w_{\perp}^2}\Big(-(p_{\perp}^2+|\mathbf{q}-\mathbf{p}|^2-w_{\perp}^2)\\
&\qquad\qquad \Big[[b_{\eta}(\mathbf{p}), b_{\eta}( \mathbf{q}-\mathbf{p})], b_{\perp}(\mathbf{k}-\mathbf{q})\Big]+2(\mathbf{p}\cdot\mathbf{q}-p_{\perp}^2-q_{\perp}^2) \Big[[b_{\perp}(\mathbf{p}), b_{\perp}(\mathbf{q}-\mathbf{p})], b_{\perp}(\mathbf{k}-\mathbf{q})\Big] \Big)\\
&\qquad\times \left\{(|\mathbf{k}-\mathbf{q}|^2-w_{\perp}^2) \int_{-\pi}^{\pi}\frac{d\theta}{2\pi} \frac{1}{u^2_{\perp}}(1-J_0(u_{\perp}\tau))-(|\mathbf{k}-\mathbf{q}|^2-q_{\perp}^2) \int_{-\pi}^{\pi}\frac{d\phi'}{2\pi} \frac{1}{w_{\perp}^{\prime 2}}(1-J_0(w'_{\perp}\tau))\right\}\\
&+\int\frac{d^2\mathbf{q}}{(2\pi)^2} \frac{d^2\mathbf{p}}{(2\pi)^2} \frac{\mathbf{q}\times \mathbf{k}}{|\mathbf{k}-\mathbf{q}|^2q_{\perp}^2}  \frac{\mathbf{q}\cdot(\mathbf{q}-2\mathbf{p})}{4p_{\perp}^2 |\mathbf{q}-\mathbf{p}|^2}\int_{-\pi}^{\pi} \frac{d\phi}{2\pi}\frac{1}{w_{\perp}^2}\Big(2\mathbf{p}\cdot(\mathbf{q}-\mathbf{p})  \left[[b_{\perp}(\mathbf{p}), b_{\perp}(\mathbf{q}-\mathbf{p})],b_{\perp}(\mathbf{k}-\mathbf{q})\right] \\
&\qquad\qquad- (q_{\perp}^2-w_{\perp}^2-2\mathbf{p}\cdot(\mathbf{q}-\mathbf{p})) \left[[b_{\eta}(\mathbf{p}), b_{\eta}(\mathbf{q}-\mathbf{p})], b_{\perp}(\mathbf{k}-\mathbf{q})\right]  \Big)\\
&\qquad\qquad \times \left\{(w_{\perp}^2-|\mathbf{k}-\mathbf{q}|^2)\int_{-\pi}^{\pi}\frac{d\theta}{2\pi} \frac{1}{u^2_{\perp}}(1-J_0(u_{\perp}\tau))   + (1-J_0(|\mathbf{k}-\mathbf{q}|\tau) )\right\}\\
&+\int\frac{d^2\mathbf{q}}{(2\pi)^2}\frac{d^2\mathbf{p}}{(2\pi)}  \frac{\mathbf{q}\times \mathbf{p}}{p_{\perp}^2|\mathbf{q}-\mathbf{p}|^2}\Big[[b_{\perp}(\mathbf{p}), b_{\eta}(\mathbf{q}-\mathbf{p})], b_{\eta}(\mathbf{k}-\mathbf{q})\Big] \int_{-\pi}^{\pi}\frac{d\phi}{2\pi}\frac{\mathbf{q}\cdot(\mathbf{q}-2\mathbf{p})+w_{\perp}^2}{q_{\perp}^2-w_{\perp}^2}\\
&\quad \times \left\{\frac{w_{\perp}^2-|\mathbf{k}-\mathbf{q}|^2}{w_{\perp}|\mathbf{k}-\mathbf{q}|} \int _{-\pi}^{\pi}\frac{d\theta}{2\pi} \frac{\cos\theta}{u^2_{\perp}}(1-J_0(u_{\perp}\tau))-\frac{q_{\perp}^2-|\mathbf{k}-\mathbf{q}|^2}{q_{\perp}|\mathbf{k}-\mathbf{q}|} \int _{-\pi}^{\pi}\frac{d\phi'}{2\pi} \frac{\cos\phi'}{w_{\perp}^{\prime 2}}(1-J_0(w'_{\perp}\tau))\right\}\,.\\
\end{split}
\end{equation}

Let us  summarize and comment on the general procedures for solving the classical Yang-Mills equations perturbatively in the dilute-dense regime. At each fixed order $g^{2m+1}$, the dynamical fields $\beta^{(2m+1)}(\tau, \mathbf{k})$ and $\beta^{(2m+1)}_{\perp}(\tau, \mathbf{k})$ satisfy the inhomogeneous Bessel differential equations of orders one and zero, respectively. They are solved using the well-established method of variation of parameters. The success of this method  relies on recombining the time-dependent factors in the source terms $S_{\eta}^{(2m+1)}$ and $S_{\perp}^{(2m+1)}$ so that $S_{\eta}^{(2m+1)}$ only depends on $J_1(c\tau)$ and $S_{\perp}^{(2m+1)}$ only depends on $J_0(c\tau)$. Owing to the Graf' s formula, this is always possible.  
 As for $\Lambda^{(2m+1)}(\tau, \mathbf{k})$, it satisfies first order differential equation with source term $S_{\Lambda}^{(2m+1)}$. All one needs to do is express the time-dependent factors in $S_{\Lambda}^{(2m+1)}(\tau, \mathbf{k})$ in terms of Bessel functions $J_1(c\tau)$.  Each time Graf's formula is used, an extra angular integral is introduced. Unfortunately, the number of terms in the solutions increases dramatically as one goes to higher and higher perturbative orders and the problem quickly becomes unmanageable analytically.

\section{Discussions and Outlooks}
\label{sec:discussion_outlooks}
In this paper, we have presented the first step towards completing the calculations of  the first saturation corrections to physical observables in high energy proton-nucleus collisions. We solved the classical Yang-Mills equations in the dilute-dense regime beyond the leading order. We explicitly constructed the order-$g^3$ and order-$g^5$ solutions. 
The main results are presented in Eqs. \eqref{eq:final_sol_beta_g3}, \eqref{eq:Lambda_g3_final_sol}, \eqref{eq:beta_perp_g3_final_sol}, \eqref{eq:beta_g5_final_sol}, \eqref{eq:betaperp_g5} and \eqref{eq:Lambda_g5_final_sol}. The major mathematical technique that makes the analytic solutions possible is Graf's formula, which expresses product of two Bessel functions in terms of  an angular integral of one Bessel function. As a consistence check of our main results, when the target Wilson line $U(\mathbf{x})=1$, the gluon fields vanish $\beta(\tau, \mathbf{k})=0$ and $\beta_i(\tau, \mathbf{k})=0$, i.e.  there is no gluon production in the absence of scattering as expected. 

There are a few apparent features of  the solutions. First of all, the time dependent factors in the longitudinal gluon field
\begin{equation} 
\beta(\tau, \mathbf{k}) = \beta^{(1)}(\tau, \mathbf{k}) + \beta^{(3)}(\tau, \mathbf{k}) + \beta^{(5)}(\tau, \mathbf{k})+\ldots
\end{equation}
are uniquely determined by one single type of Bessel functions $J_1( a\tau)$ although the values of the argument $a$ might be different at different orders.   On the other hand, the time dependent factors in the transverse field 
\begin{equation}
\beta_i(\tau, \mathbf{k}) = \beta^{(1)}_i(\tau, \mathbf{k}) + \beta^{(3)}_i(\tau, \mathbf{k}) + \beta^{(5)}_i(\tau, \mathbf{k})+\ldots
\end{equation}
are completely determined by $J_0(a\tau)$ again with possible different arguments $a$.  

Second, the order-$g$ solutions $\beta^{(1)}(\tau, \mathbf{k}), \beta^{(1)}_i(\tau, \mathbf{k})$ do not involve mutual interactions of the ``valence'' color charges in the proton. Each color charge in the proton independently scatter on the target. On the other hand,  the order-$g^3$ solutions $\beta^{(3)}(\tau, \mathbf{k}), \beta^{(3)}_i(\tau, \mathbf{k})$ represent interactions of \textit{two color charges} in the proton. Their interactions can happen before or after the collisions with the target. The order-$g^5$ solutions $\beta^{(5)}(\tau, \mathbf{k}), \beta^{(5)}_i(\tau, \mathbf{k})$ represent interactions of \textit{three color charges} in the proton.  Their interactions likewise can happen before or after the collisions with the target. It can also be that two color charges interact with each other before scattering on the target and then interact with the third color charge only after the collisions with the target. 

Another important feature is related to the gauge dependence of the concepts of initial state effects and final state effects as defined on the basis of the field.  In the main context of the paper, the solutions are given in the initial time Coulomb sub gauge. In appendix \ref{ap:noncoulomb_solutions},  we present  the order-$g$ and order-$g^3$ solutions in the non-Coulomb sub gauge. These two sets of solutions are related by a gauge transformation.  Performing direct comparison of these two sets of solutions, would convince the reader that  some final state effects in the non-Coulomb subgauge become the initial state effects in the Coulomb subgauge.

On the other hand, physical observables are independent of gauge transformations. With the solutions of the classical Yang-Mills equations at hand, one can calculate several interesting physical quantities that can be constructed from the classical gluon fields. For example, the the energy-momentum tensor of the gluon fields produced in high energy pA collisions by
\begin{equation}
T^{\mu\nu}(\tau, \mathbf{x}) = F^{\mu\lambda}(\tau, \mathbf{x}) F_{\lambda}^{\,\,\,\,\nu}(\tau, \mathbf{x}) + \frac{1}{4} g^{\mu\nu}F^{\kappa\lambda}(\tau, \mathbf{x})F_{\kappa\lambda}(\tau, \mathbf{x}).
\end{equation}
Tracing over the color matrix is understood in the above definition like $AB=A^aB^a=\mathrm{Tr}(AB)$ with $a=1,\ldots N_c^2-1$. 

In the second paper of this series, we will calculate the single inclusive soft gluon production and double inclusive soft gluon production. These observables are constructed using the LSZ formula
\begin{equation}
\begin{split}
&\hat{a}_{\mathbf{p}}^{a \dagger}(\tau) = -i\tau \sqrt{\frac{\pi}{4}}\left( H_1^{(2)}(p_{\perp}\tau) \overleftrightarrow{\partial_{\tau}} \tilde{\beta}^a(\tau, \mathbf{p})\right),\\
&\hat{c}_{\mathbf{p}}^{a \dagger}(\tau) = - i\tau \sqrt{\frac{\pi}{4}}\left( H_0^{(2)}(p_{\perp}\tau) \overleftrightarrow{\partial_{\tau}} \beta_{\perp}^a(\tau, \mathbf{p})\right),
\end{split}
\end{equation}
and taking the limit $\tau\rightarrow \infty$. Here $H_{0}^{(2)}(x)$ and $H_{1}^{(2)}(x)$ are Hankel functions of second kind, order zero and order one, respectively. The left right derivative is defined as $f_1(x)\overleftrightarrow{\partial_x} f_2(x) = f_1(x) \partial_x f_2(x) - \partial_x f_1(x) f_2(x)$. Unlike the energy-momentum tensor which is gague invariant by construction, the single inclusive gluon production by $\hat{a}_{\mathbf{p}}^{a, \dagger}\hat{a}_{\mathbf{p}}^{a}+\hat{c}_{\mathbf{p}}^{a, \dagger}\hat{c}_{\mathbf{p}}^{a}$ are explicitly dependent on $\beta(\tau, \mathbf{k})$ and $\beta_{\perp}(\tau, \mathbf{k})$ which are gauge variant objects. Thus gauge invariance of the  single inclusive production will serve as  a non-trivial  of the derived solutions.

It should be mentioned that to solve the classical Yang-Mills equations in the \textit{dense-dense} regime, another semi-analytical approach was developed in Refs.~\cite{Fries:2006pv, Chen:2015wia}. In this method, the solutions are expanded as power series expansions in proper time $\tau$.  Although at each order in $\tau$, all the saturation effects are included, the solutions are only meaningful when the values of $\tau$ are small. This small-$\tau$ expansion method cannot be used to rigorously calculate the single inclusive gluon production by the LSZ formula which requires taking the $\tau\rightarrow \infty$ limit.  Our expansions in terms of coupling constant $g$ are valid for all the proper time but can only take into account the saturation effects order by order. Amusingly one can combine these two analytical approaches into a  double expansion in $\tau$ and $g$ which gives  a non-trivial insights into the dense-dense regime. We defer further discussion for a future publication.

\newpage
 \acknowledgments
 
 We thank A. Dumitru, A. Kovner, Yu. Kovchegov, M. Lublinsky,   L. McLerran, and R. Venugopalan for insightful discussions and collaboration on related projects. We thank H. Duan for his contribution in the exploratory stage of this project. 

We acknowledge support by the DOE Office of Nuclear Physics through Grant No. DE-SC0020081.

\appendix

\section{The non-dynamical field $\Lambda(\tau, \mathbf{x})$}
\label{sec:nondaynamical_Lambda}
In this appendix, we give a general proof that $\Lambda(\tau, \mathbf{x})$ is not a dynamical field. To be precise, we want to show that the second order differential equation for  $\Lambda(\tau, \mathbf{x})$ can be obtained from the first order constraint equation. The proof is done perturbatively using  induction. 
The perturbative expansions for the solutions are
\begin{equation}
\begin{split}
&\tilde{\beta}(\tau,\mathbf{x}) = \sum_{n=0}^{\infty}g^{2n+1} \tilde{\beta}^{(2n+1)}(\tau,\mathbf{x}),\\
&\tilde\beta_i(\tau, \mathbf{x}) = \sum_{n=0}^{\infty} g^{2n+1} \beta_i^{(2n+1)} (\tau, \mathbf{x}),\\
&\Lambda(\tau, \mathbf{x}) = \sum_{n=0}^{\infty} g^{2n+1} \Lambda^{(2n+1)}(\tau, \mathbf{x}). \\
\end{split}
\end{equation}
Substituting these expansions into the classical Yang-Mills equations in Eq.~\eqref{eq:ym_beta_betai_tilde}, 
at order-$g^{2N+1}$, the first order constraint equation is 
\begin{equation}\label{eq:constraint_order_2N+1}
\partial_{\tau} \partial^2 \Lambda^{(2N+1)} = i \sum_{M=0}^{N-1} [ \beta_i^{(2N-2M-1)}, \partial_{\tau} \beta_i^{(2M+1)}] + i \sum_{M=0}^{N-1}[ \tilde{\beta}^{(2N-2M-1)}, \partial_{\tau} \tilde{\beta}^{(2M+1)}] \,.
\end{equation}
The right hand side of this equation involves all the lower order solutions $\beta_i^{(K)}, \tilde{\beta}^{(K)}$ for $K=1, 3, \dots, 2N-1$.

The second order differential equation for the order-$g^{2N+1}$ field  is 
\begin{equation}\label{eq:secondorder_order_2N+1}
\begin{split}
&\tau^2 \partial_{\tau}^2 \partial^2 \Lambda^{(2N+1)} + \tau\partial_{\tau}\partial^2 \Lambda^{(2N+1)}\\
 = &i \tau^2 \sum_{M=0}^{N-1} [\tilde{\beta}^{(2N-2M-1)}, \partial^2 \tilde{\beta}^{(2M+1)}] + i\tau^2 \sum_{M=0}^{N-1} [\beta_i^{(2N-2M-1)}, (\partial^2 \delta_{ij}-\partial_i\partial_j) \beta_j^{(2M+1)}] \\
&+\tau^2 \sum_{M=0}^{N-1} \sum_{L=0}^{M-1} \partial_i [\tilde{\beta}^{(2N-2M-1)}, [\beta_i^{(2M-2L-1)}, \tilde{\beta}^{(2L+1)}]\\
&+ \tau^2 \sum_{M=0}^{N-1} \sum_{L=0}^{M-1} \partial_i [ \beta_j^{(2N-2M-1)}, [\beta_i^{(2M-2L-1)}, \beta_j^{(2L+1)}]]. 
\end{split}
\end{equation}
Our goal is  to prove that from the order-$g^{2N+1}$ constraint equation Eq.~\eqref{eq:constraint_order_2N+1}, using all the lower order equations of motion,  the second order differential equation Eq.~\eqref{eq:secondorder_order_2N+1} can be derived. 
Taking time derivative of Eq.~\eqref{eq:constraint_order_2N+1} w.r.t. $\tau$, one obtains
\begin{equation}\label{eq:constraint_equation_derivative}
\tau^2\partial^2_{\tau} \partial^2 \Lambda^{(2N+1)} = i \sum_{M=0}^{N-1} [ \beta_i^{(2N-2M-1)}, \tau^2\partial^2_{\tau} \beta_i^{(2M+1)}] + i \sum_{M=0}^{N-1}[ \tilde{\beta}^{(2N-2M-1)}, \tau^2\partial^2_{\tau} \tilde{\beta}^{(2M+1)}].
\end{equation}
Combining Eqs.~\eqref{eq:constraint_order_2N+1} and \eqref{eq:constraint_equation_derivative}, we  obtain
\begin{equation}\label{eq:substituteone}
\begin{split}
&\tau^2 \partial_{\tau}^2 \partial^2 \Lambda^{(2N+1)} + \tau\partial_{\tau}\partial^2 \Lambda^{(2N+1)}= i \sum_{M=0}^{N-1} [ \beta_i^{(2N-2M-1)}, \tau^2\partial^2_{\tau} \beta_i^{(2M+1)}+\tau\partial_{\tau} \beta_i^{(2M+1)}] \\
&\qquad + i \sum_{M=0}^{N-1}[ \tilde{\beta}^{(2N-2M-1)}, \tau^2\partial^2_{\tau} \tilde{\beta}^{(2M+1)}+\tau\partial_{\tau}\tilde{\beta}^{(2M+1)}]\\
\end{split}
\end{equation}
It contains lower order equations of motion. They are  (for $M=0, 1, \ldots, N-1$)
\begin{equation}
\begin{split}
&\tau^2\partial^2_{\tau} \tilde{\beta}^{(2M+1)}+\tau\partial_{\tau}\tilde{\beta}^{(2M+1)}= \tilde{\beta}^{(2M+1)} + \tau^2 \partial_i^2 \tilde{\beta}^{(2M+1)} \\
&\qquad -i \tau^2 \sum_{M'=0}^{M-1} \partial_i [\beta_i^{(2M-2M'-1)}, \tilde{\beta}^{(2M'+1)}] -i\tau^2 \sum_{M'=0}^{M-1} [\beta_i^{(2M-2M'-1)}, \partial_i \tilde{\beta}^{(2M'+1)}] \\
&\qquad -\tau^2 \sum_{M'=0}^{M-1}\sum_{N'=0}^{M'-1} [\beta_i^{(2M-2M'-1)}, [\beta_i^{2M'-2N'-1}, \tilde{\beta}^{2N'+1}]]\\
\end{split}
\end{equation}
and 
\begin{equation}
\begin{split}
&\tau^2\partial^2_{\tau} \beta_i^{(2M+1)}+\tau\partial_{\tau} \beta_i^{(2M+1)}\\
=&\tau^2 (\partial^2 \delta_{ij}-\partial_i \partial_j)\beta_j^{(2M+1)} + i\tau^2 \sum_{M'=0}^{M-1} [\tilde{\beta}^{(2M-2M'-1)}, \partial_i \tilde{\beta}^{(2M'+1)}]\\
& -i\tau^2 \sum_{M'=0}^{M-1} \partial_j [\beta_j^{(2M-2M'-1)}, \beta_i^{(2M'+1)}] -i\tau^2 \sum_{M'=0}^{M-1} [\beta_j^{(2M-2M'-1)}, \partial_j \beta_i^{(2M'+1)}-\partial_i \beta_j^{(2M'+1)}] \\
&+\tau^2 \sum_{M'=0}^{M-1}\sum_{N'=0}^{M'-1}[\tilde{\beta}^{(2M-2M'-1)}, [\beta_i^{(2M'-2N'-1)}, \tilde{\beta}^{(2N'+1)}]]\\
&+ \tau^2 \sum_{M'=0}^{M-1} \sum_{N'=0}^{M'-1}[ \beta_j^{(2M-2M'-1)}, [\beta_i^{(2M'-2N'-1)}, \beta_j^{(2N'+1)}]].\\
\end{split}
\end{equation}
Substituting these two equations in Eq.~\eqref{eq:substituteone}, we get
\begin{equation}\label{eq:explict_secondorderconstraint}
\begin{split}
&\tau^2 \partial_{\tau}^2 \partial^2 \Lambda^{(2N+1)} + \tau\partial_{\tau}\partial^2 \Lambda^{(2N+1)}\\
=& i \sum_{M=0}^{N-1} [ \beta_i^{(2N-2M-1)}, \tau^2\partial^2_{\tau} \beta_i^{(2M+1)}+\tau\partial_{\tau} \beta_i^{(2M+1)}] + i \sum_{M=0}^{N-1}[ \tilde{\beta}^{(2N-2M-1)}, \tau^2\partial^2_{\tau} \tilde{\beta}^{(2M+1)}+\tau\partial_{\tau}\tilde{\beta}^{(2M+1)}]\\
=& i \sum_{M=0}^{N-1} [ \beta_i^{(2N-2M-1)}, \tau^2(\partial^2\delta_{ij} -\partial_j\partial_j)\beta_j^{(2M+1)}]+i \sum_{M=0}^{N-1}[ \tilde{\beta}^{(2N-2M-1)}, \tau^2 \partial_i^2 \tilde{\beta}^{(2M+1)} ]\\
& -\tau^2 \sum_{M=0}^{N-1} \sum_{M'=0}^{M-1}[ \beta_i^{(2N-2M-1)}, [\tilde{\beta}^{(2M-2M'-1)}, \partial_i \tilde{\beta}^{(2M'+1)}]]\\
&+\tau^2\sum_{M=0}^{N-1} \sum_{M'=0}^{M-1} [ \beta_i^{(2N-2M-1)}, \partial_j [\beta_j^{(2M-2M'-1)}, \beta_i^{(2M'+1)}]] \\
&+ \tau^2 \sum_{M=0}^{N-1} \sum_{M'=0}^{M-1} [ \beta_i^{(2N-2M-1)},[\beta_j^{(2M-2M'-1)}, \partial_j \beta_i^{(2M'+1)}-\partial_i \beta_j^{(2M'+1)}]]\\
&+\tau^2 \sum_{M=0}^{N-1}\sum_{M'=0}^{M-1}[ \tilde{\beta}^{(2N-2M-1)}, \partial_i [\beta_i^{(2M-2M'-1)}, \tilde{\beta}^{(2M'+1)}]]\\
&+\tau^2\sum_{M=0}^{N-1}\sum_{M'=0}^{M-1}[ \tilde{\beta}^{(2N-2M-1)},[\beta_i^{(2M-2M'-1)}, \partial_i \tilde{\beta}^{(2M'+1)}]]\\
\end{split}
\end{equation}

First of all, we have used the fact that the following terms vanish
\begin{equation}
\sum_{M=0}^{N-1}[ \tilde{\beta}^{(2N-2M-1)},\tilde{\beta}^{(2M+1)}   ] = ([\tilde{\beta}, \tilde{\beta}])^{(2N)}=0
\end{equation}
\begin{align}
&\notag \sum_{M=0}^{N-1}\sum_{M'=0}^{M-1} \sum_{N'=0}^{M'-1}[\beta_i^{(2N-2M-1)}, [ \beta_j^{(2M-2M'-1)}, [\beta_i^{(2M'-2N'-1)}, \beta_j^{(2N'+1)}]]]\\ &=([\beta_i, [\beta_j, [\beta_i, \beta_j]]])^{(2N-2)} =0
\end{align}
\begin{equation}
\begin{split}
&\sum_{M=0}^{N-1} \sum_{M'=0}^{M-1}\sum_{N'=0}^{M'-1}[\beta_i^{(2N-2M-1)},[\tilde{\beta}^{(2M-2M'-1)}, [\beta_i^{(2M'-2N'-1)}, \tilde{\beta}^{(2N'+1)}]]] \\
&- \sum_{M=0}^{N-1}\sum_{M'=0}^{M-1}\sum_{N'=0}^{M'-1}[ \tilde{\beta}^{(2N-2M-1)}, [\beta_i^{(2M-2M'-1)}, [\beta_i^{2M'-2N'-1}, \tilde{\beta}^{2N'+1}]]]\\
=&([\beta_i, [\tilde{\beta}, [\beta_i, \tilde{\beta}]]] - [\tilde{\beta}, [\beta_i, [\beta_i, \tilde{\beta}]]])^{(2N-2)} =0
\end{split}
\end{equation}
Trivial algebra shows  that Eq.~\eqref{eq:explict_secondorderconstraint} reproduces Eq.~\eqref {eq:secondorder_order_2N+1}. Q.E.D.

\section{The method of variation of parameters}\label{ap:method_variation_parameter}
For the sake of completeness, we briefly review the method of solving inhomogeneous differential equations used in the main body of the paper. The method is often called variation of parameters.

For a general second order inhomogeneous differential equation 
\begin{equation}
a(x) y^{\prime\prime} + b(x) y^{\prime} + c(x) y = f(x), 
\end{equation}
there are four steps to obtain  its general solutions:
\begin{itemize}
\item find  two independent solutions $y_1(x)$ and $y_2(x)$ of the homogeneous equation 
\begin{equation}
a(x) y^{\prime\prime} + b(x) y^{\prime} + c(x) y=0\, .
\end{equation}
The general solution for the homogeneous equation is 
\begin{equation}
y_h(x) = C_1 y_1(x) + C_2y_2(x)\, , 
\end{equation}
where $C_1$ snd $C_2$ are coefficients to be fixed by the initial (and/or boundary ) conditions. 

\item  calculate the Wronskian of the two independent solutions $y_1(x), y_2(x)$,
\begin{equation}
W(x) = y_1(x) y_2^{\prime} (x) - y_1^{\prime}(x) y_2(x). 
\end{equation}

\item construct the particular solution of the inhomgeneous equation.
\begin{equation}
\begin{split}
y_p(x) =& y_1(x) \int^{x} \frac{y_2( z) (-f(z))}{a(z) W(z)} dz + y_2(x) \int^{x} \frac{y_1(z) f(z)}{a(z) W(z)} dz \\
=&\int^x \frac{ y_1(z) y_2(x) - y_1(x) y_2(z) }{W(z)} \frac{ f(z)}{a(z)} dz.\\
\end{split}
\end{equation}

\item finally, by adding together the general solution of the homogeneous equation $ y_h(x)$  and the particular solution of the inhomogeneous equation  $ y_p(x)$, obtain  the general solution for the inhomogeneous equation 
\begin{equation}
y(x) = C_1 y_1(x) + C_2y_2(x)+\int^x \frac{ y_1(z) y_2(x) - y_1(x) y_2(z) }{W(z)} \frac{ f(z)}{a(z)} dz.
\end{equation}

\end{itemize}

\section{Integrals for the products of two Bessel functions}\label{ap:Grafs_formula}
There are a few general identities expressing the product of two Bessel functions in terms of an integral of one Bessel function from Dixion and Farrar's paper in 1933 \cite{Dixon1933}. For our problem here, only integer orders of Bessel functions are involved. The more well-known Graf's formula \cite{Watson1995,Abramowitz1965} serves our purpose: 
\begin{equation}
e^{in\Psi} J_n(\omega) = \sum_{m=-\infty}^{\infty} J_{n+m}(Z) J_m(Z') e^{im\phi}
\end{equation}
where 
\begin{equation}
\omega = \sqrt{ Z^2 + Z^{\prime 2} -2ZZ'\cos{\phi}}\,.
\end{equation}
The angle $\Psi$ is defined by
\begin{equation}
\omega \cos{\Psi} = Z- Z'\cos{\phi},\qquad \omega \sin{\Psi} =  Z'\sin{\phi}.
\end{equation}
Integrating both sides of the Graf's formula using $\int_{-\pi}^{\pi} \frac{d\phi}{2\pi} e^{-im' \phi}$, one obtains
\begin{equation}
\int_{-\pi}^{\pi} \frac{d\phi}{2\pi} e^{-im' \phi} e^{in\Psi}  J_n(\omega) = J_{n+m'}(Z) J_{m'}(Z').
\end{equation}
This can be rewritten in the form which is more convenient for the applications  
\begin{equation}
J_{n+m}(az) J_m(bz) = \int_{-\pi}^{\pi} \frac{d\phi}{2\pi} e^{-im \phi} e^{in\Psi}  J_n(w z)
\end{equation}
In this case,  $w = \sqrt{b^2+a^2-2ab\cos{\phi}}$ is symmetric with respect to $a, b$. and  
\begin{equation}
e^{i\Psi} = \frac{(a-b\cos{\phi}) + i(b \sin{\phi})}{\sqrt{b^2+a^2-2ab\cos{\phi}}}.
\end{equation}

Here are a few examples that were used in the main body of the paper:
\begin{itemize}
\item [$\underline{J_0(Z)J_0(Z')}$:]

In this case $m=0, n=0$, thus we obtain 
\begin{equation}
J_0(az)J_0(bz) = \int_{-\pi}^{\pi} \frac{d\phi}{2\pi} J_0(wz).
\end{equation}

\item [$\underline{J_0(Z)J_1(Z')}$:]

Here we can choose between two combinations of $n$ and $m$. 
For $n=1, m=-1$, we obtain
\begin{equation}
J_0(az) J_1(bz) = -\int_{-\pi}^{\pi} \frac{d\phi}{2\pi} e^{i\phi} e^{i\Psi} J_1(wz)\,.
\end{equation}
For $m=0, n=1$,
\begin{equation}
J_1(az)J_0(bz) = \int_{-\pi}^{\pi} \frac{d\phi}{2\pi} e^{i\Psi} J_1(wz)\,.
\end{equation}
Therefore we have two integral representation for this product 
\begin{equation}
J_0(az) J_1(bz) = -\int_{-\pi}^{\pi} \frac{d\phi}{2\pi} e^{i\phi} e^{i\Psi} J_1(wz) = \int_{-\pi}^{\pi} \frac{d\phi}{2\pi} e^{i\Psi^{\prime}} J_1(wz)
\end{equation}
Here $\Psi^{\prime}$ is related to $\Psi$ by the exchange $a\leftrightarrow b$, 
\begin{equation}
e^{i\Psi^{\prime}} = \frac{(b-a\cos{\phi}) + i(a \sin{\phi})}{\sqrt{b^2+a^2-2ab\cos{\phi}}}.
\end{equation}
It can be easily checked that $\phi + \pi + \Psi = \Psi^{\prime}$.  Thus the two integral representations are equivalent. Additionally, since $w$ is an even function of $\phi$, only $\phi-$even part of the exponentials contributes: 
\begin{equation}
\begin{split}
J_0(az) J_1(bz) =&2 \int_{0}^{\pi} \frac{d\phi}{2\pi} \cos{\Psi^{\prime}} J_1(wz)\\
=&2\int_{|b-a|}^{|b+a|} d w\frac{ b^2-a^2+w^2}{b \sqrt{ ((a+b)^2-w^2)(w^2-(a-b)^2)}} J_1(w z).
\end{split}
\end{equation} 
In the second equality, we have changed the integration variable from $\phi$ to $w$
using 
\begin{equation}
\cos\phi = \frac{ a^2+b^2 -w^2}{2ab},
\end{equation}
\begin{equation}
d\phi = \frac{wdw}{ab \sin{\phi}} = \frac{2wdw}{\sqrt{ 4a^2b^2-(a^2+b^2-w^2)^2}}.
\end{equation}
\begin{equation}
\cos\Psi^{\prime} = \frac{b- a\cos{\phi}}{w} = \frac{b^2-a^2+w^2}{2bw}.
\end{equation}

\item [$\underline {J_1(Z)J_1(Z')}$:] 

Setting  $n=0, m=1$, we get 
\begin{equation}
J_1(az) J_1(bz) =  \int_{-\pi}^{\pi} \frac{d\phi}{2\pi} e^{-i \phi}J_0(wz).
\end{equation}
Alternatively  for $m=-1, n=0$,
\begin{equation}
J_1(az) J_1(bz) =  \int_{-\pi}^{\pi} \frac{d\phi}{2\pi} e^{i \phi}J_0(wz).
\end{equation}
These two expressions are equivalent. Since $w$ is even function of $\phi$, only the even part of $e^{i\phi}$ contributes to the integral:
\begin{equation}
J_1(az) J_1(bz) =  2\int_{0}^{\pi} \frac{d\phi}{2\pi} \cos{\phi}J_0(wz).
\end{equation}

On the other hand, for $n=2, m=-1$,
\begin{equation}
J_1(az) J_1(bz)= -\int_{-\pi}^{\pi} \frac{d\phi}{2\pi} e^{i \phi} e^{i2\Psi}  J_2(w z).
\end{equation}
The case of  $n=-2, m=1$ will yield the same expression. 
We can  further simplify to get 
\begin{equation}
J_1(az) J_1(bz)= -2\int_{0}^{\pi} \frac{d\phi}{2\pi} \cos{(\phi+2\Psi)}  J_2(w z).
\end{equation}

\item [$\underline{J_1(Z)J_2(Z')}$:] 

There are two ways to express $J_1(Z)J_2(Z')$ in terms of an integral of $J_1(Z'')$. 
For  $n=1, m=-2$ or $n=-1, m=2$, one obtains
\begin{equation}
J_1(az) J_2(bz) = -\int_{-\pi}^{\pi} \frac{d\phi}{2\pi} e^{ 2i\phi} e^{i\Psi} J_1(wx).
\end{equation}
For $n=1, m=1$ or $n=-1, m=-1$, 
\begin{equation}
J_2(az)J_1(bz) =  \int_{-\pi}^{\pi} \frac{d\phi}{2\pi} e^{ -i\phi} e^{i\Psi} J_1(wx).
\end{equation}
Again, it seems like we have two expressions
\begin{equation}
J_1(az) J_2(bz) = -\int_{-\pi}^{\pi} \frac{d\phi}{2\pi} e^{ 2i\phi} e^{i\Psi} J_1(wx) = \int_{-\pi}^{\pi} \frac{d\phi}{2\pi} e^{ -i\phi} e^{i\Psi'} J_1(wx),
\end{equation}
but they are equivalent due to  $\cos{(\phi-\Psi^{\prime})} = -\cos{(2\phi+\Psi)}$.

\end{itemize}

\section{Solutions in the non-Coulomb subgauge}\label{ap:noncoulomb_solutions}
In this appendix, we present the solutions of the Yang-Mills equation at order-$g$ and order-$g^3$ in the non-Coulomb sub-gauge that is defined  by $U(\mathbf{x})$. 
Initial conditions in this gauge has been discussed in Sec.~\ref{sec:subgauge_Ux}. 

Order-$g$ solutions
\begin{equation}
\begin{split}
&\zeta^{(1)}(\tau, \mathbf{k} ) = b_{\eta}(\mathbf{k}) \frac{J_1(k_{\perp}\tau)}{k_{\perp}\tau},\\
&\zeta^{(1)}_i (\tau, \mathbf{k}) = \frac{-i\epsilon^{il}\mathbf{k}_l}{k_{\perp}^2} b_{\perp}(\mathbf{k}) J_0(k_{\perp}\tau) - i\mathbf{k}_i \Lambda(\mathbf{k})\\
\end{split}
\end{equation}
with the initial conditions 
\begin{equation}
\begin{split}
&b_{\eta}(\mathbf{k}) = 2\zeta^{(1)}(\tau=0, \mathbf{k}),\\
&b_{\perp}(\mathbf{k}) = -i \epsilon^{ij}\mathbf{k}_i \zeta_j^{(1)}(\tau=0, \mathbf{k}),\\
&\Lambda(\mathbf{k}) =\frac{ik_i}{k^2} \zeta_i^{(1)}(\tau=0, \mathbf{k}).  \\
\end{split}
\end{equation}
Here $b_{\eta}(\mathbf{k})$ and $b_{\perp}(\mathbf{k})$ are exactly the same as those in Eq.~\eqref{eq:beta_bperp_coulombgauge}. In the non-Coulomb subgauge, $\partial_i \zeta_i \neq 0$, so $\Lambda(\mathbf{k})$ contributes to the initial condition. 

The order-$g^3$ solutions are
\begin{equation}\label{eq:zeta_g3_sol}
\begin{split}
\zeta^{(3)}(\tau,\mathbf{k}) 
=&2\zeta^{(3)}(\tau=0,\mathbf{k}) \frac{J_1(k_{\perp}\tau)}{k_{\perp}\tau}+i \int \frac{d^2\mathbf{p}}{(2\pi)} \Bigg( \Big[\Lambda(\mathbf{k}-\mathbf{p}), b_{\eta}(\mathbf{p})\Big]\left(\frac{J_1 (p_{\perp}\tau)}{p_{\perp}\tau} - \frac{J_1(k_{\perp}\tau)}{k_{\perp}\tau}\right)\\
&-\frac{\mathbf{k}\times \mathbf{p}}{p_{\perp}^2|\mathbf{k}-\mathbf{p}|^2}\Big[b_{\perp}(\mathbf{k}-\mathbf{p}), b_{\eta}(\mathbf{p})\Big]
\int_{-\pi}^{\pi}\frac{d\phi}{2\pi} \left(-1+ \frac{2\mathbf{k}\cdot\mathbf{p}}{k_{\perp}^2-w_{\perp}^2}\right) \left(\frac{J_1(w_{\perp}\tau)}{w_{\perp}\tau} - \frac{J_1(k_{\perp}\tau)}{k_{\perp}\tau}\right)\Bigg),
\end{split}
\end{equation}
\begin{equation}\label{eq:zeta_perp_g3_sol}
\begin{split}
\zeta^{(3)}_{\perp}(\tau, \mathbf{k}) 
 = &\zeta^{(3)}_{\perp}(\tau=0, \mathbf{k})J_0(k_{\perp}\tau) -\frac{i}{k_{\perp}} \int \frac{d^2\mathbf{p}}{(2\pi)^2}\int_{-\pi}^{\pi}\frac{d\phi}{2\pi} \frac{1}{k_{\perp}^2-w_{\perp}^2}
 \\ &\times
 \frac{( \mathbf{k}\times \mathbf{p})}{2p^2_{\perp}|\mathbf{k}-\mathbf{p}|^2}\Big((p_{\perp}^2 + |\mathbf{k}-\mathbf{p}|^2 -w_{\perp}^2) \Big[b_{\eta}(\mathbf{p}), b_{\eta}( \mathbf{k}-\mathbf{p})\Big] \\
 &+2(\mathbf{p}\cdot\mathbf{k}-p_{\perp}^2-k_{\perp}^2) \Big[b_{\perp}(\mathbf{p}), b_{\perp}(\mathbf{k}-\mathbf{p})\Big]\Big)\left( J_0(w_{\perp}\tau)-J_0(k_{\perp}\tau)\right)  \\
&+\frac{(\mathbf{k}\cdot \mathbf{p})}{ p_{\perp}^2}\Big[b_{\perp}(\mathbf{p}), \Lambda(\mathbf{k}-\mathbf{p})\Big] (J_0(p_{\perp}\tau)-J_0(k_{\perp}\tau)) \\
&+\frac{1}{2}(\mathbf{p}\times \mathbf{k}) \Big[\Lambda(\mathbf{p}), \Lambda(\mathbf{k}-\mathbf{p})\Big](1-J_0(k_{\perp}\tau)),\\
\end{split}
\end{equation}
\begin{equation}\label{eq:Lambda_g3_sol_coulomb}
\begin{split}
\Lambda^{(3)}(\tau, \mathbf{k}) 
=&\Lambda^{(3)}(\tau=0, \mathbf{k}) - \frac{i}{k_{\perp}^2} \int\frac{d^2\mathbf{p}}{(2\pi)^2}  \int_{-\pi}^{\pi} \frac{d\phi}{2\pi}\frac{\mathbf{k}\cdot(\mathbf{k}-2\mathbf{p})}{4w^2_{\perp}p_{\perp}^2 |\mathbf{k}-\mathbf{p}|^2}\\
&\Big(2\mathbf{p}\cdot(\mathbf{k}-\mathbf{p}) \left[b_{\perp}(\mathbf{p}), b_{\perp}(\mathbf{k}-\mathbf{p})\right] - (k_{\perp}^2-w_{\perp}^2-2\mathbf{p}(\mathbf{k}-\mathbf{p})) \left[b_{\eta}(\mathbf{p}), b_{\eta}(\mathbf{k}-\mathbf{p})\right]  \Big)\\
&\times (1- J_0(w_{\perp}\tau)) +\frac{\mathbf{p}\times\mathbf{k}}{ |\mathbf{k}-\mathbf{p}|^2}\left[\Lambda(\mathbf{p}), b_{\perp}(\mathbf{k}-\mathbf{p})\right] (1-J_0( |\mathbf{k}-\mathbf{p}|\tau)).\\
\end{split}
\end{equation}

Comparing with the solutions in the \textit{initial time Coulomb sub gauge} given by Eqs.~\eqref{eq:final_sol_beta_g3}, \eqref{eq:Lambda_g3_final_sol} and \eqref{eq:beta_perp_g3_final_sol}, the terms containing $\Lambda(\mathbf{p})$ in Eqs.~\eqref{eq:zeta_g3_sol}, \eqref{eq:zeta_perp_g3_sol}, \eqref{eq:Lambda_g3_sol_coulomb}, which represent final state interactions in the \textit{non-Coulomb sub gague}, are shifted to become initial state effects in Eqs.~\eqref{eq:final_sol_beta_g3}, \eqref{eq:Lambda_g3_final_sol} and \eqref{eq:beta_perp_g3_final_sol}

\bibliography{spires}

\end{document}